\documentclass[iop]{emulateapj}
\usepackage{hyperref}

\def\HII{{\ion{H}{2}}}

\def\4959_5007{[\ion{O}{3}]~$\lambda \lambda$4959,5007}
\def\OIII49595007{[\ion{O}{3}]~$\lambda \lambda 4959,5007$}
\def\ratioR23{([\ion{O}{2}]~$\lambda \lambda$3727,9 + [\ion{O}{3}]~$\lambda\lambda$4959,5007)/H$\beta$}
\def\R23{${\rm R}_{23}$}
\def\dS23{${\rm S}_{23}$}
\def\lOI{[\ion{O}{1}]~$\lambda$6300}

\def\lOIII{[\ion{O}{3}]~$\lambda$5007}
\def\lNII{[{\ion{N}{2}}]~$\lambda$6584}
\def\lSII{[\ion{S}{2}]~$\lambda \lambda$6717,31}

\def\NII{[{\ion{N}{2}}]}

\def\ratioS23{([\ion{S}{2}]~$\lambda \lambda$6717,31 +[\ion{S}{3}]~$\lambda\lambda$9069,9532)/H$\beta$}
\def\NIIHa{[\ion{N}{2}]/H$\alpha$}

\def\SII{[{\ion{S}{2}}]}

\def\Hb{{H$\beta$}}
\def\O4363{[{\ion{O}{3}}]~$\lambda$4363}
\def\OIII{[{\ion{O}{3}}]}

\def\Ha{{H$\alpha$}}

\newcommand{\lzifu} {{\scshape lzifu}}
\newcommand{\pywifes} {{\scshape pyWiFeS}}
\newcommand{\ppxf} {{\scshape ppxf}}
\newcommand{\PII} {{Paper~II}} 

\slugcomment{Accepted for publication in ApJSS on August 7, 2017}

\shorttitle{S7: Full Galaxy Survey Data Release}
\shortauthors{A. D. Thomas et al.}

\begin{document}


\title{Probing the Physics of Narrow Line Regions in Active Galaxies IV: \\ Full data release of The Siding Spring Southern Seyfert Spectroscopic Snapshot Survey (S7)}


\author{
Adam D. Thomas\altaffilmark{1},
Michael A. Dopita\altaffilmark{1},
Prajval Shastri\altaffilmark{2},
Rebecca Davies\altaffilmark{1,3},
Elise Hampton\altaffilmark{1},
Lisa Kewley\altaffilmark{1},
Julie Banfield\altaffilmark{1,4},
Brent Groves\altaffilmark{1},
Bethan L. James\altaffilmark{5},
Chichuan Jin\altaffilmark{3},
St\'ephanie Juneau\altaffilmark{6},
Preeti Kharb\altaffilmark{7},
Lalitha Sairam\altaffilmark{2},
Julia Scharw\"achter\altaffilmark{8},
P. Shalima\altaffilmark{9},
M. N. Sundar\altaffilmark{10},
Ralph Sutherland\altaffilmark{1} \&
Ingyin Zaw\altaffilmark{11}
}
\email{adam.thomas@anu.edu.au}

\altaffiltext{1}{RSAA, Australian National University, Cotter Road, Weston Creek, ACT 2611, Australia}
\altaffiltext{2}{Indian Institute of Astrophysics, Sarjapur Road, Bengaluru 560034, India}
\altaffiltext{3}{Max-Planck-Institut f\"ur Extraterrestrische Physik, Garching, Germany}
\altaffiltext{4}{ARC Centre of Excellence for All-Sky Astrophysics (CAASTRO)}
\altaffiltext{5}{Space Telescope Science Institute, 3700 San Martin Drive, Baltimore, MD 21218, USA}
\altaffiltext{6}{CEA-Saclay, DSM/IRFU/SAp, 91191 Gif-sur-Yvette, France}
\altaffiltext{7}{National Centre for Radio Astrophysics - Tata Institute of Fundamental Research, Pune University Campus, Post Bag 3, Ganeshkhind Pune 411007, India}
\altaffiltext{8}{Gemini Observatory, Northern Operations Center, 670 N. A’ohoku Place, Hilo, Hawaii 96720, USA}
\altaffiltext{9}{Regional Institute of Education, Manasagangotri, Mysore 570006, India}
\altaffiltext{10}{Jain University, 3rd Block Jayanagar, Bengaluru 560011, India}
\altaffiltext{11}{New York University (Abu Dhabi), 70 Washington Sq. S, New York, NY 10012, USA}

\begin{abstract}
We present the second and final data release of the \emph{Siding Spring Southern Seyfert Spectroscopic Snapshot Survey} (S7).  Data are presented for 63 new galaxies not included in the first data release, and we provide 2D emission-line fitting products for the full S7 sample of 131 galaxies.  The S7 uses the WiFeS instrument on the ANU 2.3m telescope to obtain spectra with a spectral resolution of $R = 7000$ in the red (540 - 700\,nm) and $R = 3000$ in the blue (350 - 570\,nm), over an integral field of $25 \times 38$~arcsec$^2$ with $1 \times 1$~arcsec$^2$ spatial pixels.  The S7 contains both the largest sample of active galaxies and the highest spectral resolution of any comparable integral field survey to date.  The emission-line fitting products include line fluxes, velocities and velocity dispersions across the WiFeS field of view, and an artificial neural network has been used to determine the optimal number of Gaussian kinematic components for emission-lines in each spaxel.  Broad Balmer lines are subtracted from the spectra of nuclear spatial pixels in Seyfert~1 galaxies before fitting the narrow lines.  We bin nuclear spectra and measure reddening-corrected nuclear fluxes of strong narrow lines for each galaxy.  The nuclear spectra are classified on optical diagnostic diagrams, where the strength of the coronal line [\ion{Fe}{7}]\,$\lambda6087$ is shown to be correlated with \OIII / \Hb.  Maps revealing gas excitation and kinematics are included for the entire sample, and we provide notes on the newly observed objects.
\end{abstract}

\keywords{galaxies:abundances, galaxies:active, galaxies:Seyfert,  galaxies:ISM, galaxies:jets, surveys}

\section{Introduction}
\label{sec:intro}

Supermassive black holes (SMBHs) with masses of ${\sim}10^6 - 10^9$~M$_\odot$ lie at the hearts of all massive galaxies.  The close relationship between host galaxy properties and the black hole mass \citep[e.g.][]{Magorrian_1998, Ferrarese_Merritt_2000_SMBH_sigma_scaling, McConnell_2013_SMBH_host_scaling} has been interpreted as implying that black holes and their host galaxies `co-evolve' \citep{Kormendy_Ho_2013_review}; the black holes affect their host galaxies when they are `active' as Active Galactic Nuclei (AGN), and the host galaxies control the supply of gas needed to grow the black hole during an active phase.  Activity in galactic nuclei is split into two regimes - bolometric luminosities below approximately 0.01 Eddington, associated with radio galaxies and low-luminosity AGN (LLAGN), and luminosities above 0.01 Eddington and approaching or even exceeding Eddington, associated with `radiative mode' Seyfert galaxies \citep{Heckman_Best_2014_AGN_review}.  In Seyfert galaxies an accretion disk around the central SMBH produces ionizing radiation that photoionizes the ISM in the immediate surrounds and potentially the galaxy at large.  The `duty cycle' of a Seyfert nucleus, or the fraction of the time it is active, is estimated to be ${\sim}5 - 10\%$ \citep[e.g.][]{Diamond-Stanic_Rieke_2012_Sy_sample}.  

Seyfert galaxies are observed to have a wide variety of emission features such as narrow and/or broad emission lines with varying polarization properties.  The apparent diversity of AGN is explained to some extent by the `unified model' of AGN \citep{Antonucci_1993_AGN_unified}.  This simple axially-symmetric geometrical model postulates that if the orientation of the axis of the system is close to being along the line of sight, then we are able to view the extremely Doppler-broadened emission lines from the broad-line region (BLR) and possibly the accretion disk emission - we observe a Seyfert~1 (Sy\,1).  On the other hand, if our line of sight is closer to equatorial, a `torus' of dusty material obscures these features, and instead we only observe the excited gas in clouds in the wider narrow-line region (NLR) - we observe a Seyfert~2 (Sy\,2).

The EUV radiation from the SMBH accretion structure that photoionizes the NLR during a Seyfert episode may ionize kiloparsec-scale regions of the host galaxy in an `extended NLR' (ENLR).  An ENLR often has the shape of a double `ionization cone'.  Additionally a Seyfert nucleus affects the ISM of the host galaxy by driving outflows into the surrounding material.  Measured emission-line flux ratios of Seyfert (E)NLRs occupy a limited region in the standard optical diagnostic diagrams \citep{BPT_1981, 1987VO}, which show the \lOIII /\Hb\ ratio against \lNII /\Ha , \lSII /\Ha\ and \lOI /\Ha.  The very similar line ratios between Seyfert galaxies occur because the ENLR is generally radiation-pressure dominated \citep{Dopita_2002_Dusty_NLRs, Groves_2004_NLR_I, Groves_2004_NLR_II}; an equilibration occurs due to radiation pressure compressing gas at the ionization front such that the local ionization parameter is held effectively constant and the optical emission-line spectrum is not strongly affected by the ionization parameter.

Approximately $10\%$ of galaxies in the local universe host a Seyfert nucleus \citep{Ho_2008_AGN_review}.  The most prevalent class of non-star forming nuclear activity in the local universe are the low-ionization nuclear emission-line regions (LINERs) \citep{Heckman_1980_LINERs}, which are detected in ${\sim}20-30\%$ of galaxies \citep{Ho_2008_AGN_review}.  LINER emission is often thought to be associated with low luminosity AGN activity \citep[e.g.][]{Storchi-Bergmann97, Eracleous01, Ulvestad01, Ho01}, but may also be associated with other physical processes including shock excitation \citep[e.g.][]{Heckman_1980_LINERs, Dopita95, Lipari04, Monreal-Ibero06, Rich11, Ho14} and photoionization from post-asymptotic giant branch stars \citep[e.g.][]{Binette94, Kehrig12, Belfiore_2016_MaNGA_LIERs}.

\subsection{Integral field spectroscopy of Seyferts and LINERs}

Spatially resolved spectroscopy of Seyfert and LINER hosts provides an important tool for understanding the effects of a Seyfert/LINER period of SMBH growth on the host galaxy.  Integral field data reveals information about the impact of an active nucleus on the host galaxy through the spatial distribution of emission-line flux ratios (indicative of shocks or photoionization), kinematic signatures evident in emission-line profiles, and stellar populations.

Recently \citet{Karouzos_2016_GMOS_outflows, Karouzos_2016_erratum} investigated AGN-driven outflows in a sample of 6 luminous Seyfert~2 galaxies observed with the Gemini Multi-Object Spectrograph (GMOS) on the Gemini South Telescope.  The authors identify outflowing emission-line gas and kinematically measure the size of the outflows.

\citet{Ricci_2014_IFU_I, Ricci_2014_IFU_II, Ricci_2015_IFU_II_erratum, Ricci_2015_IFU_III} study a sample of 10 AGN in early-type galaxies using GMOS.  The authors use PCA Tomography and emission-line fitting to investigate LINER emission and the circum-nuclear gas excitation and morphology in their sample.

A recent example of a large sample of galaxies studied with optical IFU data is the work presented in \citet{Rich_2014_WiFeS_GOALS_shocks, Rich_2015_WiFeS_GOALS}.  In this work the authors study a sample of 27 nearby luminous and ultra-luminous infrared galaxies ((U)LIRGs) using the WiFeS instrument on the ANU 2.3~m telescope.  The authors find that the proportion of the observed H$\alpha$ emission-line flux attributable to shock excitation may increase to as much as one half in late-stage mergers.  Three-quarters of the objects in their sample would appear to have an AGN contribution to gas excitation in nuclear spectra using a na\"{\i}ve classification, due the strong effect of shock excitation on emission-line ratios.

A sample of 17 luminous local Seyfert~2 galaxies was studied with IFU data by \citet{McElroy_2015_IFU} using the SPIRAL fiber array instrument on the Anglo-Australian Telescope.  The authors find that observed emission line profiles and flux ratios are suggestive of outflows and shocks being driven by the Seyfert nuclei.

An ongoing project to observe a large sample of Seyfert galaxies with IFUs is the Close AGN Reference Survey (CARS) \citep{Rothberg_2015_CARS,McElroy_2016_CARS, Husemann_2016_CARS_MARK1018}, which will consist of 39 local Sy\,1 galaxies at $0.01 < z < 0.06$.  This project uses the Multi Unit Spectroscopic Explorer (MUSE) instrument on the Very Large Telescope (VLT) to obtain optical integral field data over a $1' \times 1'$ field of view, and aims at a multi-wavelength study of unobscured AGN in the nearby Universe.

The VIRUS-P Exploration of Nearby Galaxies (VENGA) survey is an ongoing integral field survey of 30 nearby spiral galaxies, using the VIRUS-P IFU (an array of 4\farcs24 optical fibres with a $1\farcm7 \times 1\farcm7$ field of view) on the Harlan J. Smith  telescope  at  McDonald  Observatory \citep{Blanc_2013_VENGA}.  The VENGA data has a spectral resolution of ${\sim}5$\,\AA\ over a $3600 - 6800$\,\AA\ range and the sample includes the famous AGN NGC\,1068.

Observations of AGN using NIR IFUs on 8\,m-class telescopes are able to probe Seyfert galaxies at $z \sim 2$.  \citet{Forster_Schreiber_2014_SINS_AGN} attribute strong nuclear outflows in eight massive $z \sim 2$ star-forming galaxies to AGN.  The authors use data from the SINS/zC-SINF survey \citep{Forster_Schreiber_2009_SINS, Mancini_2011_zC_SINF}, which uses the Spectrograph for INtegral Field Observations in the Near Infrared (SINFONI) instrument on Unit Telescope~4 at the VLT.

Another IFU providing very high spatial resolution in the IR is the  OH-Suppressing Infra-Red Imaging Spectrograph (OSIRIS) mounted on Keck~1.  The ongoing Keck OSIRIS Nearby AGN (KONA) survey \citep{Hicks_2016_KONA} aims to observe the nuclei of 40 Seyfert galaxies to probe molecular and ionized gas flows and stellar dynamics in the immediate environs of the SMBHs.  

The Near-Infrared Integral Field Spectrometer (NIFS) on the Gemini North telescope has been used to similarly investigate gas feeding and feedback in local Seyfert galaxies.  Flux distributions and kinematics for both H$_2$ and ionized gas, including both forbidden and recombination lines, give information about the complex environment of AGN.  Data for various Seyfert galaxies provide evidence of molecular inflows traced by H$_2$, ionized outflows traced by [\ion{Fe}{2}] and lines from other ionized species, as well as the morphological relationship between radio outflows and optical/IR line emission \citep[e.g.][]{Riffel_2008, Riffel_2009, Storchi-Bergmann_2009, Storchi-Bergmann_2010, Riffel_Storchi-Bergman_2011a, Riffel_Storchi-Bergman_2011b, Riffel_2013}.


\subsection{Large integral field galaxy surveys}

The astronomical community is currently entering a golden age of integral field unit (IFU) galaxy surveys.  It is now routine to study IFU data for samples of dozens of galaxies, and samples of hundreds of galaxies are becoming commonplace.  In this section we survey the largest samples of galaxies that are now being studied with IFU data.  The studies described above occur in the context of these ongoing next-generation massive IFU surveys, which do not target AGN specifically, but will nevertheless contain large numbers of AGN.

The Calar Alto Legacy Integral Field Area (CALIFA) survey \citep{Sanchez_2012_CALIFA_survey, Sanchez_2016_CALIFA_DR3} is a recently completed integral field survey of ${\sim}600$ local galaxies with $0.005 < z < 0.03$.  The observations with the 3.5~m telescope at the Calar Alto observatory used a hexagonal optical fiber bundle consisting of 331 fibers each with a 2.7 arcsec diameter.  The subsample presented in the second data release \citep{Garcia-Benito_2015_CALIFA_DR2} of 257 galaxies contained at least 12 Seyfert galaxies and 48 LINER galaxies based on optical diagnostic diagram classifications of 3~arcsec-aperture nuclear spectra \citep{Singh_2013_CALIFA_LINERs}.  The CALIFA data has a spectral resolution of $R = 850$ and $R = 1650$ in two different optical configurations, with the majority of galaxies being observed with both configurations.

\citet{Singh_2013_CALIFA_LINERs} use CALIFA data to study a sample of 48 galaxies with LINER-like emission.  The authors show that radial emission-line surface brightness profiles are not consistent with photoionization by AGN illumination and argue that extended LINER-like emission is widely attributable to ionizing radiation from post-AGB stars.

The SAMI Galaxy Survey \citep{Bryant_2015_SAMI, Sharp_2015_SAMI, Allen_2015_SAMI_EDR} is an ongoing survey of ${\sim}2000 - 3000$ galaxies at $z < 0.12$.  The survey uses the Sydney-AAO Multi-object Integral field spectrograph (SAMI) \citep{Croom_2012_SAMI} on the Anglo-Australian Telescope (AAT) to observe 12 galaxies simultaneously using fiber bundles (`hexabundles').  Assuming that ${\sim}10\%$ of galaxies in the local universe are Seyferts \citep{Ho_2008_AGN_review}, we expect that the SAMI sample will ultimately contain approximately 200-300 Seyfert galaxies.  The SAMI data has a spectral resolution of $R = 1700$ in the blue (3700 -- 5700\,\AA) and $R = 4500$ in the red (6300 -- 7400\,\AA).

Recently \citet{Ho_2016_SAMI_winds} used a sample of 40 edge-on disk galaxies from the SAMI Galaxy Survey to explore the nature of excited extraplanar gas.  The authors used both gas kinematics and emission-line flux ratios to classify spectra as being from galaxies dominated by diffuse ionized gas through to those dominated by galactic winds.  The wind galaxies were found to have enhanced star formation rate (SFR) surface densities and to have had bursts of star formation in the recent past.

Data from the SAMI Galaxy Survey was also recently used to study two AGN in which the narrow emission lines were at a different line-of-sight velocity to the systemic velocity \citep{Allen_2015_offset_AGN}.  The authors concluded that understanding the cause of kinematically offset emission lines requires integral field spectroscopy, and that the observed kinematics were not due to binary SMBHs.

The Mapping Nearby Galaxies at Apache Point Observatory (MaNGA) survey \citep{Bundy_2015_MaNGA_overview} is one component of the ongoing fourth iteration of the Sloan Digital Sky Survey (SDSS-IV), and aims to study $10^4$ nearby galaxies using a multi-IFU instrument also making use of fibre bundles.  Again assuming that ${\sim}10\%$ of galaxies in the local universe are Seyferts, the MaNGA data will ultimately contain on the order of 1000 Seyferts.  The spectral resolution of the MaNGA data ranges from a minimum of $R \sim 1400$ in the blue to a maximum of $R \sim 2600$ in the red.

Recently \citet{Belfiore_2016_MaNGA_LIERs} used MaNGA data for 646 galaxies to study low-ionization emission-line regions (LIERs), finding that this emission occurs not just in the centers of galaxies where it is associated with LINERs, but that it also occurs extended throughout galaxies.  The authors conclude that LIER emission is powered by post-AGB stars since the line emission follows the continuum surface brightness and the EW(H$\alpha$) is nearly constant for LIER emission.

Large IFU galaxy surveys are also now being performed in the IR.  An ongoing high-redshift IFU galaxy survey is KMOS$^\mathrm{3D}$, which has a target sample of over 600 galaxies at $0.7 < z < 2.7$ and uses the K-band Multi-Object Spectrograph (KMOS) on the VLT \citep{Wisnioski_2015_KMOS3D}.  Another high-redshift survey using KMOS is the KMOS Redshift One Spectroscopic Survey (KROSS), which aims to observe 795 typical star-forming galaxies at $z = 0.8 - 1.0$ \citep{Stott_2016_KROSS}.  The KMOS$^\mathrm{3D}$ survey includes AGN in its target catalogue, but AGN were explicitly excluded from the KROSS survey.

We refer the interested reader to Section~6 in \citet{Bundy_2015_MaNGA_overview} for a detailed comparison of the largest optical integral field galaxy surveys.

Clearly, studying large samples of galaxies with IFUs is now the norm as opposed to the exception in optical astronomy.

\subsection{The S7}

The \emph{Siding Spring Southern Seyfert Spectroscopic Snapshot Survey} (S7) \citep[][hereafter \PII]{Dopita_2015_S7_II} was performed over 2013 - 2016 using the Wide Field Spectrograph (WiFeS) on the ANU 2.3~m telescope at Siding Spring Observatory.  The WiFeS instrument is mounted on a Nasmyth platform and has a field of view of 38~$\times$~25~arcsec$^2$ composed of a grid of 1~$\times$~1~arcsec$^2$ spatial pixels (`spaxels').  The instrument design and performance are described in \citet{Dopita_2007_WiFeS_I} and \citet{Dopita_2007_WiFeS_II} respectively.

The S7 is an important addition to the IFU surveys discussed above because of its high spectral resolution, large number of active galaxies (${\sim}130$), and high physical spatial resolution due to the very low redshift of the sample.  The spectral resolution is $R = 7000$ in the red (FWHM $\sim 40$\,km\,s$^{-1}$ over $540 - 700$\,nm) and $R = 3000$ in the blue (FWHM $\sim 100$\,km\,s$^{-1}$ over $350 - 570$\,nm).  The high spectral resolution allows studies of independent velocity components in emission lines, and the high spatial resolution allow separation of regions ionized by the AGN, star formation and shocks, as well as analysis of the morphology of ENLRs.

The S7 sample was selected from the \citet{Veron_2006_12ed, Veron_2010_13ed} catalogs of AGN.  The galaxies were mostly selected to be radio-detected, nearby, southern, and away from the plane of the Galaxy.  More information on the sample is provided in Section~\ref{sec:S7_sample}.


A number of studies utilizing S7 data have already been published or are nearly complete. In \citet{Dopita_2014_S7_I} (hereafter Paper~I), the authors used the optical AGN ENLR and \HII\ region spectra of the nearby spiral galaxy NGC~5427 \mbox{(z = 0.009)} to constrain the shape of the AGN EUV spectrum, the mass of the black hole and the bolometric luminosity of the AGN. The EUV spectra of AGN encode important information about AGN accretion physics, the mechanical energy input of the central AGN, and the formation of broad- and narrow-line regions, but the EUV spectrum may not be observed directly. The relative intensities of the emission lines in optical ENLR spectra are sensitive to the shape of the EUV spectrum but are also strongly dependent on the chemical abundances. The authors of Paper~I measured the abundances of individual \HII\ regions in the disk of NGC~5427 using the ratios of strong emission lines, calculated the gas phase metallicity gradient and subsequently determined the nuclear metallicity to be approximately three times the solar value (3~$Z_\odot$). The \textsc{MAPPINGS IV} photoionization code was then used to generate a grid of emission line ratios for a 3~$Z_\odot$ nebula illuminated by model Seyfert radiation fields for a range of black hole masses and Eddington ratios. The ENLR spectrum of NGC~5427 is best fit by a model with a black hole mass of \mbox{5$\times$10$^7$ M$_\odot$} and a bolometric luminosity of \mbox{$\log(L_{bol})$ = 44.3 $\pm$ 0.1 erg s$^{-1}$} (approximately $10\%$ of the Eddington luminosity).

\citet{Thomas_2016_OXAF} develop a new model of the ionizing EUV spectrum from Seyfert galaxies, {\sc oxaf}, designed for photoionization modeling of NLR clouds.  The authors explore how the spectral-shape parameters of the ionizing spectrum affect predicted emission-line ratios.  Predicted line ratios are shown to be consistent with `pure-AGN' Seyfert line ratios, with an example being the `pure-AGN' line-ratios at the top of the `mixing sequence' between `pure-\HII' and `pure-AGN' line-ratios on a standard optical diagnostic diagram of S7 data for the Seyfert galaxy NGC~1365.

Although the majority of the S7 galaxies are Seyferts, there are also a number of interesting LINER galaxies in the sample.  The authors of \citet{Dopita_2015_S7_III} (Paper~III) combined S7 integral field data with \textit{Hubble Space Telescope} Faint Object Spectrograph (HST~FOS) spectroscopy to conduct a detailed study of the dynamical structure and the origin of the LINER emission in the nearby galaxy NGC~1052 \mbox{(z = 0.005)}. Along the major axis of the galaxy they find evidence for a dusty, turbulent accretion flow forming a small-scale accretion disk. Along the minor axis of the galaxy buoyant bubbles of plasma are rising into the extended interstellar medium, forming a large scale outflow and ionization cone. Part of the outflow region is ionized by the AGN, but there are clear signatures of shock excitation in the inner accretion disk and the region surrounding the radio jet. The emission-line properties of NGC~1052 can be modeled with a `double shock' in which the accretion flow first passes through an accretion shock (with a velocity of 150~km~s$^{-1}$) in the presence of hard X-ray radiation, and the accretion disk is then processed by a pair of cocoon shocks (with velocities of 200-300~km~s$^{-1}$) driven by the ram pressure of the radio jet. The emission-line model is sensitive to the ratio of the cocoon shock emission to the accretion shock emission but is virtually insensitive to the shock velocities. This type of model may therefore offer a generic explanation for the excitation of LINER galaxies.

\citet{Davies16} used S7 data to investigate the relative significance of gas pressure and radiation pressure across AGN ENLRs. The accretion energy of a Seyfert nucleus couples to the gas in the host galaxy primarily through radiation pressure that can drive galaxy-scale outflows, potentially removing large quantities of molecular gas from the galaxy disk. \citet{Davies16} showed that there exist two distinct types of AGN fraction sequences (`starburst-AGN mixing sequences') on standard emission line ratio diagnostic diagrams. In some galaxies, radiation pressure is dominant across the ENLR and the ionization parameter remains constant. In other galaxies, radiation pressure is important in the nuclear regions but gas pressure becomes dominant at larger radii as the ionization parameter in the ENLR decreases. Where radiation pressure is dominant, the AGN regulates the density of the interstellar medium on kiloparsec scales and may therefore have a direct impact on outflows and/or star formation far beyond the zone of gravitational influence of the black hole. Both radiation pressure and gas pressure dominated ENLRs are dynamically active with evidence for outflows, indicating that radiation pressure may be an important source of feedback even when it is not dominant over the entire ENLR.

The spectra of Seyfert galaxies are often dominated by strong emission lines that contain information about the rates of star formation and AGN accretion and the physical properties of the ionized gas. However, these properties are degenerate with the relative contributions of different ionization mechanisms to the line emission. \citet{Davies16b} demonstrated how integral field data of Seyfert galaxies can be used to spatially and spectrally separate emission associated with star formation and AGN activity. The spatially resolved spectra of the S7 galaxies NGC~5728 and NGC~7679 form clear starburst-AGN mixing sequences on standard emission line ratio diagnostic diagrams. The emission line luminosities of the majority of the spectra along each mixing sequence can be reproduced by linear combinations of the star formation and AGN dominated `basis spectra' that form the endpoints of the mixing sequence. \citet{Davies16b} separate the \Ha, \Hb, \NII, \SII\ and \OIII\ luminosities of all spectra along each mixing sequence into contributions from star formation and AGN activity. The SFRs and AGN bolometric luminosities calculated from the decomposed flux measurements are mostly consistent with independent estimates from data at other wavelengths. The recovered star forming and AGN components also have distinct spatial distributions that trace structures seen in high resolution imaging of the galaxies. 

A very similar technique (varying only in the method for selecting the basis spectra) can be used to separate emission associated with star formation, shock excitation and AGN activity in galaxies where all three ionization mechanisms are prominent. Davies~et~al.~(2016c, submitted) isolate the contributions of star formation, shock excitation and AGN activity to the strong emission lines across the central \mbox{3$\times$3 kpc$^2$} region of the nearby spiral galaxy NGC~613 (\mbox{z = 0.005}). The star formation component traces the B-band stellar continuum emission, and the AGN component forms an ionization cone. The SFR and AGN bolometric luminosity calculated from the decomposed emission line maps are consistent with independent estimates from data at other wavelengths. The shock component traces the outer boundary of the AGN ionization cone and may be produced by outflowing material interacting with the surrounding interstellar medium. Our new decomposition technique makes it possible to determine the properties of star formation, shock excitation and AGN activity from optical spectra, without contamination from other ionization mechanisms.

We have carried out radio observations with the aim of searching for kiloparsec-scale radio outflows in Seyfert galaxies and probing their interactions with ENLR clouds.  A sub-sample of 38 S7 Seyfert and LINER galaxies was observed at 610 and 1390~MHz with the Giant Metrewave Radio Telescope (GMRT), and a subsample of 6 galaxies was observed at 5 and 9~Ghz with the Australia Telescope Compact Array (ATCA). The GMRT targets lie within the declination range of $\pm$10 degrees, not readily accessible to ATCA, and have integrated flux densities greater than 20~mJy at 1.4~GHz. The sources span the redshift range of 0.004\,--\,0.02, which translates to spatial scales of $\sim$200 to $\sim$750~parsec at the GMRT resolution of $\sim$2~arcsec at 1.4~GHz. We have detected kiloparsec-scale radio structures (KSRs) in a majority of the GMRT sub-sample; this is a much higher fraction than previously observed in the literature. The lower frequency observations at 610~MHz are also resulting in detections of host galaxy disk emission in a much larger fraction of targets compared to observations at 1.4 and 5~GHz in the literature. The 610\,--\,1390~MHz spectral index images are providing useful and unique information about contributions from the AGN and star formation to the radio emission. Results from this study will be presented in a forthcoming paper by Kharb~et~al.~(2017).

Recently \citet{Scharwachter_2016_NGC6300} gave an update on the S7 and presented a mosiac of 6 WiFeS pointings on the Seyfert galaxy NGC\,6300.  Emission lines in the ENLR showed a correlation between \NIIHa\ and line width.

Future work on S7 data will include a study of the forbidden high-ionization emission lines observed in the nuclei of some galaxies in the sample, among other projects.

In this work we present data for the full S7 sample including value-added products, to assist the astronomical community in taking advantage of the enormous potential of the data set.

In Section~\ref{sec:S7_sample} we describe the properties of the full S7 sample, including the global properties of the galaxies.  In Section~\ref{sec:obs_DR} we describe the observations and data reduction.  In Section~\ref{sec:products} we present the S7 data products that we are making available to the community.  In Section~\ref{sec:nuc_spec_results} we discuss S7 nuclear spectra and the related emission-line analyses and AGN classifications.  Section~\ref{sec:spatial_properties} presents maps of the spatial properties of the S7 galaxies over the observed field of view, showing the gas ionization and kinematics.  In Section~\ref{sec:individual_objects} we describe basic properties evident in the S7 data for individual galaxies not already discussed in \PII.  Our conclusions are presented in Section~\ref{sec:conclusions}.

\vskip 1cm
\section{The S7 sample}
\label{sec:S7_sample}

\begin{figure}
\begin{centering}
\includegraphics[scale=0.46]{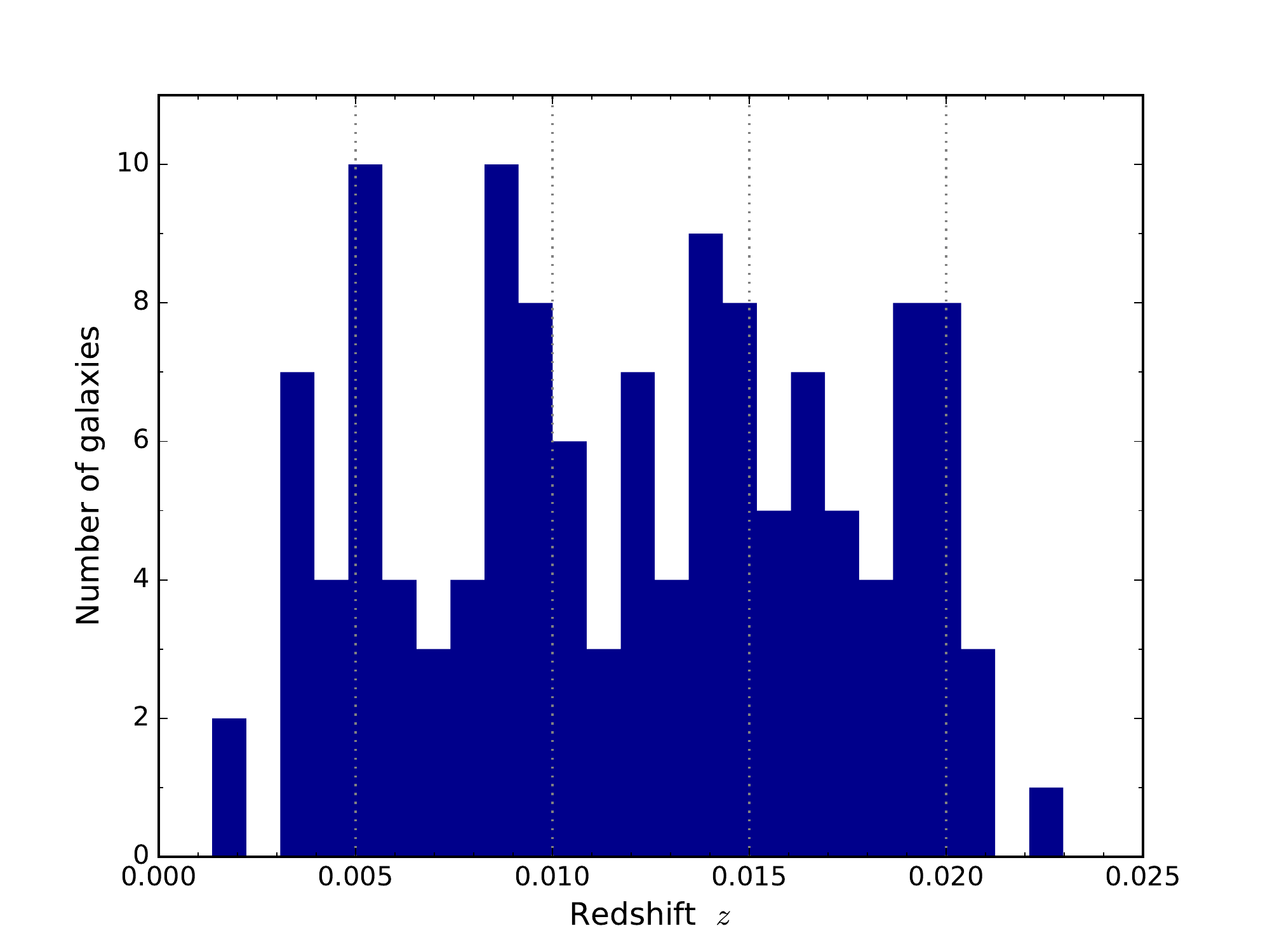}
\end{centering}
\caption{The distribution of redshifts of the galaxies in the S7 sample.  The QSO PKS~0056-572 at $z = 0.679$ is excluded.\\}\label{fig:redshifts}
\end{figure}

The S7 sample consists of 131 nearby active galaxies, mostly with detected 20~cm radio emission.  The sample was selected from the AGN catalog of \citet{Veron_2006_12ed, Veron_2010_13ed}, to fulfill the following criteria: 
\begin{itemize}
\item Declination below +10$^\circ$, to ensure acceptable zenith distances.  Atmospheric dispersion differs by nearly 6~arcsec between 3500 and 7000~\AA\ at a zenith distance of 60$^\circ$, which is a substantial fraction of the WiFeS field of view.  The highest declination in the sample is $\delta = +8.9^\circ$ for NGC~7469.
\item Galactic latitude more than 20$^\circ$ away from zero, to avoid severe extinction and a high density of foreground stars along the plane of the Galaxy.  This restriction was relaxed for 18 galaxies in the sample that were known to have an interesting ENLR.  Of these, only the four galaxies NGC~6300 ($b = -14.05^\circ$), NGC~6221 ($b = -9.57^\circ$), ESO~138-G01 ($b = -9.44^\circ$) and ESO~138-G34 ($b = -7.10^\circ$) have $|b| < 15^\circ$.
\item A measured radio flux density at 20~cm of larger than approximately 20~mJy from \citet{Veron_2006_12ed, Veron_2010_13ed} for most galaxies with declinations above $\delta = -40$.  The V{\'e}ron-Cetty \& V{\'e}ron data is from the NRAO VLA Sky Survey \citep[NVSS;][]{Condon_1998_NVSS_survey} or from the Faint Images of the Radio Sky at Twenty centimeters (FIRST) survey \citep{Becker_1995_FIRST}, and is available only for objects with $\delta \gtrsim -40$.  Of the 131 S7 galaxies, 104 have a declinaton above -40, and of these, 88 have a measured radio flux density of larger than 20~mJy.  Of the 27 galaxies with declination below -40, only two had 20~cm radio fluxes available in the V{\'e}ron-Cetty \& V{\'e}ron catalog.
\item Redshift $z \lesssim 0.02$, to achieve a physical spatial resolution of ${\sim} 400$~pc~arcsec$^{-1}$ or better.  This spatial resolution allows detailed spatial analysis of the ENLR.  An additional consequence of the close redshift cutoff is that the diagnostic [\ion{S}{2}] doublet is below the limit of the $R = 7000$ red WiFeS grating.  The distribution of redshifts in the sample is shown in Figure~\ref{fig:redshifts}.  Apart from the $z = 0.679$ QSO PKS 0056-572 (included in the sample due to an incorrect redshift in the V{\'e}ron-Cetty \& V{\'e}ron catalog), there are 7 galaxies in the sample with $z > 0.02$, with NGC~5252 having the largest redshift of $z = 0.023$.  The galaxies with $z > 0.02$ were included either due to imprecise redshifts in the V{\'e}ron-Cetty \& V{\'e}ron catalog or as filler targets.
\end{itemize}

In this work we present data cubes for 63 new galaxies.  Data cubes for 68 of the 131 S7 galaxies were already presented in \PII.  Additionally in this work we provide emission-line fitting products and nuclear spectra for all 131 galaxies.

\subsection{Global galaxy properties of the S7 sample}
\label{sec:global_sample_props}

The full S7 sample of 131 galaxies consists of approximately 18 Seyfert~1 galaxies, 77 Seyfert~2 galaxies, 27 LINER galaxies, 11 galaxies showing signs of gas excitation by star formation only, at least 4 galaxies showing strong post-starburst signatures, 1 radio elliptical galaxy showing little line emission (3C\,278), and one Seyfert~1 QSO (PKS\,0056-572).  At least 34 of the Seyfert~2 galaxies also show signs of star formation in either nuclear spectra or in other parts of the galaxy.  The classifications of the galaxies based upon S7 nuclear spectra are presented in Section~\ref{sec:nuc_spec_results}.

Morphological classifications from HyperLeda suggest that the S7 sample consists of approximately 71 spiral galaxies, 35 lenticular galaxies, 19 ellipticals and an irregular galaxy (ESO\,350-IG38).  A small number of S7 galaxies are undergoing interactions, for example both galaxies in the interacting pair NGC\,833 and NGC\,835 are included in the sample.

Stellar masses and SFRs were estimated for the S7 galaxies using IR photometry.  Stellar masses were calculated from WISE W1 (3.4~$\mu$m) luminosities using a constant mass-to-light ratio of 0.7, as described in \citet{Cluver_2014_GAMA_MIR}. The SFRs were calculated using the 60 and 100~$\mu$m flux densities, using the formula from \citet{Helou_1988_IRAS} to convert the flux densities into the FIR flux, the conversion of FIR luminosity to total IR luminosity from \citet{Calzetti_2000_Dust} ($L_{\rm IR} = 1.75 \times L_{\rm FIR}$), and the IR SFR calibration from \cite{Kennicutt_1998_SFRs}.

\begin{figure}
\begin{centering}
\includegraphics[width=0.49\textwidth]{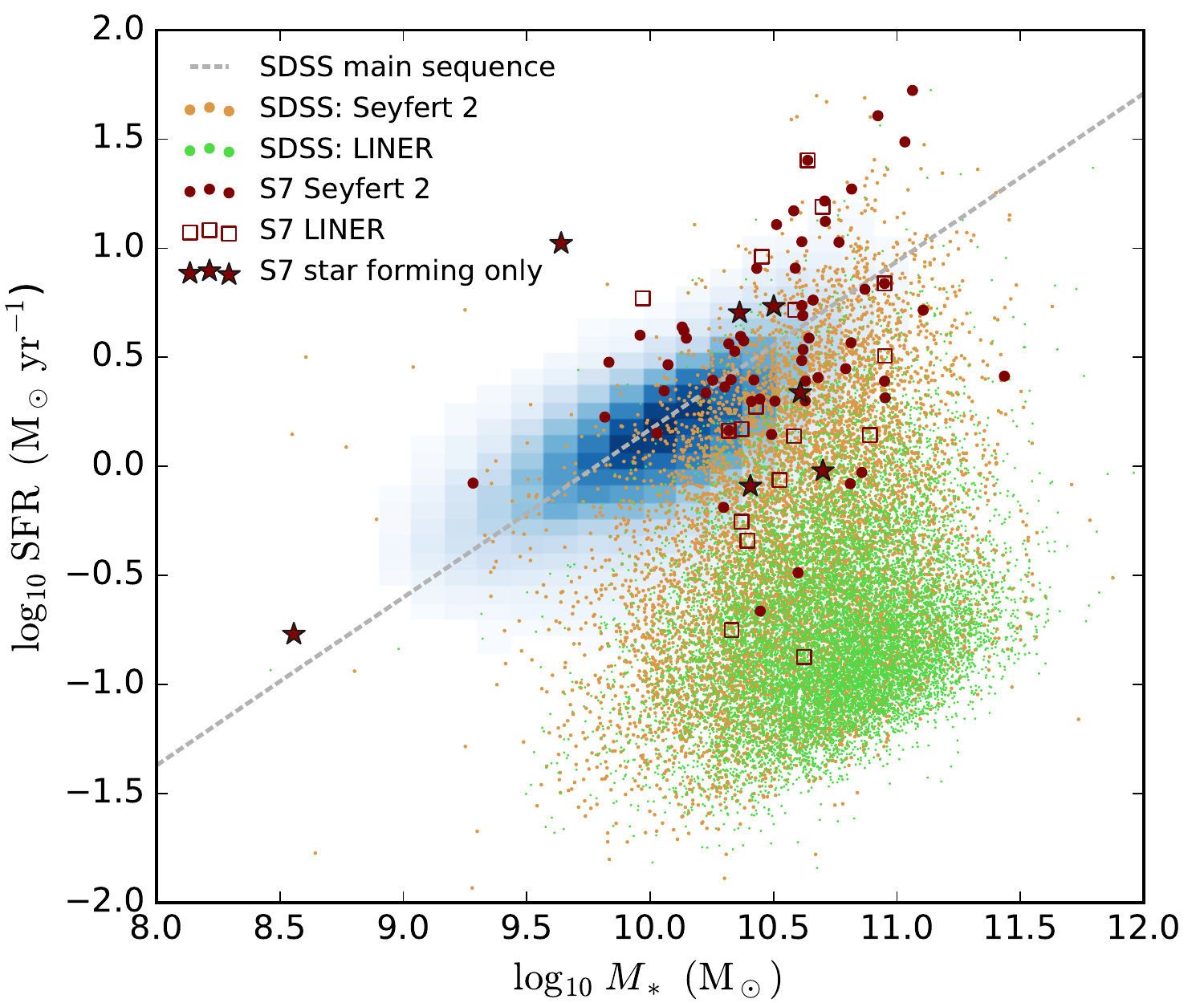}
\end{centering}
\caption{The position of the S7 galaxies relative to the star-forming main sequence.  The SDSS classifications are from \citet{Leslie_2016_MS}.  The blue grid shows the SDSS star-forming galaxies, and the dotted line shows the best-fit SDSS main sequence for $z \sim 0$ star-forming galaxies from Equation~5 in \citet{Elbaz_2007_MS}.}\label{fig:nuc_SFR_Mstar}
\end{figure}

Figure~\ref{fig:nuc_SFR_Mstar} shows where the S7 galaxies lie relative to the star-forming main sequence of local galaxies.  The S7 galaxies lie scattered along the top end of the main sequence and extend through the `green valley' -- the location of a large proportion of Seyfert~2 galaxies in the local universe -- and into the `red cloud', where a large proportion of local LINERs are found.  The S7 `star-forming only'-classified galaxies lie approximately along the main sequence, as expected.  The LINER and Seyfert S7 galaxies that lie above the SDSS main sequence in Figure~\ref{fig:nuc_SFR_Mstar} are likely galaxies in which the IR luminosity is significantly enhanced by AGN contamination.

\vskip 2cm
\section{Observations and data reduction}
\label{sec:obs_DR}

In this section we describe the observations and reduction of the S7 data.

\subsection{Observations}
\label{sec:observations}

Galaxies were observed with a single pointing centered on the nucleus, with the exception of NGC\,5252 and NGC\,5643, for which the observations consisted of two mosaiced pointings with the mosaic centered on the nucleus.  The observing strategy generally involved alternating between sky and object targets, with three exposures being taken on the object then only one exposure for the corresponding sky.  The observing plans were designed so that the sky frames from both before and after the three exposures on a galaxy target could usually be used for sky subtraction when reducing the galaxy data.  Spectrophotometric calibration was performed using spectrophotometric standard stars\footnote{\url{http://www.mso.anu.edu.au/~bessell/FTP/Bohlin2013/GO12813.html}} observed at the beginning and the end of each night (conditions permitting); additionally telluric standard stars with smooth spectra observed once or twice a night were used to better correct for telluric absorption features.

The observing log for the 63 new galaxies not presented in \PII\ is presented in Table~\ref{table:Obs}.  This table supplements Table~1 in \PII; the full table is available as part of the catalogue provided in this data release.  Table~\ref{table:Obs} shows the galaxies arranged in order of RA, showing the galaxy coordinates, redshift, classifications, observing date, WiFeS rotator PA, total exposure time, seeing FWHM, and a note if a galaxy was observed in non-photometric conditions.  Sixteen galaxies in the full sample were observed more than once, but only the data for the `best' observation is presented here.  The `best' observation was chosen based on a combination of the seeing, the total exposure time, and whether or not the conditions were photometric.

\begin{deluxetable*}{lcccccccllc}
	\centering
	\tabletypesize{\scriptsize}
	\tablecaption{Log of the new S7 observations \label{table:Obs}}
	\tablewidth{525pt}
	\tablehead{
		\colhead{Object} & \colhead{RA}       & \colhead{Dec}      &  \colhead{$z$} & \colhead{Type$^{\rm a}$} & \colhead{Type}    & \colhead{PA}          & \colhead{ Date of}     & \colhead{Exposure} & \colhead{Seeing}\\
		\colhead{Name}   & \colhead{ (J2000)} & \colhead{ (J2000)} &  \colhead{}    & \colhead{}                & \colhead{(Veron)} & \colhead{$(^\circ)$} & \colhead{ Observation} & \colhead{Time}     & \colhead{FWHM}\\
		\colhead{}       & \colhead{}         & \colhead{}         &  \colhead{}    & \colhead{}                & \colhead{}        & \colhead{}            & \colhead{}             & \colhead{(s)}      & \colhead{(arcsec)}
	}
	\startdata
	MARK938 &   00 11 06.60 &  -12 06 27.00 & 0.0196 &             SB + Sy 2 &      S2 &                 0 &            2013-08-03 &                     3000 &             2.0 \\
	ESO350-IG38 &   00 36 52.90 &  -33 33 29.16 & 0.0206 &                    SB &      H2 &                90 &            2014-08-22 &                     2400 &             1.6 \\
	NGC1068 &   02 42 40.70 &  +00 00 47.16 & 0.0038 &                  Sy 2 &     S1h &                45 &            2014-08-23 &                     2400 &             1.2 \\
	IC1858 &   02 49 08.50 &  -31 17 21.84 & 0.0202 &                    SB &      S3 &                 0 &            2014-08-23 &                     2400 &             1.2 \\
	NGC1194 &   03 03 49.10 &  -01 06 13.00 & 0.0136 &             SB + Sy 2 &    S1.9 &               135 &            2014-08-22 &                     2400 &             1.5 \\
	NGC1217 &   03 06 05.90 &  -39 02 09.96 & 0.0210 &                 LINER &      S3 &                45 &            2016-02-09 &                     2700 &             1.5 \\
	NGC1266 &   03 16 00.70 &  -02 25 37.92 & 0.0072 &                 LINER &      S? &                90 &            2014-08-24 &                     2700 &             1.0 \\
	NGC1320 &   03 24 48.70 &  -03 02 32.00 & 0.0089 &                  Sy 2 &      S2 &               315 &            2014-08-24 &                     2700 &             1.0 \\
	NGC1346 &   03 30 13.20 &  -05 32 35.00 & 0.0135 &                    SB &      S1 &                90 &            2013-11-02 &                     3000 &             1.4 \\
	NGC1365 &   03 33 36.41 &  -36 08 24.00 & 0.0055 &                  Sy 1 &    S1.8 &                60 &            2016-02-10 &                     2700 &             1.1 \\
	ESO420-G13 &   04 13 49.70 &  -32 00 24.12 & 0.0121 &             SB + Sy 2 &      S2 &                90 &            2016-02-09 &                     2700 &             1.7 \\
	NGC1672 &   04 45 42.19 &  -59 14 51.00 & 0.0044 &             SB + Sy 2 &       S &                90 &            2016-02-07 &                     2700 &             1.9 \\
	NGC1667 &   04 48 37.20 &  -06 19 12.00 & 0.0152 &                  Sy 2 &      S2 &                 0 &  2016-02-08$^{\rm b}$ &                     2700 &             1.6 \\
	UGC3255 &   05 09 50.21 &  +07 28 59.16 & 0.0189 &                 LINER &      S2 &                20 &  2016-02-07$^{\rm b}$ &                     2700 &             2.4 \\
	ESO362-G08 &   05 11 09.00 &  -34 23 35.00 & 0.0157 &            PSB + Sy 2 &      S2 &                 0 &  2016-02-06$^{\rm b}$ &                     2700 &             1.8 \\
	ESO362-G18 &   05 19 35.80 &  -32 39 27.00 & 0.0124 &                  Sy 1 &    S1.5 &               135 &            2016-02-10 &                     2700 &             1.1 \\
	NGC2110 &   05 52 11.40 &  -07 27 23.04 & 0.0078 &                  Sy 2 &     S1h &                 0 &            2016-02-08 &                     2700 &             1.4 \\
	NGC2217 &   06 21 39.70 &  -27 14 00.96 & 0.0054 &                 LINER &      S3 &                90 &            2016-02-10 &                     2700 &             1.3 \\
	ESO565-G19 &   09 34 43.80 &  -21 55 41.88 & 0.0163 &         LINER or Sy 2 &      S2 &                 0 &            2016-02-09 &                     2700 &             2.2 \\
	NGC2945 &   09 37 41.09 &  -22 02 06.00 & 0.0154 &                 LINER &       S &                90 &            2016-02-08 &                     2700 &             1.8 \\
	NGC2974 &   09 42 33.30 &  -03 41 57.00 & 0.0063 &             SB + Sy 2 &      S2 &                45 &  2016-02-05$^{\rm b}$ &                     1800 &             2.0 \\
	ESO500-G34 &   10 24 31.49 &  -23 33 10.08 & 0.0122 &       SB + PSB + Sy 2 &      S2 &                 0 &            2016-02-07 &                     2700 &             1.8 \\
	NGC3281 &   10 31 52.10 &  -34 51 12.96 & 0.0116 &                  Sy 2 &      S2 &               135 &            2016-02-09 &                     2700 &             1.7 \\
	MCG-02-27-009 &   10 35 27.30 &  -14 07 47.00 & 0.0151 &         LINER or Sy 2 &      S2 &                90 &            2016-02-05 &                     2700 &             1.8 \\
	NGC3312 &   10 37 02.50 &  -27 33 56.16 & 0.0096 &                 LINER &      S3 &                 0 &            2016-02-08 &                     2700 &             1.4 \\
	NGC3390 &   10 48 04.39 &  -31 31 59.88 & 0.0102 &                    SB &      S3 &                 0 &            2016-02-07 &                     2700 &             1.5 \\
	NGC3393 &   10 48 23.40 &  -25 09 43.92 & 0.0125 &                  Sy 2 &      S2 &                45 &            2016-02-08 &                     2700 &             1.5 \\
	NGC3831 &   11 43 18.60 &  -12 52 42.00 & 0.0176 &                 LINER &      S2 &                30 &            2016-02-09 &                     2700 &             2.4 \\
	NGC3858 &   11 45 11.70 &  -09 18 50.00 & 0.0191 &                    SB &      S2 &                45 &            2016-02-08 &                     2700 &             1.6 \\
	NGC4418 &   12 26 54.70 &  -00 52 41.88 & 0.0073 &                    SB &      S2 &                60 &            2016-05-04 &                     2700 &             1.5 \\
	NGC4472 &   12 29 46.80 &  +08 00 02.16 & 0.0033 &                    SB &      S2 &                 0 &            2016-05-05 &                     2700 &             1.2 \\
	NGC4507 &   12 35 36.50 &  -39 54 33.12 & 0.0118 &                  Sy 2 &     S1h &                60 &            2016-02-10 &                     2400 &             1.2 \\
	NGC4593 &   12 39 39.40 &  -05 20 39.00 & 0.0083 &                  Sy 1 &    S1.0 &                45 &  2016-02-05$^{\rm b}$ &                     2700 &             2.0 \\
	NGC4594 &   12 39 59.30 &  -11 37 23.16 & 0.0037 &             SB + Sy 2 &      S3 &                90 &            2016-02-07 &                     2700 &             2.0 \\
	IC3639 &   12 40 52.90 &  -36 45 20.88 & 0.0109 &                  Sy 2 &     S1h &               135 &            2016-02-10 &                     2700 &             1.3 \\
	NGC4636 &   12 42 49.99 &  +02 41 16.08 & 0.0031 &                 LINER &     S3b &                45 &            2014-04-08 &                     1800 &             1.3 \\
	NGC4691 &   12 48 13.01 &  -03 19 59.16 & 0.0037 &                    SB &      S1 &                90 &            2016-05-04 &                     2700 &             1.4 \\
	NGC4696 &   12 48 49.30 &  -41 18 39.96 & 0.0099 &                 LINER &      S3 &                90 &  2016-05-03$^{\rm b}$ &                      900 &             2.0 \\
	ESO443-G17 &   12 57 44.90 &  -29 45 59.04 & 0.0102 &             SB + Sy 2 &      H2 &                 0 &            2016-02-10 &                     2700 &             1.4 \\
	NGC4845 &   12 58 01.20 &  +01 34 32.88 & 0.0041 &             SB + Sy 2 &      S2 &                90 &            2016-05-04 &                     2700 &             1.3 \\
	NGC4939 &   13 04 14.30 &  -10 20 23.00 & 0.0104 &                  Sy 2 &      S2 &                 0 &            2016-05-06 &                     2100 &             1.4 \\
	ESO323-G77 &   13 06 26.21 &  -40 24 51.84 & 0.0156 &                  Sy 1 &    S1.2 &                45 &            2016-02-10 &                     2700 &             1.2 \\
	NGC4968 &   13 07 05.40 &  -23 40 36.84 & 0.0099 &                  Sy 2 &      S2 &                45 &            2016-05-06 &                     2700 &             1.3 \\
	PKS1306-241 &   13 08 41.81 &  -24 22 57.00 & 0.0139 &             SB + Sy 2 &      S2 &                90 &            2016-05-05 &                     2700 &             1.3 \\
	ARK402 &   13 08 50.11 &  -00 49 01.92 & 0.0178 &             SB + Sy 2 &      S2 &                90 &            2016-05-04 &                     2700 &             1.5 \\
	NGC4990 &   13 09 17.30 &  -05 16 22.08 & 0.0106 &             SB + Sy 2 &      H2 &                90 &            2016-05-05 &                     2700 &             1.0 \\
	MCG-03-34-064 &   13 22 24.50 &  -16 43 41.88 & 0.0165 &                  Sy 1 &     S1h &                90 &            2016-05-04 &                     2700 &             1.3 \\
	NGC5128 &   13 25 28.01 &  -43 01 00.12 & 0.0018 &                  Sy 2 &      B? &                90 &            2016-05-06 &                     1200 &             1.4 \\
	NGC5135 &   13 25 43.99 &  -29 50 02.04 & 0.0137 &            PSB + Sy 2 &      S2 &               135 &  2016-05-06$^{\rm b}$ &                     2400 &             1.6 \\
	ESO383-G35 &  13 35 53.70  &  -34 17 44.00 & 0.0077 &                  Sy 1 &    S1.2 &                90 &            2016-05-06 &                     2100 &             1.7 \\
	NGC5252 &   13 38 15.89 &  +04 32 33.00 & 0.0230 &                  Sy 2 &      S2 &                 0 &            2016-05-06 &  2400; 4800$^\mathrm{c}$ &             1.4 \\
	IC4329A &   13 49 19.30 &  -30 18 33.84 & 0.0161 &                  Sy 1 &    S1.2 &                45 &            2016-05-04 &                     2700 &             1.4 \\
	NGC5427 &   14 03 25.90 &  -06 01 50.16 & 0.0092 &                  Sy 2 &      S2 &                90 &  2016-05-06$^{\rm b}$ &                     2400 &             1.8 \\
	NGC5643 &   14 32 40.70 &  -44 10 27.84 & 0.0040 &                  Sy 2 &      S2 &                 0 &            2016-05-04 &  1980; 3960$^\mathrm{c}$ &             1.4 \\
	PKS1521-300 &   15 24 33.41 &  -30 12 20.88 & 0.0195 &             SB + Sy 2 &      S1 &                90 &            2016-05-03 &                     2700 &             1.9 \\
	NGC5990 &   15 46 16.49 &  +02 24 56.16 & 0.0128 &       SB + PSB + Sy 2 &      S2 &                90 &  2016-05-03$^{\rm b}$ &                     2700 &             1.9 \\
	NGC6328 &   17 23 40.90 &  -65 00 36.00 & 0.0144 &             SB + Sy 2 &      S3 &                 0 &            2014-08-22 &                     2400 &             1.3 \\
	IC4777 &   18 48 11.30 &  -53 08 51.00 & 0.0187 &                  Sy 2 &      S2 &               135 &            2016-05-06 &                     2700 &             1.7 \\
	ESO460-G09 &   19 30 26.60 &  -31 52 38.00 & 0.0196 &             SB + Sy 2 &       S &                90 &            2016-05-03 &                     2700 &             1.8 \\
	MCG-02-51-008 &   20 17 06.31 &  -12 05 51.00 & 0.0186 &             SB + Sy 2 &      S3 &                90 &            2014-08-24 &                     2700 &             1.3 \\
	IC4995 &   20 19 58.99 &  -52 37 18.84 & 0.0161 &                  Sy 2 &      S2 &                 0 &            2014-08-22 &                     2400 &             1.3 \\
	NGC6936 &   20 35 56.30 &  -25 16 48.00 & 0.0195 &             SB + Sy 2 &      S3 &                 0 &            2014-08-22 &                     2400 &             1.9 \\
	MCG-04-49-001 &   20 46 37.20 &  -23 37 49.08 & 0.0201 &                 LINER &      S3 &                 0 &            2014-08-23 &                     2400 &             1.9 \\
	ESO287-G42 &   21 38 07.90 &  -42 36 18.00 & 0.0187 &                  Sy 1 &    S1.8 &               120 &  2016-05-06$^{\rm b}$ &                     2700 &             1.7 \\
	NGC7172 &   22 02 01.90 &  -31 52 08.04 & 0.0087 &             SB + Sy 2 &      S2 &                90 &            2014-08-23 &                     2400 &             1.3 \\
	IC5169 &   22 10 10.00 &  -36 05 19.00 & 0.0104 &             SB + Sy 2 &      S2 &                30 &            2016-05-04 &                     2700 &             1.2 \\
	IC1481 &   23 19 25.30 &  +05 54 20.88 & 0.0204 &             SB + Sy 2 &      S3 &                90 &            2014-08-24 &                     2700 &             1.3 \\
	NGC7682 &   23 29 03.91 &  +03 31 59.88 & 0.0171 &                  Sy 2 &     S1h &                 0 &            2014-08-24 &                     2700 &             1.0 \\
	\enddata
	\newline
	\tablenotetext{a}{SB = Starburst, PSB = Post-starburst}
	\tablenotetext{b}{Conditions were not photometric}
	\tablenotetext{c}{Effective exposure time doubled in mosaic overlap region.}
\end{deluxetable*}

Table~\ref{table:Obs} also includes NGC\,1346, NGC\,3281 and NGC\,4636, which were not included in the corresponding table (Table~1) in \PII.

We note that NGC\,7552 has been removed from the S7 sample since the first data release.  It was included in Table~1 of \PII\ but is a star-forming galaxy that had very poor quality observations.

The seeing FWHM measurements presented in Table~\ref{table:Obs} were recorded in the observer's log during the relevant observations.  The observer judged appropriate values for the seeing FWHM using time series of periodic measurements of the guide star PSF during the relevant exposures.

\subsection{Reduction of raw observations}
\label{sec:pywifes}

In this section we describe the reduction of the raw data, the registration to determine sky coordinates, and the expected accuracy of the WCS, flux calibration and wavelength calibration.

The raw frames from the WiFeS CCDs were combined and reduced with the \pywifes\ data reduction pipeline, which is described in detail in \citet{Childress_2014_PyWiFeS}.  The data reduction procedure corrects for the detector bias levels and for cosmic ray strikes.  It uses arc lamp frames taken on each night of observing to perform wavelength calibration, uses observations of spectrophotometric stars to perform flux calibration, and uses observations of telluric standard stars to correct for telluric absorption.  Corrections for Differential Atmospheric Refraction \citep[DAR;][]{Filippenko_1982_DAR} are applied when forming the spectral cubes.

The \pywifes\ code initializes variances with photon counting noise and detector read noise after subtracting the overscan, repairing bad pixels, and subtracting a global `superbias' (combined bias frame).  Full statistical propagation of uncertainties is then applied over subsequent pipeline steps, in particular the subtraction of sky frames, co-adding of multiple frames and flat-fielding.  \pywifes\ does not account for any covariance between adjacent spectra in a slitlet.

The output of the data reduction pipeline is two independently processed `data cube' files, one for each of the red and blue arms of the WiFeS instrument.  The files are provided for every S7 galaxy as part of this data release.  Each file contains both a flux cube and a variance cube; the data format is described in more detail in Section~\ref{sec:prod_data_cubes}.

Registration of the data cubes was performed by summing the red cubes in the wavelength direction and cross-correlating the resulting synthetic pseudo-continuum images with DSS2 r-band images.  The recorded telescope position angle was assumed to be correct, and the cross-correlation took advantage of the fact that both the S7 and DSS2 data consisted of 1~arcsec square pixels.  The results were manually inspected for each galaxy, and typically the registration (which was not attempted to sub-pixel accuracy) was clearly valid to within 1-2~arcsec.  Very nearby galaxies often had overexposed centres in the DSS2 images (indicated in the `S7\_WCS\_flags' field in the catalogue table and in the `WCS\_QLTY' FITS header keyword in the data products), and for the few worst objects the resulting WCS simply used the telescope pointing model, and may be incorrect by tens of seconds of arc.

We investigated the consistency of the flux calibration by comparing duplicate observations of four galaxies - NGC\,7496, NGC\,6915, NGC\,835, and ESO330-G11.  These objects were each observed twice in photometric conditions, once without dedicated sky exposures for sky subtraction and once with the dedicated exposures.  Binned nuclear spectra were compared between observations to give an indication of the quality of the flux calibration.  We found that the spectra were always systematically higher or lower in flux between the two observations, with the differences being consistent between the red and blue arms.  For NGC\,6915 and ESO330-G11 the duplicate observations were a year apart over $2013 - 2014$. The flux differences between the two observations were $15 - 20\%$, with the 2013 observations higher than the 2014 observations for both galaxies.  The repeat observations of NGC\,7496 and NGC\,835 were separated by three months in 2013. The flux differences between the two observations were $4 - 9\,$per cent for these galaxies.  Overall these differences could plausibly be accounted for by the different sky-subtraction methods and slightly different apertures between the observations.  Unfortunately it is not possible to systematically compare S7 fluxes to existing data due the difficulty in matching apertures, spatial resolution, spectral resolution, and the all-sky nature of the target set.

The accuracy of the wavelength calibration produced by \pywifes\ is discussed in \citet{Childress_2014_PyWiFeS}.  The authors show that the dominant uncertainty is the systematic uncertainty due to temperature drift in the instrument between an arc frame being taken and observations of the target.  An error of approximately $1 - 2$\,\AA\ resulted from using an arc frame taken in the early afternoon to calibrate an arc frame taken at the end of the following night.  The wavelength calibration errors for the S7 data products will be significantly smaller than $1 - 2$\,\AA\ because arc frames were taken on 2 or 3 separate occasions during each S7 observing night.  We note that if the error is as high as 0.5\,\AA\ for any S7 galaxies, the resulting dispersion in recession velocities between the blue and red extremes of the observed spectra would be only ${\sim}20$\,km\,s$^{-1}$.

We note that the S7 gas velocity maps in the Appendix display a symmetry between blueshifted and redshifted emission (relative to systemic velocity).  The symmetry occurs because the NED redshifts used in the line fitting are generally well matched to the systemic recession velocities associated with the calibrated S7 wavelengths.  The maps required adjustment for only ${\sim}18$ galaxies, with required offsets ranging from $-90$ to $+80$\,km\,s$^{-1}$.  The S7 wavelength vectors are reliable and we expect the offsets to generally be attributable to the external redshifts used in line fitting.

\subsection{Deriving emission-line products}
\label{sec:fitting}

\subsubsection{Fitting emission lines and the stellar continuum}
\label{sec:lzifu}

Emission lines for both data cubes and nuclear spectra were fit using \lzifu\footnote{\url{https://github.com/hoiting/LZIFU}} \citep[`Lazy-IFU';][]{Ho_2016_LZIFU}.  The \lzifu\ code was designed to perform emission-line fitting on IFU data cubes.  It first fits the underlying continuum in each spectrum using stellar spectral templates before fitting the emission lines using multiple Gaussians.  The \lzifu\ code models the stellar continuum using the penalized pixel-fitting routine \citep[\ppxf;][]{Cappellari_Emsellem_2004_ppxf}, and was configured to use the theoretical single stellar population (SSP) spectra from \citet{Gonzalez_Delgado_2005_SSP} based on the Padova isochrones.  A 5th degree additive Legendre polynomial is fitted simultaneously to the stellar spectral templates.

After the continuum is subtracted from a spectrum, \lzifu\ fits the emission lines using the Levenberg-Marquardt least-squares algorithm as implemented in {\sc mpfit} \citep{Markwardt_2009_MPFIT}.  Emission lines are fitted three times, with each of one, two and three Gaussian components.  For a given component (1, 2 or 3), the velocity and velocity dispersion of all emission lines are constrained to be the same.  In the output the components are ordered by velocity dispersion, with component~1 being the narrowest.  The \lzifu\ code accounts for instrumental dispersion when measuring the velocity dispersion.

We expect spatial variation in the fitting results to be generally smooth.  To encourage the code to produce spatially smooth output, we used the ability of \lzifu\ to iteratively spatially smooth the fitting results and refit the data.  The output fit parameters were smoothed using a kernel of width 5 spaxels and each spaxel was then refit using the `smoothed' parameter values as initial parameter guesses.  Two iterations of this smoothen-refit procedure were performed.

The \lzifu\ output contains errors on the emission-line fluxes,
velocities and velocity dispersions derived from the fitting errors outputted by {\sc mpfit}.  The {\sc mpfit} fitting errors in turn depend on the data cube variances.  The covariances reported by {\sc mpfit} are propagated into the \lzifu\ flux errors.  The \lzifu\ errors do not account for uncertainty in the subtracted stellar continuum, and as a consequence the errors in the Balmer line fluxes are generally underestimated when the underlying stellar absorption is strong compared to the line emission.  The errors produced by \lzifu\ are compared to Monte Carlo simulations and discussed in detail in \citet{Ho_2016_LZIFU}.

The \lzifu\ output is provided to the community and is described in detail in Section~\ref{sec:prod_emission_lines}.

\subsubsection{Selecting the number of kinematic components in emission lines}
\label{sec:lzcomp}

The optimal number of Gaussians needed to describe the emission line features across a galaxy differs from spaxel to spaxel.  Each galaxy is fit with 1, 2 and 3 Gaussian components using \lzifu.  The optimal number of Gaussian components (i.e.\ 1, 2 or 3) is determined for each spaxel separately using an Artificial Neural Network (ANN) trained by astronomers.  The selection of the number of components must be automated because manual classification is impractical for the ${\sim}10^4 - 10^5$ spaxels requiring consideration in the S7.

The ANN, \texttt{LZComp}\footnote{\url{https://github.com/EliseHampton/LZComp}} (`Lazy-Components'), is described in detail in Hampton et al. (2017; submitted). \texttt{LZComp} is a supervised machine learning algorithm that has been trained using manual classifications produced by three astronomers from the S7 team.  The training and testing, described in Hampton et al. (2017; submitted), determined \texttt{LZComp} to be equivalent to using an astronomer to make the decisions on the numbers of Gaussian components required for the emission line features in the S7 survey.

An extension in the FITS file output of \texttt{LZComp} indicates the number of components selected in each spaxel. The product FITS files are described in Section~\ref{sec:prod_emission_lines}.

\subsection{Emission-line fitting of Seyfert~1 galaxies }
\label{sec:Sy1_fitting}

Galaxies with nuclei showing very broad components in Balmer lines required special treatment when performing emission-line fitting.  Fitting using \lzifu\ often fails on Balmer lines with extremely broad components, especially when the profiles are asymmetrical.  In the most extreme objects, broad \Ha\ wings extend past the \SII\ doublet; these nuclei cannot be fit using the same method as Sy\,2s.  Both the Balmer line widths and the relative flux in the broad components vary continuously over the S7 sample, so there is no clear delineation between the Seyfert~1 nuclei and other nuclei.

We identified spaxels with problematically broad Balmer components by checking for a broad \Ha\ wing redward of the \NII-\Ha\ complex in each spaxel of each galaxy.  The method essentially consisted of the following considerations.  The difference between the median of the spectrum in the region 10 -- 40\,\AA\ redward of \NII\,$\lambda6583$ and the continuum background was calculated.  Where this flux difference was more than 1.8 times the interquartile range in the continuum (a measure of the noise), the spaxel was flagged as having a broad component.  As a S/N cut, spaxels were not considered if the height of the peak of the \NII-\Ha\ complex above the continuum was less than 7 times the continuum interquartile range.  Appropriate values for the tolerances were found by experimentation.

There were 14 galaxies in the S7 sample that were identified using this method as having very broad and prominent Balmer line components in nuclear spaxels, which were ESO\,323-G77,
ESO\,362-G18, ESO\,383-G35, Fairall\,49, Fairall\,51, IC\,4329A, Mark\,1239, NGC\,1365, NGC\,2617, NGC\,3783, NGC\,4593, NGC\,6860, NGC\,7213 and NGC\,7469.  For these galaxies, the broad lines were sufficiently broad and/or dominant over the narrow components that it was necessary to subtract out the broad components before fitting the data cubes and nuclear spectra.  A total of 316 spaxels over the 14 galaxies had broad-line subtraction.  Another six galaxies identified by the method described above had a broad component that was judged to be too narrow or weak to merit special subtraction.

Subtracting out broad \Ha\ and \Hb\ lines from the nuclear spaxels was accomplished by approximately following the method outlined in \cite{Greene_Ho_2005_broad_fitting}.  Firstly, 1- and 2-component \lzifu\ fits were made to the \SII$\,\lambda\lambda$\,6716,\,6731 lines to constrain the velocity and velocity dispersion of the narrow components.  This important step was used to reduce the number of free parameters when fitting the \NII-\Ha\ complex.  The \NII-\Ha\ complex was fit using both narrow Gaussians (kinematics set by the \SII\ fits) and broad Gaussians (minimum $\sigma_v = 1000$\,km\,s$^{-1}$).  Fits were performed with up to 4 or 5 broad components for \Ha\ and up to 3 or 4 narrow components for \Ha\ and both \NII\ lines.  Corresponding narrow components all had the same velocity and velocity dispersion and the \NII$\lambda$\,6583/\NII$\lambda$\,6548 flux ratio was fixed to 3.  For some galaxies an additional broad Lorentzian \Ha\ component was used.  The fit with the lowest reduced-$\chi^2$ was used for each spaxel.

When fitting \OIII-\Hb, some differences in the approach were necessary compared to the method used when fitting the \NII-\Ha\ complex. Again multiple broad Gaussians were used to build up the profile of the broad Balmer line, however often it was necessary to use only one narrow component for \Hb\ while using 3, 4 or 5 narrow components for \OIII\,$\lambda\lambda4959,5007$.  Additionally only the first (narrowest) component of two-component \SII\ fits were used to constrain an \OIII-\Hb\ narrow component, unlike for \NII-\Ha\ in which both \SII\ fit components were used.  We note that any \ion{Fe}{2} emission in this region was not considered.  Hence some \ion{Fe}{2} flux may have been subtracted in the \Hb\ broad component (contributing to the red wing) and may have contaminated the red wing of \OIII$\,\lambda 5007$.

Data cubes were produced with the fitted \Ha\ and \Hb\ broad lines subtracted, and \lzifu\ was then used to fit the narrow lines that remained in the data.  The narrow-line parameters resulting from the simultaneous fits to the broad lines were not used.

The decompositions into broad- and narrow-spectra are presented in Figures~\ref{fig:nuc_spec_Sy1_A} to \ref{fig:nuc_spec_Sy1_C} in the Appendix and are discussed in Section~\ref{sec:nuc_spec_plots}.

Although the method used to find the spectrum of the broad components was generally successful and resulted in satisfactory measured narrow-line fluxes in most cases, there are systematic issues with the approach described above.  Firstly, an obvious limitation is that there is no physical reason why the narrow- and broad- line regions need to be entirely distinct as opposed to continuously varying, and hence dividing the emission-line flux into `broad' and `narrow' parts may in some cases be arbitrary.  The line profiles sometimes contained little information that enabled splitting into `broad' and `narrow' parts, a problem that was most severe where the narrow components were particularly broad and in the innermost nuclear spaxels when the broad components were very dominant.  Often manual tweaks to the number of components and fitting constraints were necessary to ensure that the broad-narrow division `looked' appropriate, even if the tweaking produced a fit that was statistically slightly worse.

A second issue is that the modeled and measured line profiles were sometimes consistently mismatched in a particular part of the spectrum.  This issue was associated with the combination of model line profiles and the fitting strategy.  A relatively frequent problem was  caused by the broad line fit slightly overestimating the flux around  the red wing of \NII$\lambda$6583, (e.g.\ NGC\,2617 in Figure~\ref{fig:nuc_spec_Sy1_A}; where the narrow-line spectrum show a `dip' from the over-subtraction of the broad component).  Another occasional issue was the broad component fit overestimating the flux in the `flux dip' between \Ha\ and \NII$\lambda6583$.

Despite these systematic issues, fitting narrow lines to the broad component-subtracted cubes produced reasonable results, as seen, for example, in the maps for NGC\,4593 (Figure~\ref{fig:spatial_6}) and IC\,4329A and (Figure~\ref{fig:spatial_7}) in the appendix.  The maps show fitted emission-line velocities that are spatially consistent with nearby spaxels that did not have broad-component subtraction (spaxels outside the green outline).

As a result of the systematic errors described above, users of the data should carefully consider the appropriate number of narrow-line components that are required in the nuclear spaxels of Seyfert~1 galaxies.  Users may decide to consider the narrow-line fluxes measured in broad-component-subtracted spaxels to be approximate lower bounds, and/or only use the narrowest component in multiple-component fits.  These caveats are also discussed in Section~\ref{sec:prod_emission_lines}.

\subsection{Extraction and analysis of nuclear spectra}
\label{sec:nuc_spec_reduction}

The circum-nuclear spectra presented in this work were extracted from the data cubes using an aperture chosen to approximate the circular, 3\,arcsec-diameter aperture of an SDSS fibre.  A slightly larger diameter of ${\sim}4$\,arcsec was chosen because the S7 targets are very nearby compared to the SDSS sample.  There were 13 whole spaxels and zero fractional spaxels in the bin region, which was centered on the `nuclear' spaxel of each galaxy.  The bin region was the union of the following three shapes: a $3 \times 3$ square centered on the nuclear spaxel, a column of 5 spaxels symmetrical about the nuclear spaxel, extending 2 spaxels above and 2 below, and a row of 5 spaxels symmetrical about the nuclear spaxel, extending 2 to the left and 2 to the right.  Masks showing the aperture shape and location are included in the relevant FITS files (Section~\ref{sec:prod_nuc_spec}).  The spectral variances were summed over the same aperture.

Emission-line fluxes for the nuclear spectra given in this work and in \PII\ were measured using \lzifu\ as described above.  Spatial smoothing and refitting to obtain spatially smooth results was not applicable for the single nuclear spectra.  One-component fits were used for all galaxies except NGC\,7469, for which the narrowest component of a two-component fit was used.  The \lzifu\ fitting of the stellar continuum failed in one or both of the red and blue for 9 galaxies, and manually-provided linear continua were used instead.  This is indicated in the `Linear\_continuum?' field in the relevant tables in the data release.  Both the \lzifu\ fluxes and flux errors are provided as part of the data release.

Nuclear fluxes for each galaxy were corrected for reddening following \citet{Vogt_2013_HCGs_I}, but only for initial Balmer decrements of $F_{{\rm H}\alpha} / F_{{\rm H}\beta} > 3.1$.  Where it was applied, the reddening correction was set to produce a target Balmer decrement of $F_{{\rm H}\alpha} / F_{{\rm H}\beta} = 3.1$.  For the Seyfert~1 nuclei with broad-component subtraction, the narrow component fluxes were used.  Errors in the H$\alpha$ and H$\beta$ fluxes were propagated through the reddening correction into the de-reddened fluxes.

\vspace{1.5cm}
\section{Data products}
\label{sec:products}

The data products are available to download from the All-Sky Virtual Observatory\footnote{\url{http://datacentral.aao.gov.au/asvo/surveys/s7/}} or the S7 website\footnote{\url{https://miocene.anu.edu.au/S7/Data\_release\_2}}, and consist of the following:
\begin{enumerate}
	\item A catalogue of the full S7 sample, including ancillary data merged from NED and HyperLeda.  The catalogue includes the information presented in Table~\ref{table:Obs} for each galaxy.
	\item Data cubes for both the red and blue arms of WiFeS, and additional cubes with broad lines subtracted from nuclear spectra for 14 Seyfert~1 galaxies (Section~\ref{sec:Sy1_fitting})
	\item \lzifu\ emission-line fitting output for 1-, 2- and 3-component fits (Section~\ref{sec:lzifu})
	\item \lzifu\ emission-line fitting output where the optimal number of components for each spaxel has been selected by the \texttt{LZComp} Artificial Neural Network (Section~\ref{sec:lzcomp})
	\item Nuclear spectra found by summing the data cubes over a 4~arcsec aperture, and additional broad and narrow-line nuclear spectra for the 14 Seyfert galaxies with special treatment
	\item A table of extinction-corrected nuclear emission-line fluxes measured using \lzifu\ fits to the nuclear spectra (an extended version of Table~\ref{table:nuc_fit_fluxes} in the Appendix), and a corresponding table of flux errors
	\item A table of associated extinctions, physical aperture sizes, and nuclear \Hb\ and \OIII\ fluxes and luminosities (an extended version of Table~\ref{table:nuc_fit_luminosities} in the Appendix)
\end{enumerate}

The spectra and cubes are provided in the FITS file format \citep{Pence_2010_FITS}.  The products are described in further detail in the following sections.  The products related to the nuclear spectra are described in Section~\ref{sec:nuc_spec_results}.

\subsection{Data cubes}
\label{sec:prod_data_cubes}

The data cubes have a spatial size of $25 \times 38$ spatial pixels (`spaxels'), where each spaxel is $1 \times 1$~arcsec$^2$.  The red cubes cover 5400 to 7000\,\AA\ in the wavelength dimension in uniform increments of 0.44\,\AA.  The blue cubes cover 3500 to 5700\,\AA\ in the wavelength dimension in uniform increments of 0.77\,\AA.  In the FITS files for each cube, the 0th extension contains the flux cube, and the 1st extension contains the variance cube.

Additionally, data cubes with the broad \Ha\ and \Hb\ components subtracted from the nuclear spaxels are also provided for the 14 Seyfert~1 galaxies that had broad-component subtraction (Section~\ref{sec:Sy1_fitting}).  For NGC\,7213 only, the blue cube with broad-component subtraction is the same as the cube without this subtraction, because \Ha\ but not \Hb\ required the broad-component subtraction.

For both the normal and broad-component subtracted data cubes for these 14 galaxies, an additional extension is provided in the FITS files.  This extension contains a mask to indicate which nuclear spaxels had broad-component subtraction.

\subsection{Emission line fits}
\label{sec:prod_emission_lines}

Four FITS files containing emission-line fit information are provided for each galaxy.  Three of the FITS files correspond to each of the 1-, 2- and 3-component Gaussian fits to emission lines produced by \lzifu\ (Section~\ref{sec:lzifu}).  A fourth FITS file contains the `best' number of components, as determined by the neural network {\sc lzcomp} (Section~\ref{sec:lzcomp}).

Each FITS file containing emission line fit data consists of multiple extensions.  The extensions available in each \lzifu\ product FITS file are detailed in Table~\ref{table:lzifu_exts}.  Most of the extensions contain arrays with measured emission-line fluxes or the errors in these fluxes.  Some extensions contain additional information such as the emission-line gas velocity relative to the input systemic redshift (`V'; in km\,s$^{-1}$), or the corresponding velocity dispersion (`VDISP'; in km\,s$^{-1}$).

\begin{deluxetable}{lccc}
	\centering
	\tabletypesize{\scriptsize}
	\tablecaption{Indicative extension indices and `EXTNAME' extension header values in \lzifu\ product FITS files. \label{table:lzifu_exts}}
	\tablewidth{230pt}
	\tablehead{
		\colhead{Name} & \colhead{Index$^1$} & \colhead{Errors } & \colhead{Errors }\\
		\colhead{} & \colhead{} & \colhead{name} & \colhead{index}
	}
	\startdata
	B\_CONTINUUM &          1 &             - &             - \\
	R\_CONTINUUM &          2 &             - &             - \\
	B\_RESIDFIT &          3 &             - &             - \\
	R\_RESIDFIT &          4 &             - &             - \\
	B\_CONT\_MASK &          5 &             - &             - \\
	R\_CONT\_MASK &          6 &             - &             - \\
	B\_LINE &          7 &             - &             - \\
	R\_LINE &          8 &             - &             - \\
	R\_LINE\_COMP1 &          9 &             - &             - \\
	R\_LINE\_COMP2 &         10 &             - &             - \\
	R\_LINE\_COMP3 &         11 &             - &             - \\
	STAR\_V &         12 &             - &             - \\
	STAR\_VDISP &         13 &             - &             - \\
	CONT\_CHI2 &         14 &             - &             - \\
	V &         15 &         V\_ERR &            16 \\
	VDISP &         17 &     VDISP\_ERR &            18 \\
	CHI2 &         19 &             - &             - \\
	DOF &         20 &             - &             - \\
	OII3726 &         21 &   OII3726\_ERR &            22 \\
	OII3729 &         23 &   OII3729\_ERR &            24 \\
	OIII4363 &         25 &  OIII4363\_ERR &            26 \\
	HBETA &         27 &     HBETA\_ERR &            28 \\
	OIII4959 &         29 &  OIII4959\_ERR &            30 \\
	OIII5007 &         31 &  OIII5007\_ERR &            32 \\
	OI6300 &         33 &    OI6300\_ERR &            34 \\
	NII6548 &         35 &   NII6548\_ERR &            36 \\
	HALPHA &         37 &    HALPHA\_ERR &            38 \\
	NII6583 &         39 &   NII6583\_ERR &            40 \\
	SII6716 &         41 &   SII6716\_ERR &            42 \\
	SII6731 &         43 &   SII6731\_ERR &            44 \\
	COMP\_PREDICTIONS &         45 &             - &             - \\
	\enddata
	\newline
	\tablenotetext{1}{The Header Data Unit (HDU) with index 0 contains header information but no data.}
\end{deluxetable}

In all four emission-line fit files for a galaxy, each extension containing emission-line fluxes contains a data cube.  Two dimensions of the cube are the spatial dimensions, with slices in the third dimension corresponding to kinematic components.  The 0th slice contains the total emission-line flux summed over all the kinematic components, and the 1st, 2nd and 3rd slices contain fluxes for the 1st, 2nd and 3rd kinematic components, where applicable.  A cube of fluxes for the $n$-component fit contains $n + 1$ slices.  The components in each spaxel are ordered by velocity dispersion, with the 1st and 3rd components corresponding to the narrowest and broadest Gaussians respectively.  In the `best' number of components file, the `COMP\_PREDICTIONS' extension is a mask showing whether \texttt{LZComp} chose 1, 2, or 3 components as the optimal number in each spaxel.

The provided emission-line fluxes are subject to the following caveats:
\begin{itemize}
\item The [\ion{O}{2}]\,$\lambda\lambda 3726,29$ and [\ion{O}{3}]\,$\lambda\,4363$ fluxes are provided as a sum of all components only.  Due to reddening and lower S/N in the blue compared to the red, the component breakdowns for these lines are not reliable.
\item The total H$\beta$ flux is reapportioned between components to be consistent with the flux-ratios between components for H$\alpha$, to ensure consistent Balmer decrements. 
\item The fitting for the 14 Seyfert~1 galaxies that had special treatment was performed on the broad component-subtracted cubes.  The fluxes inside the `broad-component subtraction' regions of these galaxies are subject to substantial systematic uncertainties and should be considered on a case-by-case basis.  In particular, care needs to be taken in choosing the number of kinematic components to use.  The data for the `best' number of components is not necessarily the most appropriate.  Users of this data should read the caveats in Section~\ref{sec:Sy1_fitting} and inspect Figures~\ref{fig:nuc_spec_Sy1_A} to \ref{fig:nuc_spec_Sy1_C} in the Appendix or the broad component-subtracted cubes to determine how many kinematic components are appropriate to use for the broad component-subtracted narrow lines.  Higher-order kinematic components may describe residuals of the subtraction as opposed to narrow-line emission.
\end{itemize}

The \lzifu\ output for 3C278 is missing some emission lines, however this galaxy has very little optical line emission.

\subsection{Binned circum-nuclear spectra}
\label{sec:prod_nuc_spec}

Two FITS files containing nuclear spectra are provided for each galaxy, corresponding to the red and blue arms of WiFeS.  The nuclear spectra are binned over an approximately circular, ${\sim}4$~arcsec (4-spaxel) diameter aperture around the nucleus (Section~\ref{sec:nuc_spec_reduction}).  The 0th extension of each FITS file contains the 1D flux array, the 1st extension contains the corresponding variance array, and the 2nd extension contains a mask identifying the spaxels that were binned to produce the nuclear spectrum.

For the 14 Seyfert~1 galaxies with special treatment, an additional nuclear spectrum file is provided (for each of the red and blue), which is the ${\sim}4$-arcsec nuclear spectrum binned from the broad component-subtracted cubes.

Figures~\ref{fig:nuc_spec_Sy1_A} to \ref{fig:nuc_spec_Sy1_C} in the Appendix compare nuclear spectra before and after broad component-subtraction for the relevant galaxies.  These figures show nuclear spectra binned only over the spaxels with special treatment (instead of the ${\sim}4$-arcsec aperture) and are further described in Section~\ref{sec:nuc_spec_plots} below.

Emission-line fluxes measured in the nuclear spectra are described in Section~\ref{sec:nuc_fluxes} below.

\vspace{1.5cm}
\section{Nuclear spectra}
\label{sec:nuc_spec_results}

In this section we present the results of the \lzifu\ fitting of the nuclear spectra.

\subsection{Emission-line fluxes}
\label{sec:nuc_fluxes}

Table~\ref{table:nuc_fit_fluxes} in the Appendix presents extinction-corrected nuclear emission-line fluxes.  The full table is available as a data product in this data release (Section~\ref{sec:products}), along with a corresponding table of flux errors.  The flux table contains measurements of nuclear emission-line fluxes for 39 emission lines for the S7 sample of 131 galaxies.  Table~\ref{table:nuc_fit_fluxes} corresponds to Table~4 in \PII\ but Table~\ref{table:nuc_fit_fluxes} in this work has data for the full sample.  The fluxes for the 14 Seyfert~1 galaxies with special treatment should be used with some caution.  The \Ha\ and \Hb\ fluxes are narrow-component fluxes, and users of this data should be aware of the systematic effects discussed in Section~\ref{sec:Sy1_fitting}.  Additionally the higher-order Balmer lines did not have broad-component subtraction so these fluxes may be unreliable or contain significant broad-component contamination.

Table~\ref{table:nuc_fit_luminosities} in the Appendix presents extinctions, physical aperture sizes, \Hb\ and \OIII\ fluxes and corresponding luminosities from the nuclear spectra for the newly observed S7 galaxies.  A full table with data for the entire S7 sample is available in this data release (Section~\ref{sec:products}).  Table~2 in \PII\ presents similar data as Table~\ref{table:nuc_fit_luminosities} in this work but for the galaxies in the first data release.  The data show that the nuclear extinction in the sample is commonly $A_V \sim 1 - 3$\, mag, and the \OIII\ luminosities vary over approximately four orders of magnitude ($L_{\mbox{\OIII}} \sim 10^{38} - 10^{42}$~erg\,s$^{-1}$).  The caveats associated with fluxes measured in the 14 broad component-subtracted Seyfert~1 spectra also apply to the \Hb\ and \OIII\ fluxes and luminosities in Table~\ref{table:nuc_fit_luminosities}.

\subsection{Nuclear classifications}
\label{sec:nuc_class}

In this section we present the classifications of the nuclear spectra using the optical diagnostic diagram divisions of \citet{Kewley_2006_AGN_hosts}.  Our IFU data ensures that analyses of S7 galaxies do not need to be constrained by single nuclear spectra, but the classifications derived from nuclear spectra are nevertheless informative.  We note that the physical aperture covered by the 4\,arcsec nuclear spectra varies from ${\sim}100$\,pc to ${\sim}1.9$\,kpc over the S7 sample, and as a result the nuclear spectra probe widely varying fractions of galaxies in the sample.

\begin{figure*}
\begin{centering}
\includegraphics[scale=0.85]{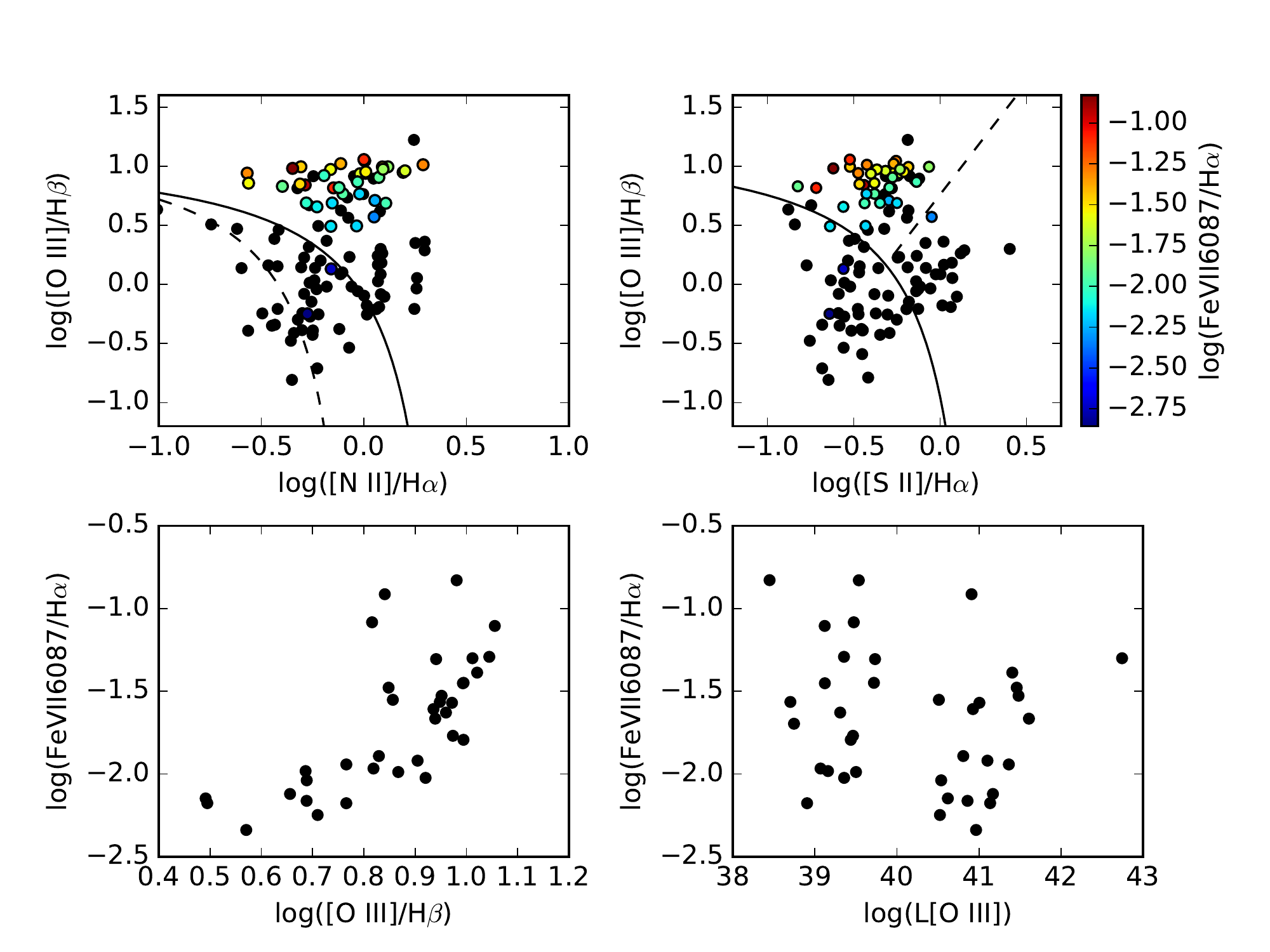}
\end{centering}
\caption{Top row: Optical diagnostic diagrams for nuclear spectra for the S7 sample. The solid lines \citep{Kewley_2001_starburst} and the dashed line in the top-left panel \citep{Kauffmann_2003_AGN} divide \HII\ regions (below) and nebulae with higher ionization states (above).  The dashed line in the top-right panel \citep{Kewley_2006_AGN_hosts} divides LINERs (below) and Seyferts (above).  Black data points represent galaxies which have negligible nuclear [\ion{Fe}{7}]\,$\lambda 6087$ flux.  Bottom row: The relative strength of the [\ion{Fe}{7}]\,$\lambda 6087$ coronal line (for galaxies where it is detected) plotted against the \OIII\,$\lambda 5007 / $\Hb\ diagnostic ratio (left) and plotted against the \OIII\ luminosity (right).  There is a clear trend of increasing [\ion{Fe}{7}]\,$\lambda 6087 / $\Ha\ with increasing \OIII\,$\lambda 5007 / $\Hb.}\label{fig:nuc_BPT}
\end{figure*}

Optical diagnostic diagrams showing classifications of nuclear spectra for the S7 galaxies are presented in the upper two panels in Figure~\ref{fig:nuc_BPT}.  Although many galaxies lie in the `Seyfert' and `LINER' regions of the diagnostic diagrams, a substantial fraction of the galaxies lie in the composite region of the \NII-\Ha\ diagram (above the \citet{Kauffmann_2003_AGN} empirical extreme-starburst line but below the \citet{Kewley_2001_starburst} theoretical extreme-starburst line).  These spectra are contaminated by light from \HII\ regions, either because of significant star formation close to the active nuclei which is unresolved in the S7 cubes, or because of the summing of spaxels in the S7 cubes to produce the nuclear spectra.  Several galaxies lie below the empirical starburst line, and are therefore classified as star-forming and not AGN.  These galaxies include NGC\,4691, which is not an AGN, the misclassified Seyfert\,1 galaxy Fairall\,51, some galaxies such as ESO350-IG38 that have among the highest redshifts in the sample and are therefore suffering strong aperture effects, and NGC\,4472, which is an early-type galaxy with very little line emission.

We note that the nuclear emission-line ratios result in BPT classifications that are broadly consistent with the source catalogue \citep{Veron_2010_13ed}.  In some cases the higher-quality S7 data results in an improvement over an incorrect prior classification, e.g.\ NGC\,4691, which is classified as a Seyfert\,1 in the \citet{Veron_2010_13ed} catalogue.

Figure~\ref{fig:nuc_BPT} also shows an interesting relationship between the strength of the forbidden high-ionization (coronal) emission line [\ion{Fe}{7}]\,$\lambda 6087$ relative to \Ha\ and the diagnostic line ratio \OIII\,$\lambda 5007$ / \Hb.  The coronal line strength shows a clear positive correlation with \OIII\, / \Hb, with the correlation being visible in both the bottom-left panel and the coloring of the data in the upper panels.  We expect the coronal line emission to occur much closer to the central engine of the AGN than the bulk of the \OIII\ and \Hb\ emission, so this correlation may be partly or entirely a consequence of orientation effects \cite[c.f.][]{Rose_2015_CLiF, Glidden_2016_CLiF}.  The theoretical grids published by \citet{Davies16} show that the objects with highest [\ion{Fe}{7}] are located in the region of highest ionization parameter. In these objects, radiation pressure will be dominant, which may produce observable dynamical effects in the ENLR. Objects with extreme radiation pressures will naturally produce extensive Compton heated zones which favor the production of observable coronal lines.

The bottom-right panel of Figure~\ref{fig:nuc_BPT} shows that the relative coronal line strength is not correlated with the \OIII\ luminosity.  This non-correlation supports the idea that orientation effects may be important in producing the correlation with \OIII\, / \Hb.

\subsection{Plots of nuclear spectra}
\label{sec:nuc_spec_plots}

Figures~\ref{fig:nuc_spec_Sy1_A} to \ref{fig:nuc_spec_Sy1_C} in the Appendix show nuclear spectra of the 14 Seyfert galaxies that were identified as having very broad H emission requiring special treatment in emission-line fitting (Section~\ref{sec:Sy1_fitting}).  The figures show spectra summed over the nuclear spaxels identified as having particularly prominent and/or broad emission in Balmer broad-components.  For each galaxy the decomposition of the total spectrum into `narrow' and `broad' components is shown.  The spectra show the diversity of broad-line profiles in the sample and the nature of the residuals remaining after subtraction of the model broad components.

Figure~\ref{fig:nuc_spec} in the Appendix shows nuclear spectra for a selection of galaxies from the new observations.  The varying line ratios are evident between the star-forming, LINER and Seyfert galaxies.  An artefact of the subtraction of the strong [\ion{O}{1}]$\,\lambda 5577$ sky line is visible in all but one of the spectra.

\vspace{1.5cm}
\section{Spatial properties}
\label{sec:spatial_properties}

The spatial properties of the ionized gas in the entire sample of S7 galaxies is illustrated in Figures~\ref{fig:spatial_1} to \ref{fig:spatial_11} in the Appendix.  The galaxies are ordered by RA.  For each galaxy the panels show the following (from left to right):
\begin{itemize}
\item A $118\arcsec \times 180\arcsec$ DSS2 r-band image, with an outline indicating the size, position and orientation of the WiFeS field of view.  North is towards the top of each image, and east is to the left.

\item A three-color image showing the ionization state of the gas over the WiFeS field of view, with colors chosen to distinguish ENLRs in Seyferts from \HII\ regions.  The red channel is \Ha, the green channel is \lNII, and the blue channel is \lOIII.  In high-metallicity \HII\ regions found in Seyfert nuclei, \Ha\ is generally weaker than \lNII\ and \lOIII\ is generally weak, so these regions appear red, gold, or yellow.  The NLR gas has strong \lOIII\ and generally higher \NIIHa, so appears blue, turquoise or green.  For NGC~5253, the bright lines were saturated so the combination of lines used was H$\gamma$, \lNII\ and [\ion{O}{3}]~$\lambda$4959.  A yellow arrow in the top-left indicates the direction of north.  This panel is blank for the QSO PKS~0056-572 and the elliptical galaxies with little line emission NGC\,4472 and 3C\,278.

\item A map of the line-of-sight emission-line gas velocity over the WiFeS field of view, from the single-component \lzifu\ fits (Section~\ref{sec:lzifu}). For ${\sim}18$ galaxies a constant offset was applied to the data to ensure that the zero of the velocity maps corresponded to the nuclear emission-line velocity.  A black circle displayed in the bottom-right shows the size of the seeing FWHM as recorded by the observer.  For the 14 Seyfert~1 galaxies with particularly broad and prominent Balmer lines that required special treatment (Section~\ref{sec:Sy1_fitting}), the green outline delineates the nuclear region where broad components were subtracted before fitting the narrow lines.

\item A map of the line-of-sight emission-line gas velocity dispersion over the WiFeS field of view, from the single-component \lzifu\ fits.  The \lzifu\ fitting corrects the velocity dispersion measurements for the instrumental dispersion in both the red and the blue arms of WiFeS.
\end{itemize}

The figures demonstrate the wide variety of emission-line gas morphology and kinematics found in the sample, including regular rotation (smooth `spider diagrams', e.g.\ NGC\,1346), bi-polar ionization cones delineated by enhanced velocity dispersion (e.g.\ NGC\,7582), star-forming rings (e.g.\ NGC\,5728), an apparently counter-rotating core (NGC\,1068), and `ridges' of enhanced velocity dispersion following spiral arms (NGC\,6926).  Detailed notes are presented on individual objects in Section~\ref{sec:individual_objects}.

\vspace{1.5cm}
\section{Notes on Individual Objects}
\label{sec:individual_objects}

Section~5 of \PII\ presented notes on individual galaxies.  In this section we do the same for the galaxies of interest that were not included in the first data release or were otherwise not described in \PII.  The notes are based on inspection of the maps in the Appendix, the S7 spectral cubes (using {\tt QFitsView}) and ancillary data such as HST observations (where stated).  Galaxies are listed in order of RA.\\

\noindent{\bf MARK938:} A well-studied recent merger, with distinct tidal tails. The nature of the excitation has been controversial - \citet{Veron-Cetty86}  classify it as a Sy\,2 but \citet{Thean00} argue that it should be classified as a starburst rather than a Seyfert. However, our data clearly show it to be a post-starburst galaxy with an A-star dominant continuum spectrum in the blue (very deep Balmer absorption), while the red spectrum is consistent with shock-excited emission. This is visible in all parts of the main galaxy. \newline
{\bf ESO 350-IG38:} Sy\,2 with jet-like ENLR extending $\sim 8$~arcsec from the nucleus at $\mathrm{PA} \sim 210^\circ$. The counter-`jet' is also dimly visible.\newline
{\bf IC 1858: } This poorly studied highly-inclined galaxy shows a spectrum dominated by old stars. Weak LINER-like emission is visible in the red. The \ion{Na}{1} absorption is deep, broad and symmetric, possibly indicating an outflow. \newline
{\bf NGC 1194: } This galaxy had been classified as a Sy\,1, but in our data we see no evidence of broad Balmer lines, and it is now classified as a Sy\,2.  \citet{Schmitt03} imaged this galaxy with HST and found the \OIII\ emission to be only mildy extended over $\sim 2$~arcsec.  However, we observe an X-shaped ENLR extending over $\sim 36$~arcsec in \NII. The opening angle is $30^\circ$, and the $\mathrm{PA} = 170^\circ$. The southern lobe is blueshifted, and the northern lobe is redshifted. There is no sign of recent star formation.\newline
{\bf NGC 1217: } A LINER nucleus with complex dynamics extended within a very smooth elliptical ring of \HII\ regions in fast rotation,\newline
{\bf NGC 1266: } This galaxy shows an intermediate-age stellar spectrum, suggesting that it is post-starburst. It has a prominent  ENLR with a complex, dynamically active, double-lobe morphology extending over $\sim 20$~arcsec at $\mathrm{PA} = 0^\circ$, roughly orthogonal in PA to the underlying disk. The southern lobe of the ENLR is blueshifted, and both lobes have large velocity broadening. The emission spectrum appears more shock-like close to the nuclear dust lane. The [\ion{N}{1}]\,$\lambda\lambda 5158,5200$ doublet reaches an extraordinary strength here, surpassing \Hb\ and rivaling \OIII $\lambda 5007$. The \NII\ lines are also unusually strong, suggesting an enhancement in the nitrogen abundance, and/or the presence of many low-velocity shocks.  This unusual galaxy has a massive AGN-driven molecular outflow and has been shown to have strongly-suppressed star formation \citep{Alatalo_2015_NGC1266}.  The S7 WiFeS data also reveals a prominent galaxy-wide outflow seen in Na\,D doublet absorption that will be the subject of a detailed analysis by Rupke et al. (2017, in prep.).\newline
{\bf NGC 1320: } This high-inclination galaxy has been imaged in \OIII\ and \NII\ using HST by \citet{Ferruit00}. It has a Sy\,2 nucleus, a compact ENLR, and has active star formation largely confined to the narrow inner NW spiral arm. The rotation curve is regular.\newline
{\bf NGC 1346: } This galaxy appears to be interacting with MCG-01-09-041, located 1.6 arc min. away. A high-excitation nucleus is embedded in a disk of \HII\ regions of high metallicity with an undisturbed rotational signature. It is not clear whether the nucleus is simply a local knot of high star formation activity, rather than being a \emph{bona fide} Seyfert.\newline
{\bf NGC 1365: } Sy\,1 nucleus with strong broad \Ha\ and traces of broad \Hb\ with a number of apparently high-metallicity \HII\ regions in a ring-like structure surrounding it. Some regions of the field have very strong post-starburst A-star signatures but little \HII\ activity. The starburst ring is highly inclined with respect to the outer spiral features (which can be traced leading into the inner ring). A brilliant bubble-like ENLR extends across the WiFeS field orthogonally to the starburst ring at $\mathrm{PA} \sim 120^\circ$. The brightest part of this bubble was detected and a dynamical model was presented by \citet{Hjelm96}. The ENLR is also faintly visible on other side of galaxy, and is equally extended ($>15$~arcsec). \newline
{\bf ESO 420-G13: } In \OIII\ there is a one-side ENLR ($\mathrm{PA} \sim 30^\circ$) embedded in a galaxy with a strong post-starburst (A-star) signature. The ENLR can be traced to the edge of the WiFeS field. Close to the nucleus, and on the SW side, there is a bright region of star formation. This probably forms a confining ring to the ENLR. At \Ha\ + \NII\, the reddened counter-cone of the ENLR is visible. This also extends to the edge of the field.\newline
{\bf NGC 1672: } A weak LINER nucleus embedded in a star-forming ring of intermediate-metallicity \HII\ regions first imaged by \citet{Evans96}. \newline
{\bf NGC 1667: } Fairly compact Sy\,2 nucleus surrounded by a ring + spiral arms of \HII\ regions. These are fairly metal-rich and share the strong rotational signature of the galaxy.\newline
{\bf NGC 1808: } An active starburst galaxy with a starburst ring, a warped inner disk, and deep Na\,D absorption.  There is a prominent galaxy-wide outflow that will be the subject of a detailed analysis by Rupke et al. (2017, in prep.).  The nucleus has enhanced \NII\, but no other sign of being a Sy\,2. The \HII\ regions in this galaxy have been catalogued by \citet{Tsvetanov95}.\newline
{\bf UGC 3255: } A heavily reddened Sy\,2 nucleus embedded in an edge-on disk with active star formation. There is evidence of a one-sided ionization cone at $\mathrm{PA}\sim 300^\circ$ with large velocity dispersion at \NII.\newline
{\bf ESO 362-G08: } Obscured Sy\,2 nucleus in post-starburst galaxy seen at high inclination. Very deep \ion{Na}{1} D-line absorption is seen. An amazing high-excitation ``Green Bean" emission region is seen with Sy\,2 emission signature on NE side of nucleus; looks to be a lower abundance minor galaxy blueshifted with respect to the main galaxy. The Green Bean displays a separate rotational signature, confirming that it is a dynamically distinct entity. This region had been first described by \citet{Fraquelli00}, but was interpreted in terms of the ejection of an ENLR from the main galaxy. \newline
{\bf ESO 362-G18: } A somewhat extended velocity-broadened Sy\,2 ENLR (imaged with HST by \citet{Schmitt03}) is seen surrounded by complexes of \HII\ regions and with other \HII\ regions visible across the field. The galaxy has a spiral structure with smooth rotation. In \Ha\ the presence of a Sy\,1 nucleus is evident.\newline
{\bf NGC 2110: } This Sy\,2 galaxy has a very extended two-sided curved ENLR that extends across the WiFeS field (38~arcsec), emerging from the nuclear region at $\mathrm{PA} \sim 0^\circ$. The ENLR displays pronounced bi-polar outflow. The ENLR is much more extended than was revealed by HST \OIII\ imaging, which shows a narrow, 1-arcsec long jet-like feature at $\mathrm{PA} = 340^\circ$, and a weaker feature at $\mathrm{PA} = 160^\circ$ \citep{Mulchaey94}. There is no evidence of active star formation or \HII\ regions.\newline
{\bf NGC 2217: } This is a barred spiral (SBa) seen almost face-on. It displays LINER-like emission and very deep and broad Na D line absorption suggestive of an outflow. A bar-like nuclear emission region is surrounded by a $\Theta$-shaped ``double-bubble" of LINER emission extending right across the WiFeS field ($\sim 38$~arcsec). The gas motions suggest a strongly warped disk, with high velocity dispersion seen within the bubble-like cavities. The distribution and kinematics of this ionized gas has been studied by \citet{Bettoni90}. \newline
{\bf ESO 565-G19: } This poorly studied galaxy contains a LINER nucleus displaying complex knotty dynamics in both \Ha\ and \NII. \HII\ region emission can be detected almost across the full field, and the galaxy displays a smooth rotational signature.\newline
{\bf NGC 2945: } A LINER nucleus with a strange velocity broadened one-sided LINER jet some 12~arcsec long at $\mathrm{PA} = 200^\circ$. There is also a star-forming ring in rapid rotation with major axis diameter 18~arcsec. \newline
{\bf NGC 2974: } This E-type galaxy contains a velocity broadened LINER nucleus embedded in a  faintly emitting disk with a strong rotational signature. The Na~D absorption is deep and wide. The dynamics of this galaxy were studied by \citet{Bettoni92}. In addition there is SAURON data and detailed modeling of both stars and gas by \citet{Krajnovic05}, and HST data has been obtained by \citet{Emsellem03}.\newline
{\bf NGC 2992: } This well-studied interacting edge-on galaxy contains a spectacular double-sided ionization cone with wide opening angle, studied in detail by \citet{Allen99}. \newline
{\bf NGC 3100: } This S0 galaxy displays deep and broad Na~D absorption around the nucleus. A bipolar outflow of gas with a LINER spectrum is seen aligned with the minor axis of the galaxy. In this respect, it is like NGC\,1052. \newline
{\bf ESO 500-G34: } Contains a dynamically active, Sy\,2-like ionization cone ENLR extending $\sim 22$~arcsec E-W along its brightest part. The galaxy itself has a post-starburst spectrum and a number of \HII\ regions are embedded in the plane, which displays a regular rotation curve.\newline
{\bf NGC 3281: } This Sy\,2 galaxy was studied by \citet{Storchi-Bergmann92}, and imaged by HST \citep{Schmitt03}. Our data show a very extended V-shaped ENLR extending north from the centre with an opening angle of $40^\circ$, and a more diffuse bubble-like structure on the other side. The ENLR extent is greater than the 38~arcsec field of WiFeS.  The internal dynamics are suggestive of an outflow. No \HII\ regions are visible. \newline
{\bf MCG-02-27-009: } Displays a faint LINER nucleus embedded in ring of \HII\ regions. \newline
{\bf NGC 3312: } This galaxy has a LINER nucleus with rotation, with a curved (accretion?) stream. Looks very like NGC\,2945. The spectra show deep \ion{Na}{1} absorption. \newline
{\bf NGC 3393: } The HST \OIII\ image of this Seyfert~2 galaxy by \citet{Schmitt03} shows an S-shaped ENLR, extended by 5.3~arcsec (1410~pc) along P.A. = 65~deg and 2.8~arcsec (740~pc) along the perpendicular direction. In our WiFeS \OIII\ image, this is simply the inner part of a much more extended X-shaped double-sided ionization cone with enhanced intensity along the major axis. This can be traced across the full $38 \times 25$~arcsec WiFeS field. The ENLR is embedded in a  faint ring of \HII\ regions with rotation. The \HII\ regions in this galaxy have been catalogued by \citet{Tsvetanov95}. The axis of the ENLR seems to be orthogonal to the rotation.\newline
{\bf NGC 3831: } This poorly studied galaxy shows a very broad component underlying H$\alpha$, so it is probably an obscured Sy\,1.  There is deep \ion{Na}{1} absorption, no signs of recent star formation, and the stellar population is dominated by old stars. \newline
{\bf NGC 3858: } Although classified as a Sy\,2 by \citet{Maia03}, we find a LINER nucleus embedded in a steeply-inclined star-forming ring with a  strong rotational signature. There is a second broken ring of \HII\ regions further out. \newline
{\bf NGC 4507: } This is a bright, Compton-thick \citep{Bianchi06} hidden broad-line region \citep{Moran00} Sy\,2 imaged with HST by \citet{Schmitt03}, who finds an ENLR elongated along $\mathrm{PA} = -35^\circ$ in the inner ${\sim}2$~arcsec. We find a highly dynamically active ENLR with multiple line components close to the nucleus extending both to the SE and NW of the nucleus over a total extent of $\sim 20$~arcsec. Some faint \HII\ regions are also visible in the field. At larger radii there is a ring of bright \HII\ regions cataloged by \citet{Tsvetanov95}.  \newline
{\bf NGC 4594: } The Sombrero galaxy. It displays LINER emission extended across the WiFeS field along the major axis of the galaxy with ubiquitous deep and broad \ion{Na}{1} absorption. There is clear evidence of a broad component underlying \Ha\ at the nucleus. This broad component may be time variable since \citet{Ho97} did not detect it, but a broad component was found in an earlier paper by these authors \citep{Filippenko85}. The extended LINER emission is seen in rotation.\newline
{\bf IC 3639: }Has a Sy\,2 nucleus with an ENLR which can be traced out to a radius of ${\sim}4$\,arcsec. This is embedded in a rich complex of well-resolved \HII\ regions (catalogued by \citet{Tsvetanov95}) that extend all over the field. The region surrounding the nucleus shows a post-starburst continuum spectrum dominated by A-type stars. \newline
{\bf NGC 4636: } A LINER nucleus in an early-type galaxy with a rotating circum-nuclear gas disk. \newline
{\bf NGC 4691: } This galaxy is dominated by metal rich \HII\ regions with an underlying post-starburst continuum. There is no sign of an AGN. \newline
{\bf NGC 4696: } A very extensive and dynamically rich fan-shaped LINER.  The opening angle is $90^\circ$, and the axis is at $\mathrm{PA} = 180^\circ$. Emission can be traced to the edge of the WiFeS field. In some regions the \NII\ and \Ha\ lines are split into two velocity components, while the outer regions are characterised by narrow FWHM. This object has been previously studied in detail with WiFeS by \citet{Farage10}, which showed that the filaments are shock heated remnants of a minor merger.\newline
{\bf ESO 443-G17: } A compact \HII\ disk in very rapid and smooth rotation. \citet{Kewley00} detected a compact radio core in this warm IR galaxy.\newline
{\bf NGC 4845: } This steeply-inclined Sy\,2 galaxy has a magnificent ionization cone with an opening angle of $120^\circ$ emerging perpendicularly from a circum-nuclear disk of \HII\ regions with its axis at $\mathrm{PA} = 350^\circ$. The counter ionization cone is faintly visible, mainly in \NII\ because of the high extinction. Together, both ionization cones fill the WiFeS field of view. The dynamics of the ENLR are dominated by the underlying rotation of the galaxy disk.\newline
{\bf NGC 4939: } A compact Sy\,2 nucleus is seen in \OIII\ but it is clearly extended by 16~arcsec along the bar ($\mathrm{PA} = 5^\circ$) in \NII. However, with HST, a biconical ionization region is seen both in \OIII\ and in the UV extended perpendicularly to the disk of the galaxy \citep{Martini03,Munoz07}.  There are a number of faint \HII\ regions in the field. The \HII\ regions in this galaxy have been catalogued by \citet{Tsvetanov95}.\newline
{\bf ESO 323-G77: } This is a partially obscured Sy\,1 with polarised emission close to the nucleus \citep{Schmid03}. We find a two-sided ionization cone with a remarkably narrow opening angle ($\sim 20^\circ$) extending N-S across the field. The inner region of this galaxy was mapped with HST by \citet{Mulchaey96}. There is evidence of active star formation activity close to the nucleus (strong \Hb\ emission and deep absorption).\newline
{\bf NGC 4968: } This Sy\,2  galaxy has been imaged in \OIII\ and \NII\ using HST by \citet{Ferruit00} and in \OIII\ by \citet{Schmitt03}. With HST there is a fan-shaped ENLR extended towards the SE by ${\sim}1.5$\,arcsec. This  can be traced out to 7\,arcsec in our data, and the counter ionization cone is also faintly visible. Previous ground-based data by \citet{Pogge89} did not detect any extended \OIII\ emission.\newline
{\bf ARK 402: } A LINER nucleus embedded in a rotating disk.\newline
{\bf NGC 4990: } This galaxy has a bright starburst nucleus and is not an AGN.\newline
{\bf MCG-03-34-064: } Although classified by \citet{Aguero94} as a Sy\,1.8, we find no evidence that the nucleus is anything other than a LINER. The nucleus is surrounded by a rapidly rotating ring of \HII\ regions. \newline
{\bf NGC 5128: } The Centaurus A galaxy. The highly obscured Sy\,2 nucleus and ENLR is seen embedded in a complex of \HII\ regions.\newline
{\bf NGC 5135: } A Sy\,2 with a double-sided ionization cone with opening angle $\sim 40^\circ$.  The ENLR can be traced over the WiFeS field ($\sim 30$ arc sec). \HII\ regions are present and a post-starburst stellar continuum can be traced along the major axis of the bar - c.f. \citet{Gonzalez98,Gonzalez01}. \newline
{\bf ESO 383-G35: } This E/S0 galaxy contains a Sy\,1 with coronal line emission. It has been imaged in \OIII\ and \NII\ using HST by \citet{Ferruit00}. We find an extensive ENLR that can be traced along the major axis of the galaxy.\newline
{\bf NGC 5252: } This galaxy displays a spectacular bi-conical ionization structure with arcs, first described by \citet{Tadhunter89}. In the bright inner region, strong line splitting is visible. A spectral analysis of this galaxy was performed by \citet{Acosta96}; both their data and our data provides evidence for the presence of a broad component to \Ha\ in the nucleus. \newline
{\bf IC 4329A: } A bright Sy\,1 nucleus is surrounded by a compact dynamically active ENLR seen by \citet{Colbert96}. We find a fainter, more dynamically quiescent, X-shaped and limb-brightened ionization cone with an opening angle of $30^\circ$, which extends right across the 38\,arcsec WiFeS field.\newline
{\bf NGC 5427: } This galaxy is tidally interacting with NGC 5426.  A detailed description of the ENLR in this galaxy and its properties using WiFeS data is given in Paper~I. \newline
{\bf NGC 5643: } Since this Compton-thick galaxy was known to have a spectacular ENLR \citep{Schmitt94}, we obtained a 2 field mosaic in order to map it completely. Nonetheless with our greater sensitivity, we find the \OIII\ emission can be traced right to the edge of the field (over 48\,arcsec N-S and 38\,arcsec E-W). On the southern side the brightest part of the ENLR is knotty and curved. The \HII\ regions in this galaxy have been catalogued by \citet{Tsvetanov95}.\newline
{\bf NGC 5990: } A bright post-starbust galaxy with a compact Sy\,2 nucleus. There is a bright \HII\ region complex to the NE of the nucleus, apparently of high metallicity.\newline
{\bf NGC 6328: } This, the closest known GPS galaxy \citep{Tingay97}, has a bright, spatially extended and velocity broadened LINER nucleus. \newline
{\bf IC 4777: } A Sy\,2 nucleus embedded in an elliptical star-forming ring.\newline
{\bf ESO 460-G09: }A LINER nucleus is surrounded by a number of \HII\ region complexes. \newline
{\bf MCG-02-51-008: } This poorly studied galaxy has a bright Sy\,2 nucleus with an extensive two-sided ENLR emerging at $\mathrm{PA} = 110^\circ$, but twisting and expanding towards the S on the E side, and towards the N on the W side.  This is embedded in a rich star-forming disk with bright \HII\ region complexes.\newline
{\bf NGC 6936: } A bright, extensive and rotating LINER disk with very prominent and wide Na~D-line absorption on the nucleus.\newline
{\bf MCG-04-49-001: } This galaxy has a very rapidly rotating disk with a pronounced warp. It shows a LINER nucleus surrounded by a ring of bright \HII\ regions.\newline
{\bf ESO 287-G42: } A LINER nucleus with two spiral arms rich in \HII\ regions. There is a hint of a broad component underlying H$\alpha$.\newline
{\bf NGC 7172: } A two-sided ionization cone with opening angle $120^\circ$ emerges in the polar direction from a highly dust-obscured nucleus in this highly inclined galaxy, a member of the compact group HCG90. The southern cone is the brighter. A line of \HII\ regions is visible on either side of the central E-W dust lane. \newline
{\bf IC 5169: } A two-sided ionization cone with opening angle $130^\circ$ emerges in the polar direction from a bright, rapidly rotating circum-nuclear \HII\ starburst ring. This is surrounded at larger radius by a second ring of fainter \HII\ regions with lower rotational velocities. The intimate connection of the ENLR with the inner starburst ring explains the ambiguities in the earlier classification of the galaxy - see the discussion by \citet{Zaw09}.\newline
{\bf IC 1481: } This extraordinary galaxy has a post-starburst circum-nuclear continuum spectrum with apparently large reddening at the Sy\,2 nucleus itself. The ENLR is very extensive, its brightest portions forming a figure of eight extending ${\sim}13$\,arcsec at $\mathrm{PA} = 50^\circ$ either side of the nucleus. Additional knots of emission are visible to the edges of the field.\newline

\vspace{1.5cm}
\section{Conclusions}
\label{sec:conclusions}

We have presented the second and final data release of the {\it Siding Spring Southern Seyfert Spectroscopic Snapshot Survey} (S7).  The S7 uses the WiFeS instrument on the ANU 2.3\,m telescope at Siding Spring Observatory to obtain integral field data with a spectral resolution of $R = 3000$ in the blue and $R = 7000$ in the red.  The final S7 sample consists of 131 galaxies, including 63 galaxies observed since the first data release.  There are approximately 18 Seyfert~1 galaxies, 77 Seyfert~2 galaxies, 27 LINERs, and 11 starbursts.  The sample was selected to consist of nearby galaxies ($z \lesssim 0.02$), which are almost all detected in the radio, and are in the southern sky ($\delta < {+}10^\circ$) at galactic latitudes of $|b| \gtrsim 15^\circ$.

The data products provided with this work include data cubes from both the blue and red arms of WiFeS, emission line fits (with 1, 2 and 3 Gaussian components as well as a `best' number of components chosen by an artificial neural network), nuclear spectra extracted using a 4~arcsec aperture, and reddening-corrected emission-line fluxes for the nuclear spectra.  Maps of gas excitation, gas line-of-sight velocity and gas line-of-sight velocity dispersion were presented for each galaxy.  Additionally we have classified the nuclear spectra using optical diagnostic diagrams, demonstrating that the strength of the coronal line [\ion{Fe}{7}]\,$\lambda6087$ is correlated with the diagnostic line ratio \OIII / \Hb.  We have also provided notes on each of the newly observed objects.

The S7 is a basis set for comparison for all future AGN surveys. The large sample size and the wide field of view, high dynamic range and high spectral resolution of the integral field data form a foundation for important insights into the physics of Active Galactic Nuclei.

M.D. and L.K. acknowledge support from ARC discovery project \#DP160103631.  Parts of this research were conducted by the Australian Research Council Centre of Excellence for All-sky Astrophysics (CAASTRO), through project number CE110001020.
Prajval Shastri acknowledges the Distinguished Visitor Fellowship at the RSAA, Australian National University where part of this work was undertaken.  This research is supported by an Australian Government Research Training Program (RTP) Scholarship.  M.N. Sundar acknowledges the Indian Institute of Astrophysics student internship.  This research has made extensive use of the National Aeronautics and Space Administration (NASA) Astrophysics Data System Bibliographic Services, the NASA/IPAC Extragalactic Database which is operated by the Jet Propulsion Laboratory, California Institute of Technology, under contract with NASA, and data, software and/or web tools obtained from the NASA High Energy Astrophysics Science Archive Research Center, a service of the Goddard Space Flight Center and the Smithsonian Astrophysical Observatory.



\begin{thebibliography}{}
\bibitem[Acosta-Pulido et al.(1996)]{Acosta96} Acosta-Pulido, J.~A., Vila-Vilaro, B., Perez-Fournon, I., Wilson, A.~S., \& Tsvetanov, Z.~I.\ 1996, \apj, 464, 177 
\bibitem[Aguero et al.(1994)]{Aguero94} Aguero, E.~L., Calderon, J.~H., Paolantonio, S., \& Suarez Boedo, E.\ 1994, \pasp, 106, 978 
\bibitem[Allen et al.(1999)]{Allen99} Allen, M.~G., Dopita, M.~A., Tsvetanov, Z.~I., \& Sutherland, R.~S.\ 1999, \apj, 511, 686 
\bibitem[Allen et al.(2015)]{Allen_2015_SAMI_EDR} Allen, J.~T., Croom, S.~M., Konstantopoulos, I.~S., et al.\ 2015, \mnras, 446, 1567
\bibitem[Allen et al.(2015)]{Allen_2015_offset_AGN} Allen, J.~T., Schaefer, A.~L., Scott, N., et al.\ 2015, \mnras, 451, 2780
\bibitem[Alatalo et al.(2015)]{Alatalo_2015_NGC1266} Alatalo, K., Lacy, M., Lanz, L., et al.\ 2015, \apj, 798, 31
\bibitem[Antonucci (1993)]{Antonucci_1993_AGN_unified} Antonucci, R.\ 1993, \araa, 31, 473
\bibitem[Baldwin, Phillips and Terlevich (1981)]{BPT_1981} Baldwin, J.~A., Phillips, M.~M., \& Terlevich, R.\ 1981, \pasp, 93, 5
\bibitem[Becker et al.(1995)]{Becker_1995_FIRST} Becker, R.~H., White, R.~L., \& Helfand, D.~J.\ 1995, \apj, 450, 559
\bibitem[Belfiore et al.(2016)]{Belfiore_2016_MaNGA_LIERs} Belfiore, F., Maiolino, R., Maraston, C., et al.\ 2016, \mnras, 461, 3111
\bibitem[Bettoni et al.(1990)]{Bettoni90} Bettoni, D., Fasano, G., \& Galletta, G.\ 1990, \aj, 99, 1789 
\bibitem[Bettoni(1992)]{Bettoni92} Bettoni, D.\ 1992, \aaps, 96, 333 
\bibitem[Bianchi et al.(2006)]{Bianchi06} Bianchi, S., Guainazzi, M., \& Chiaberge, M.\ 2006, \aap, 448, 499 
\bibitem[Binette et al.(1994)]{Binette94} Binette, L., Magris, C.~G., Stasi{\'n}ska, G., \& Bruzual, A.~G.\ 1994, \aap, 292, 13
\bibitem[Blanc et al.(2013)]{Blanc_2013_VENGA} Blanc, G.~A., Weinzirl, T., Song, M., et al.\ 2013, \aj, 145, 138
\bibitem[Bryant et al.(2015)]{Bryant_2015_SAMI} Bryant, J.~J., Owers, M.~S., Robotham, A.~S.~G., et al.\ 2015, \mnras, 447, 2857
\bibitem[Bundy et al.(2015)]{Bundy_2015_MaNGA_overview} Bundy, K., Bershady, M.~A., Law, D.~R., et al.\ 2015, \apj, 798, 7
\bibitem[Calzetti et al.(2000)]{Calzetti_2000_Dust} Calzetti, D., Armus, L., Bohlin, R.~C., et al.\ 2000, \apj, 533, 682
\bibitem[Cappellari \& Emsellem(2004)]{Cappellari_Emsellem_2004_ppxf} Cappellari, M., \& Emsellem, E.\ 2004, \pasp, 116, 138
\bibitem[Childress et al.(2014)]{Childress_2014_PyWiFeS} Childress, M.~J., Vogt, F.~P.~A., Nielsen, J., \& Sharp, R.~G.\ 2014, \apss, 349, 617
\bibitem[Cluver et al.(2014)]{Cluver_2014_GAMA_MIR} Cluver, M.~E., Jarrett, T.~H., Hopkins, A.~M., et al.\ 2014, \apj, 782, 90
\bibitem[Colbert et al.(1996)]{Colbert96} Colbert, E.~J.~M., Baum, S.~A., Gallimore, J.~F., et al.\ 1996, \apjs, 105, 75 
\bibitem[Condon et al.(1998)]{Condon_1998_NVSS_survey} Condon, J.~J., Cotton, W.~D., Greisen, E.~W., et al.\ 1998, \aj, 115, 1693
\bibitem[Croom et al.(2012)]{Croom_2012_SAMI} Croom, S.~M., Lawrence, J.~S., Bland-Hawthorn, J., et al.\ 2012, \mnras, 421, 872
\bibitem[Davies et al.(2016a)]{Davies16} Davies, R.~L., Dopita, M.~A., Kewley, L., et al.\ 2016, \apj, 824, 50
\bibitem[Davies et al.(2016b)]{Davies16b} Davies, R.~L., Groves, B., Kewley, L.~J., et al.\ 2016, \mnras, 462, 1616
\bibitem[Diamond-Stanic \& Rieke(2012)]{Diamond-Stanic_Rieke_2012_Sy_sample} Diamond-Stanic, A.~M., \& Rieke, G.~H.\ 2012, \apj, 746, 168
\bibitem[Dopita \& Sutherland(1995)]{Dopita95} Dopita, M.~A., \& Sutherland, R.~S.\ 1995, \apj, 455, 468
\bibitem[Dopita et al.(2002)]{Dopita_2002_Dusty_NLRs} Dopita, M.~A., Groves, B.~A., Sutherland, R.~S., Binette, L., \& Cecil, G.\ 2002, \apj, 572, 753
\bibitem[Dopita et al.(2007)]{Dopita_2007_WiFeS_I} Dopita, M., Hart, J., McGregor, P., et al.\ 2007, \apss, 310, 255
\bibitem[Dopita et al.(2010)]{Dopita_2007_WiFeS_II} Dopita, M., Rhee, J., Farage, C., et al.\ 2010, \apss, 327, 245
\bibitem[Dopita et al.(2014)]{Dopita_2014_S7_I} Dopita, M.~A., Scharw{\"a}chter, J., Shastri, P., et al.\ 2014, \aap, 566, A41 (Paper~I)
\bibitem[Dopita et al.(2015)]{Dopita_2015_S7_II} Dopita, M.~A., Shastri, P., Davies, R., et al.\ 2015, \apjs, 217, 12 (\PII)
\bibitem[Dopita et al.(2015)]{Dopita_2015_S7_III} Dopita, M.~A., Ho, I.-T., Dressel, L.~L., et al.\ 2015, \apj, 801, 42 (Paper~III)
\bibitem[Emsellem et al.(2003)]{Emsellem03} Emsellem, E., Goudfrooij, P., \& Ferruit, P.\ 2003, \mnras, 345, 1297 
\bibitem[Elbaz et al.(2007)]{Elbaz_2007_MS} Elbaz, D., Daddi, E., Le Borgne, D., et al.\ 2007, \aap, 468, 33 
\bibitem[Eracleous \& Halpern(2001)]{Eracleous01} Eracleous, M., \& Halpern, J.~P.\ 2001, \apj, 554, 240
\bibitem[Evans et al.(1996)]{Evans96} Evans, I.~N., Koratkar, A.~P., Storchi-Bergmann, T., et al.\ 1996, \apjs, 105, 93 
\bibitem[Farage et al.(2010)]{Farage10} Farage, C.~L., McGregor, P.~J., Dopita, M.~A., \& Bicknell, G.~V.\ 2010, \apj, 724, 267 
\bibitem[Ferrarese \& Merritt(2000)]{Ferrarese_Merritt_2000_SMBH_sigma_scaling} Ferrarese, L., \& Merritt, D.\ 2000, \apjl, 539, L9
\bibitem[Ferruit et al.(2000)]{Ferruit00} Ferruit, P., Wilson, A.~S., \& Mulchaey, J.\ 2000, \apjs, 128, 139
\bibitem[Filippenko(1982)]{Filippenko_1982_DAR} Filippenko, A.~V.\ 1982, \pasp, 94, 715 
\bibitem[Filippenko \& Sargent(1985)]{Filippenko85} Filippenko, A.~V., \& Sargent, W.~L.~W.\ 1985, \apjs, 57, 503 
\bibitem[F{\"o}rster Schreiber et al.(2009)]{Forster_Schreiber_2009_SINS} F{\"o}rster Schreiber, N.~M., Genzel, R., Bouch{\'e}, N., et al.\ 2009, \apj, 706, 1364
\bibitem[F{\"o}rster Schreiber et al.(2014)]{Forster_Schreiber_2014_SINS_AGN} F{\"o}rster Schreiber, N.~M., Genzel, R., Newman, S.~F., et al.\ 2014, \apj, 787, 38
\bibitem[Fraquelli et al.(2000)]{Fraquelli00} Fraquelli, H.~A., Storchi-Bergmann, T., \& Binette, L.\ 2000, \apj, 532, 867 
\bibitem[Garc{\'{\i}}a-Benito et al.(2015)]{Garcia-Benito_2015_CALIFA_DR2} Garc{\'{\i}}a-Benito, R., Zibetti, S., S{\'a}nchez, S.~F., et al.\ 2015, \aap, 576, A135
\bibitem[Glidden et al.(2016)]{Glidden_2016_CLiF} Glidden, A., Rose, M., Elvis, M., \& McDowell, J.\ 2016, \apj, 824, 34
\bibitem[Gonz{\'a}lez Delgado et al.(1998)]{Gonzalez98} Gonz{\'a}lez Delgado, R.~M., Heckman, T., Leitherer, C., et al.\ 1998, \apj, 505, 174 
\bibitem[Gonz{\'a}lez Delgado et al.(2001)]{Gonzalez01} Gonz{\'a}lez Delgado, R.~M., Heckman, T., \& Leitherer, C.\ 2001, \apj, 546, 845 
\bibitem[Gonz{\'a}lez Delgado et al.(2005)]{Gonzalez_Delgado_2005_SSP} Gonz{\'a}lez Delgado, R.~M., Cervi{\~n}o, M., Martins, L.~P., Leitherer, C., \& Hauschildt, P.~H.\ 2005, \mnras, 357, 945
\bibitem[Greene \& Ho(2005)]{Greene_Ho_2005_broad_fitting} Greene, J.~E., \& Ho, L.~C.\ 2005, \apj, 630, 122
\bibitem[Groves et al.(2004a)]{Groves_2004_NLR_I} Groves, B.~A., Dopita, M.~A., \& Sutherland, R.~S.\ 2004, \apjs, 153, 9
\bibitem[Groves et al.(2004b)]{Groves_2004_NLR_II} Groves, B.~A., Dopita, M.~A., \& Sutherland, R.~S.\ 2004, \apjs, 153, 75
\bibitem[Heckman(1980)]{Heckman_1980_LINERs} Heckman, T.~M.\ 1980, \aap, 87, 152
\bibitem[Heckman \& Best(2014)]{Heckman_Best_2014_AGN_review} Heckman, T.~M., \& Best, P.~N.\ 2014, \araa, 52, 589
\bibitem[Helou et al.(1988)]{Helou_1988_IRAS} Helou, G., Khan, I.~R., Malek, L., \& Boehmer, L.\ 1988, \apjs, 68, 151
\bibitem[Hicks et al.(2016)]{Hicks_2016_KONA} Hicks, E.~K.~S., M{\"u}ller-S{\'a}nchez, F., Malkan, M.~A., \& Yu, P.-C.\ 2016, Active Galactic Nuclei: What's in a Name?, 41
\bibitem[Hjelm \& Lindblad(1996)]{Hjelm96} Hjelm, M., \& Lindblad, P.~O.\ 1996, \aap, 305, 727 
\bibitem[Ho et al.(1997)]{Ho97} Ho, L.~C., Filippenko, A.~V., Sargent, W.~L.~W., \& Peng, C.~Y.\ 1997, \apjs, 112, 391 
\bibitem[Ho et al.(2001)]{Ho01} Ho, L.~C., Feigelson, E.~D., Townsley, L.~K., et al.\ 2001, \apjl, 549, L51
\bibitem[Ho(2008)]{Ho_2008_AGN_review} Ho, L.~C.\ 2008, \araa, 46, 475
\bibitem[Ho et al.(2014)]{Ho14} Ho, I.-T., Kewley, L.~J., Dopita, M.~A., et al.\ 2014, \mnras, 444, 3894
\bibitem[Ho et al.(2016)]{Ho_2016_SAMI_winds} Ho, I.-T., Medling, A.~M., Bland-Hawthorn, J., et al.\ 2016, \mnras, 457, 1257
\bibitem[Ho et al.(2016)]{Ho_2016_LZIFU} Ho, I.-T., Medling, A.~M., Groves, B., et al.\ 2016, \apss, 361, 280
\bibitem[Husemann et al.(2016)]{Husemann_2016_CARS_MARK1018} Husemann, B., Urrutia, T., Tremblay, G.~R., et al.\ 2016, \aap, 593, L9
\bibitem[Karouzos et al.(2016a)]{Karouzos_2016_GMOS_outflows} Karouzos, M., Woo, J.-H., \& Bae, H.-J.\ 2016, \apj, 819, 148
\bibitem[Karouzos et al.(2016b)]{Karouzos_2016_erratum} Karouzos, M., Woo, J.-H., \& Bae, H.-J.\ 2016, \apj, 828, 64
\bibitem[Kauffmann et al.(2003)]{Kauffmann_2003_AGN} Kauffmann, G., Heckman, T.~M., Tremonti, C., et al.\ 2003, \mnras, 346, 1055
\bibitem[Kehrig et al.(2012)]{Kehrig12} Kehrig, C., Monreal-Ibero, A., Papaderos, P., et al.\ 2012, \aap, 540, A11
\bibitem[Kennicutt(1998)]{Kennicutt_1998_SFRs} Kennicutt, R.~C., Jr.\ 1998, \araa, 36, 189
\bibitem[Kewley et al.(2000)]{Kewley00} Kewley, L.~J., Heisler, C.~A., Dopita, M.~A., et al.\ 2000, \apj, 530, 704 
\bibitem[Kewley et al.(2001)]{Kewley_2001_starburst} Kewley, L.~J., Dopita, M.~A., Sutherland, R.~S., Heisler, C.~A., \& Trevena, J.\ 2001, \apj, 556, 121
\bibitem[Kewley et al.(2006)]{Kewley_2006_AGN_hosts} Kewley, L.~J., Groves, B., Kauffmann, G., \& Heckman, T.\ 2006, \mnras, 372, 961
\bibitem[Krajnovi{\'c} et al.(2005)]{Krajnovic05} Krajnovi{\'c}, D., Cappellari, M., Emsellem, E., McDermid, R.~M., \& de Zeeuw, P.~T.\ 2005, \mnras, 357, 1113 
\bibitem[Kormendy \& Ho(2013)]{Kormendy_Ho_2013_review} Kormendy, J., \& Ho, L.~C.\ 2013, \araa, 51, 511
\bibitem[Leslie et al.(2016)]{Leslie_2016_MS} Leslie, S.~K., Kewley, L.~J., Sanders, D.~B., \& Lee, N.\ 2016, \mnras, 455, L82 
\bibitem[L{\'{\i}}pari et al.(2004)]{Lipari04} L{\'{\i}}pari, S., Mediavilla, E., Garcia-Lorenzo, B., et al.\ 2004, \mnras, 355, 641
\bibitem[Magorrian et al.(1998)]{Magorrian_1998} Magorrian, J., Tremaine, S., Richstone, D., et al.\ 1998, \aj, 115, 2285
\bibitem[Maia et al.(2003)]{Maia03} Maia, M.~A.~G., Machado, R.~S., \& Willmer, C.~N.~A.\ 2003, \aj, 126, 1750 
\bibitem[Mancini et al.(2011)]{Mancini_2011_zC_SINF} Mancini, C., F{\"o}rster Schreiber, N.~M., Renzini, A., et al.\ 2011, \apj, 743, 86
\bibitem[Markwardt(2009)]{Markwardt_2009_MPFIT} Markwardt, C.~B.\ 2009, Astronomical Data Analysis Software and Systems XVIII, 411, 251
\bibitem[Martini et al.(2003)]{Martini03} Martini, P., Regan, M.~W., Mulchaey, J.~S., \& Pogge, R.~W.\ 2003, \apjs, 146, 353 
\bibitem[McConnell \& Ma(2013)]{McConnell_2013_SMBH_host_scaling} 
McConnell, N.~J., \& Ma, C.-P.\ 2013, \apj, 764, 184
\bibitem[McElroy et al.(2015)]{McElroy_2015_IFU} McElroy, R., Croom, S.~M., Pracy, M., et al.\ 2015, \mnras, 446, 2186
\bibitem[McElroy et al.(2016)]{McElroy_2016_CARS} McElroy, R.~E., Husemann, B., Croom, S.~M., et al.\ 2016, \aap, 593, L8
\bibitem[Monreal-Ibero et al.(2006)]{Monreal-Ibero06} Monreal-Ibero, A., Arribas, S., \& Colina, L.\ 2006, \apj, 637, 138
\bibitem[Moran et al.(2000)]{Moran00} Moran, E.~C., Barth, A.~J., Kay, L.~E., \& Filippenko, A.~V.\ 2000, \apjl, 540, L73 
\bibitem[Mulchaey et al.(1994)]{Mulchaey94} Mulchaey, J.~S., Wilson, A.~S., Bower, G.~A., et al.\ 1994, \apj, 433, 625 
\bibitem[Mulchaey et al.(1996)]{Mulchaey96} Mulchaey, J.~S., Wilson, A.~S., \& Tsvetanov, Z.\ 1996, \apjs, 102, 309 
\bibitem[Mu{\~n}oz Mar{\'{\i}}n et al.(2007)]{Munoz07} Mu{\~n}oz Mar{\'{\i}}n, V.~M., Gonz{\'a}lez Delgado, R.~M., Schmitt, H.~R., et al.\ 2007, \aj, 134, 648 
\bibitem[Pence et al.(2010)]{Pence_2010_FITS} Pence, W.~D., Chiappetti, L., Page, C.~G., Shaw, R.~A., \& Stobie, E.\ 2010, \aap, 524, A42
\bibitem[Pogge(1989)]{Pogge89} Pogge, R.~W.\ 1989, \apj, 345, 730 
\bibitem[Ricci et al.(2014a)]{Ricci_2014_IFU_I} Ricci, T.~V., Steiner, J.~E., \& Menezes, R.~B.\ 2014, \mnras, 440, 2419
\bibitem[Ricci et al.(2014b)]{Ricci_2014_IFU_II} Ricci, T.~V., Steiner, J.~E., \& Menezes, R.~B.\ 2014, \mnras, 440, 2442
\bibitem[Ricci et al.(2015a)]{Ricci_2015_IFU_II_erratum} Ricci, T.~V., Steiner, J.~E., \& Menezes, R.~B.\ 2015, \mnras, 447, 1504
\bibitem[Ricci et al.(2015b)]{Ricci_2015_IFU_III} Ricci, T.~V., Steiner, J.~E., \& Menezes, R.~B.\ 2015, \mnras, 451, 3728
\bibitem[Rich et al.(2011)]{Rich11} Rich, J.~A., Kewley, L.~J., \& Dopita, M.~A.\ 2011, \apj, 734, 87
\bibitem[Rich et al.(2014)]{Rich_2014_WiFeS_GOALS_shocks} Rich, J.~A., Kewley, L.~J., \& Dopita, M.~A.\ 2014, \apjl, 781, L12
\bibitem[Rich et al.(2015)]{Rich_2015_WiFeS_GOALS} Rich, J.~A., Kewley, L.~J., \& Dopita, M.~A.\ 2015, \apjs, 221, 28
\bibitem[Riffel et al.(2008)]{Riffel_2008} Riffel, R.~A., Storchi-Bergmann, T., Winge, C., et al.\ 2008, \mnras, 385, 1129
\bibitem[Riffel et al.(2009)]{Riffel_2009} Riffel, R.~A., Storchi-Bergmann, T., Dors, O.~L., \& Winge, C.\ 2009, \mnras, 393, 783
\bibitem[Riffel \& Storchi-Bergmann(2011a)]{Riffel_Storchi-Bergman_2011a} Riffel, R.~A., \& Storchi-Bergmann, T.\ 2011, \mnras, 411, 469
\bibitem[Riffel \& Storchi-Bergmann(2011b)]{Riffel_Storchi-Bergman_2011b} Riffel, R.~A., \& Storchi-Bergmann, T.\ 2011, \mnras, 417, 2752
\bibitem[Riffel et al.(2013)]{Riffel_2013} Riffel, R.~A., Storchi-Bergmann, T., \& Winge, C.\ 2013, \mnras, 430, 2249
\bibitem[Rothberg et al.(2015)]{Rothberg_2015_CARS} Rothberg, B., Husemann, B., Busch, G., et al.\ 2015, IAU General Assembly, 22, 2257803
\bibitem[Rose et al.(2015)]{Rose_2015_CLiF} Rose, M., Elvis, M., \& Tadhunter, C.~N.\ 2015, \mnras, 448, 2900
\bibitem[S{\'a}nchez et al.(2012)]{Sanchez_2012_CALIFA_survey} S{\'a}nchez, S.~F., Kennicutt, R.~C., Gil de Paz, A., et al.\ 2012, \aap, 538, A8
\bibitem[S{\'a}nchez et al.(2016)]{Sanchez_2016_CALIFA_DR3} S{\'a}nchez, S.~F., Garc{\'{\i}}a-Benito, R., Zibetti, S., et al.\ 2016, \aap, 594, A36
\bibitem[Scharw{\"a}chter et al.(2016)]{Scharwachter_2016_NGC6300} Scharw{\"a}chter, J., Dopita, M.~A., Shastri, P., et al.\ 2016, The Universe of Digital Sky Surveys, 42, 263
\bibitem[Schmitt et al.(1994)]{Schmitt94} Schmitt, H.~R., Storchi-Bergmann, T., \& Baldwin, J.~A.\ 1994, \apj, 423, 237 
\bibitem[Schmid et al.(2003)]{Schmid03} Schmid, H.~M., Appenzeller, I., \& Burch, U.\ 2003, \aap, 404, 505 
\bibitem[Schmitt et al.(2003)]{Schmitt03} Schmitt, H.~R., Donley, J.~L., Antonucci, R.~R.~J., Hutchings, J.~B., \& Kinney, A.~L.\ 2003, \apjs, 148, 327 
\bibitem[Sharp et al.(2015)]{Sharp_2015_SAMI} Sharp, R., Allen, J.~T., Fogarty, L.~M.~R., et al.\ 2015, \mnras, 446, 1551
\bibitem[Singh et al.(2013)]{Singh_2013_CALIFA_LINERs} Singh, R., van de Ven, G., Jahnke, K., et al.\ 2013, \aap, 558, A43
\bibitem[Storchi-Bergmann et al.(1992)]{Storchi-Bergmann92} Storchi-Bergmann, T., Wilson, A.~S., \& Baldwin, J.~A.\ 1992, \apj, 396, 45 
\bibitem[Storchi-Bergmann et al.(1997)]{Storchi-Bergmann97} Storchi-Bergmann, T., Eracleous, M., Teresa Ruiz, M., et al.\ 1997, \apj, 489, 87
\bibitem[Storchi-Bergmann et al.(2009)]{Storchi-Bergmann_2009} Storchi-Bergmann, T., McGregor, P.~J., Riffel, R.~A., et al.\ 2009, \mnras, 394, 1148
\bibitem[Storchi-Bergmann et al.(2010)]{Storchi-Bergmann_2010} Storchi-Bergmann, T., Lopes, R.~D.~S., McGregor, P.~J., et al.\ 2010, \mnras, 402, 819
\bibitem[Stott et al.(2016)]{Stott_2016_KROSS} Stott, J.~P., Swinbank, A.~M., Johnson, H.~L., et al.\ 2016, \mnras, 457, 1888
\bibitem[Tadhunter \& Tsvetanov(1989)]{Tadhunter89} Tadhunter, C., \& Tsvetanov, Z.\ 1989, \nat, 341, 422 
\bibitem[Thean et al.(2000)]{Thean00} Thean, A., Pedlar, A., Kukula, M.~J., Baum, S.~A., \& O'Dea, C.~P.\ 2000, \mnras, 314, 573 
\bibitem[Thomas et al.(2016)]{Thomas_2016_OXAF} Thomas, A.~D., Groves, B.~A., Sutherland, R.~S., et al.\ 2016, \apj, 833, 266 
\bibitem[Tingay et al.(1997)]{Tingay97} Tingay, S.~J., Jauncey, D.~L., Reynolds, J.~E., et al.\ 1997, \aj, 113, 2025 
\bibitem[Tsvetanov \& Petrosian(1995)]{Tsvetanov95} Tsvetanov, Z.~I., \& Petrosian, A.~R.\ 1995, \apjs, 101, 287 
\bibitem[Ulvestad \& Ho(2001)]{Ulvestad01} Ulvestad, J.~S., \& Ho, L.~C.\ 2001, \apjl, 562, L133
\bibitem[Veron-Cetty \& Veron(1986)]{Veron-Cetty86} Veron-Cetty, M.-P., \& Veron, P.\ 1986, \aaps, 66, 335 
\bibitem[V{\'e}ron-Cetty \& V{\'e}ron(2006)]{Veron_2006_12ed} V{\'e}ron-Cetty, M.-P., \& V{\'e}ron, P.\ 2006, \aap, 455, 773
\bibitem[V{\'e}ron-Cetty \& V{\'e}ron(2010)]{Veron_2010_13ed} V{\'e}ron-Cetty, M.-P., \& V{\'e}ron, P.\ 2010, \aap, 518, A10
\bibitem[Veilleux \& Osterbrock (1987)]{1987VO} Veilleux, S., \& Osterbrock, D.~E.\ 1987, \apjs, 63, 295
\bibitem[Villar-Mart{\'{\i}}n et al.(2016)]{Villar-Martin2016} Villar-Mart{\'{\i}}n, M., Arribas, S., Emonts, B., et al.\ 2016, \mnras, 460, 130
\bibitem[Vogt et al.(2013)]{Vogt_2013_HCGs_I} Vogt, F.~P.~A., Dopita, M.~A., \& Kewley, L.~J.\ 2013, \apj, 768, 151
\bibitem[Wisnioski et al.(2015)]{Wisnioski_2015_KMOS3D} Wisnioski, E., F{\"o}rster Schreiber, N.~M., Wuyts, S., et al.\ 2015, \apj, 799, 209
\bibitem[Zaw et al.(2009)]{Zaw09} Zaw, I., Farrar, G.~R., \& Greene, J.~E.\ 2009, \apj, 696, 1218 
\end{thebibliography}
{}

\appendix

\section{Tables of measured nuclear line fluxes}

De-reddened nuclear emission-line fluxes for the newly observed galaxies are listed in Table~\ref{table:nuc_fit_fluxes}.  Luminosity distances, extinctions, physical aperture sizes and fluxes and luminosities of the \OIII\ and \Hb\ lines are given for the newly observed galaxies in Table~\ref{table:nuc_fit_luminosities}.  These tables are discussed in Section~\ref{sec:nuc_fluxes}.  Full versions of both tables are available as part of this data release, with the addition of a table corresponding to Table~\ref{table:nuc_fit_fluxes} but containing flux errors.

\clearpage

\begin{deluxetable*}{lccccccc}\centering
	\tabletypesize{\scriptsize}
	\tablecaption{Measured de-reddened nuclear emission-line fluxes for the newly observed S7 galaxies.  Here the second column presents the H$\beta$ flux and the fluxes in the remaining columns are relative to the H$\beta$ flux.  For 14 Seyfert~1 galaxies the fluxes were measured after subtraction of broad H$\alpha$ and H$\beta$ components.\label{table:nuc_fit_fluxes} }
	\tablewidth{525pt}
	\tablehead{
		\colhead{Galaxy} & \colhead{$F_{\mathrm{H}\beta}$ ($10^{-16}$ erg$\,$cm$^{-2}\,$s$^{-1}$)} & \colhead{[FeVII]$\,\lambda$3586} & \colhead{[OII]$\,\lambda$3726} & \colhead{[OII]$\,\lambda$3729} & \colhead{H$\eta$} & \colhead{[NeIII]$\,\lambda$3869} & \colhead{  ...  }
	}
	\startdata
	ARK402 &                        35 &        ...  &     1.454 &     0.695 &  0.086 &        ...  &    ...   \\
	ESO287-G42 &                        29 &        ...  &     1.535 &     0.882 &   ...  &        ...  &    ...   \\
	ESO323-G77 &                       588 &        ...  &     0.127 &     0.243 &  0.058 &       0.163 &    ...   \\
	ESO350-IG38 &                      6460 &        ...  &     1.044 &     1.234 &  0.062 &       0.259 &    ...   \\
	ESO362-G08 &                       380 &        ...  &     1.232 &     0.917 &  0.049 &       0.208 &    ...   \\
	ESO362-G18 &                       599 &       0.026 &     0.443 &     0.342 &  0.065 &       0.687 &    ...   \\
	ESO383-G35 &                       850 &       0.047 &     0.696 &     0.893 &  0.162 &       0.553 &    ...   \\
	ESO420-G13 &                      4169 &        ...  &     0.422 &     0.436 &  0.060 &       0.142 &    ...   \\
	ESO443-G17 &                      2243 &        ...  &     0.650 &     0.547 &  0.022 &        ...  &    ...   \\
	ESO460-G09 &                        74 &        ...  &     2.577 &     2.246 &   ...  &       0.232 &    ...   \\
	ESO500-G34 &                      3316 &        ...  &     1.128 &     1.528 &  0.079 &       0.348 &    ...   \\
	ESO565-G19 &                       321 &        ...  &     1.398 &     0.946 &  0.021 &       0.107 &    ...   \\
	IC1481 &                       705 &        ...  &     1.246 &     2.781 &  0.151 &       0.337 &    ...   \\
	IC1858 &                        30 &        ...  &     0.633 &      ...  &   ...  &        ...  &    ...   \\
	IC3639 &                      3033 &        ...  &     0.975 &     0.271 &  0.032 &       0.348 &    ...   \\
	IC4329A &                      1575 &        ...  &     0.396 &     0.246 &  0.058 &       0.371 &    ...   \\
	IC4777 &                       291 &        ...  &     1.306 &     1.014 &  0.026 &       0.704 &    ...   \\
	IC4995 &                       397 &       0.049 &     0.734 &     0.589 &  0.065 &       0.817 &    ...   \\
	IC5169 &                       951 &       0.139 &     0.852 &     0.769 &  0.063 &       0.101 &    ...   \\
	MARK938 &                      6812 &       0.231 &     6.870 &     2.277 &   ...  &       0.766 &    ...   \\
	MCG-02-27-009 &                        24 &        ...  &     1.295 &     1.197 &   ...  &        ...  &    ...   \\
	MCG-02-51-008 &                        98 &        ...  &     0.807 &     1.552 &  0.048 &       0.445 &    ...   \\
	MCG-03-34-064 &                      9067 &       0.124 &     0.598 &     0.857 &  0.076 &       1.523 &    ...   \\
	MCG-04-49-001 &                        30 &        ...  &     0.740 &     1.120 &   ...  &        ...  &    ...   \\
	NGC1068 &                     69965 &       0.072 &     0.814 &     0.676 &  0.002 &       1.304 &    ...   \\
	NGC1194 &                       127 &       0.229 &     0.431 &     1.210 &  0.119 &       0.234 &    ...   \\
	NGC1217 &                        72 &        ...  &     0.810 &     0.591 &   ...  &        ...  &    ...   \\
	NGC1266 &                      4667 &        ...  &     1.995 &    12.794 &   ...  &       1.049 &    ...   \\
	NGC1320 &                       587 &       0.065 &     0.377 &     0.605 &  0.064 &       0.485 &    ...   \\
	NGC1346 &                       387 &        ...  &     0.614 &     0.570 &   ...  &       0.058 &    ...   \\
	NGC1365 &                      2182 &        ...  &     0.370 &     0.507 &  0.051 &       0.275 &    ...   \\
	NGC1667 &                       515 &        ...  &     2.899 &     1.395 &   ...  &       0.379 &    ...   \\
	NGC1672 &                       402 &       0.016 &     0.453 &     0.499 &  0.072 &       0.008 &    ...   \\
	NGC2110 &                      3848 &        ...  &     2.315 &     3.266 &  0.043 &       0.830 &    ...   \\
	NGC2217 &                       220 &        ...  &     1.611 &     0.596 &   ...  &        ...  &    ...   \\
	NGC2945 &                        55 &        ...  &     2.558 &     0.740 &  0.043 &        ...  &    ...   \\
	NGC2974 &                       145 &        ...  &     1.733 &     0.333 &   ...  &        ...  &    ...   \\
	NGC3281 &                       550 &        ...  &     1.252 &     0.812 &  0.080 &       0.684 &    ...   \\
	NGC3312 &                       179 &        ...  &     2.494 &     1.980 &  0.025 &       0.041 &    ...   \\
	NGC3390 &                        10 &        ...  &     1.039 &      ...  &   ...  &        ...  &    ...   \\
	NGC3393 &                      1512 &       0.024 &     1.120 &     1.112 &  0.071 &       0.825 &    ...   \\
	NGC3831 &                        29 &        ...  &     1.172 &     1.339 &   ...  &        ...  &    ...   \\
	NGC3858 &                        18 &        ...  &     0.510 &     0.495 &   ...  &        ...  &    ...   \\
	NGC4418 &                       189 &        ...  &     2.663 &     2.865 &   ...  &       0.706 &    ...   \\
	NGC4472 &                       235 &        ...  &     0.253 &      ...  &   ...  &        ...  &    ...   \\
	NGC4507 &                      4301 &       0.072 &     1.302 &     0.928 &  0.075 &       0.952 &    ...   \\
	NGC4593 &                       168 &        ...  &     0.368 &     0.289 &  0.213 &       0.356 &    ...   \\
	NGC4594 &                       469 &        ...  &     1.031 &     0.068 &   ...  &        ...  &    ...   \\
	NGC4636 &                       119 &        ...  &     0.552 &     0.559 &  0.185 &        ...  &    ...   \\
	NGC4691 &                      3424 &        ...  &     0.825 &     0.805 &  0.071 &       0.017 &    ...   \\
	NGC4696 &                        31 &        ...  &     0.945 &     0.992 &   ...  &        ...  &    ...   \\
	NGC4845 &                      1504 &        ...  &     0.731 &     0.868 &  0.161 &       0.190 &    ...   \\
	NGC4939 &                       649 &        ...  &     1.261 &     0.791 &  0.076 &       0.852 &    ...   \\
	NGC4968 &                      2061 &        ...  &     0.970 &     0.678 &  0.084 &       1.190 &    ...   \\
	NGC4990 &                      1872 &        ...  &     1.514 &     1.339 &  0.067 &       0.092 &    ...   \\
	NGC5128 &                      2255 &        ...  &     1.384 &     1.474 &   ...  &        ...  &    ...   \\
	NGC5135 &                      2595 &        ...  &     0.808 &     0.523 &  0.055 &       0.314 &    ...   \\
	NGC5252 &                       325 &        ...  &     1.379 &     2.329 &  0.034 &       0.926 &    ...   \\
	NGC5427 &                       124 &        ...  &     1.375 &     0.781 &  0.120 &       0.792 &    ...   \\
	NGC5643 &                      1849 &       0.047 &     1.613 &     1.131 &  0.084 &       0.916 &    ...   \\
	NGC5990 &                      1996 &        ...  &     0.566 &     1.319 &   ...  &       0.317 &    ...   \\
	NGC6328 &                       512 &        ...  &     0.927 &     1.557 &  0.032 &       0.106 &    ...   \\
	NGC6936 &                        56 &        ...  &     0.948 &     0.814 &   ...  &        ...  &    ...   \\
	NGC7172 &                       143 &        ...  &     0.545 &     1.466 &   ...  &       0.204 &    ...   \\
	NGC7682 &                       485 &        ...  &     1.062 &     1.789 &  0.061 &       0.769 &    ...   \\
	PKS1306-241 &                       301 &       0.462 &     1.723 &     1.751 &   ...  &       0.437 &    ...   \\
	PKS1521-300 &                        44 &        ...  &     0.833 &     0.638 &   ...  &        ...  &    ...   \\
	UGC3255 &                      ...  &        ...  &      ...  &      ...  &   ...  &        ...  &    ...   \\
    \enddata
	\newline
\end{deluxetable*}

\clearpage

\begin{deluxetable*}{lcccccccc}
	\centering
	\tabletypesize{\scriptsize}
	\tablecaption{Derived nuclear extinctions, physical aperture sizes, and \Hb\ and \OIII\ fluxes and luminosities for the newly observed S7 galaxies \label{table:nuc_fit_luminosities}
	}\tablewidth{525pt}
	\tablehead{
		\colhead{Galaxy} & \colhead{Type} & \colhead{D$_{\rm Lum.}$} & \colhead{$A_V$} & \colhead{Aperture} & \colhead{$F_{\rm O~III}$~($10^{-16}\times$} & \colhead{$F_{\rm O~III}$~($10^{-16}\times$} & \colhead{$\log L_{H\beta}$} & \colhead{$\log L_{\rm O~III}$}\\
		\colhead{} & \colhead{} & \colhead{(Mpc)$^a$} & \colhead{(mag.)} & \colhead{(kpc)$^b$} & \colhead{erg\,cm$^{-2}\,$s$^{-1}$)} & \colhead{erg\,cm$^{-2}\,$s$^{-1}$)} & \colhead{(erg~s$^{-1})$} & \colhead{(erg~s$^{-1})$}
	}
	\startdata
	ARK402 &        SB + Seyfert 2 &       76 &      0 &         1.48 &                       40 &                      30 &               39.4 &              39.3 \\
	ESO287-G42 &             Seyfert 1 &       80 &      0 &         1.55 &                       30 &                      30 &               39.3 &              39.3 \\
	ESO323-G77$^c$ &             Seyfert 1 &       64 &      0 &         1.25 &                      590 &                    1260 &               40.5 &              40.8 \\
	ESO350-IG38 &                    SB &       88 &   1.33 &         1.71 &                     6460 &                   22890 &               41.8 &              42.3 \\
	ESO362-G08 &       PSB + Seyfert 2 &       67 &   1.72 &         1.31 &                      380 &                    2850 &               40.3 &              41.2 \\
	ESO362-G18$^c$ &             Seyfert 1 &       53 &   0.76 &         1.03 &                      600 &                    5170 &               40.3 &              41.2 \\
	ESO383-G35$^c$ &             Seyfert 1 &       33 &   1.69 &         0.64 &                      850 &                    4150 &               40.0 &              40.7 \\
	ESO420-G13 &        SB + Seyfert 2 &       51 &   2.19 &         0.99 &                     4170 &                    6990 &               41.1 &              41.3 \\
	ESO443-G17 &        SB + Seyfert 2 &       44 &   2.05 &         0.85 &                     2240 &                     710 &               40.7 &              40.2 \\
	ESO460-G09 &        SB + Seyfert 2 &       84 &   0.88 &         1.63 &                       70 &                     110 &               39.8 &              40.0 \\
	ESO500-G34 &  SB + PSB + Seyfert 2 &       52 &   3.55 &         1.02 &                     3320 &                    3640 &               41.0 &              41.1 \\
	ESO565-G19 &    LINER or Seyfert 2 &       70 &   1.05 &         1.35 &                      320 &                     740 &               40.3 &              40.6 \\
	IC1481 &        SB + Seyfert 2 &       87 &   2.55 &         1.70 &                      700 &                    1020 &               40.8 &              41.0 \\
	IC1858 &                    SB &       87 &      0 &         1.68 &                       30 &                      10 &               39.4 &              39.1 \\
	IC3639 &             Seyfert 2 &       47 &   1.95 &         0.91 &                     3030 &                   17240 &               40.9 &              41.7 \\
	IC4329A$^c$ &             Seyfert 1 &       69 &   1.65 &         1.33 &                     1580 &                    9370 &               41.0 &              41.7 \\
	IC4777 &             Seyfert 2 &       80 &   1.27 &         1.55 &                      290 &                    1850 &               40.3 &              41.2 \\
	IC4995 &             Seyfert 2 &       69 &   0.74 &         1.34 &                      400 &                    4710 &               40.4 &              41.4 \\
	IC5169 &        SB + Seyfert 2 &       44 &   2.47 &         0.86 &                      950 &                    1060 &               40.4 &              40.4 \\
	MARK938 &        SB + Seyfert 2 &       84 &   4.41 &         1.63 &                     6810 &                    8260 &               41.8 &              41.8 \\
	MCG-02-27-009 &    LINER or Seyfert 2 &       65 &   0.60 &         1.26 &                       20 &                      70 &               39.1 &              39.6 \\
	MCG-02-51-008 &        SB + Seyfert 2 &       80 &   0.95 &         1.55 &                      100 &                     420 &               39.9 &              40.5 \\
	MCG-03-34-064 &             Seyfert 1 &       71 &   1.80 &         1.37 &                     9070 &                   95060 &               41.7 &              42.8 \\
	MCG-04-49-001 &                 LINER &       86 &      0 &         1.67 &                       30 &                      20 &               39.4 &              39.3 \\
	NGC1068 &             Seyfert 2 &       16 &   1.76 &         0.32 &                    69970 &                  768300 &               41.3 &              42.4 \\
	NGC1194 &        SB + Seyfert 2 &       58 &   1.27 &         1.13 &                      130 &                     790 &               39.7 &              40.5 \\
	NGC1217 &                 LINER &       90 &      0 &         1.74 &                       70 &                     110 &               39.8 &              40.0 \\
	NGC1266 &                 LINER &       31 &   5.33 &         0.60 &                     4670 &                    6730 &               40.7 &              40.9 \\
	NGC1320 &             Seyfert 2 &       38 &   1.23 &         0.74 &                      590 &                    5480 &               40.0 &              41.0 \\
	NGC1346 &                    SB &       58 &   1.96 &         1.13 &                      390 &                     170 &               40.2 &              39.8 \\
	NGC1365$^c$ &             Seyfert 1 &       23 &   1.85 &         0.45 &                     2180 &                    4850 &               40.2 &              40.5 \\
	NGC1667 &             Seyfert 2 &       65 &   2.33 &         1.26 &                      520 &                    2830 &               40.4 &              41.2 \\
	NGC1672 &        SB + Seyfert 2 &       19 &      0 &         0.37 &                      400 &                     290 &               39.2 &              39.1 \\
	NGC2110 &             Seyfert 2 &       33 &   2.39 &         0.65 &                     3850 &                   16170 &               40.7 &              41.3 \\
	NGC2217 &                 LINER &       23 &   0.26 &         0.45 &                      220 &                     360 &               39.1 &              39.4 \\
	NGC2945 &                 LINER &       66 &      0 &         1.28 &                       60 &                      40 &               39.5 &              39.4 \\
	NGC2974 &        SB + Seyfert 2 &       27 &      0 &         0.52 &                      140 &                     110 &               39.1 &              39.0 \\
	NGC3281 &             Seyfert 2 &       46 &   1.56 &         0.89 &                      550 &                    4240 &               40.1 &              41.0 \\
	NGC3312 &                 LINER &       41 &   0.44 &         0.80 &                      180 &                     330 &               39.6 &              39.8 \\
	NGC3390 &                    SB &       44 &      0 &         0.84 &                       10 &                      10 &               38.3 &              38.1 \\
	NGC3393 &             Seyfert 2 &       54 &   0.67 &         1.04 &                     1510 &                   14260 &               40.7 &              41.7 \\
	NGC3831 &                 LINER &       75 &      0 &         1.46 &                       30 &                      50 &               39.3 &              39.5 \\
	NGC3858 &                    SB &       82 &      0 &         1.59 &                       20 &                      20 &               39.2 &              39.3 \\
	NGC4418 &                    SB &       31 &   2.88 &         0.60 &                      190 &                     120 &               39.3 &              39.1 \\
	NGC4472 &                    SB &       14 &      0 &         0.27 &                      240 &                    ...  &               38.7 &              ...  \\
	NGC4507 &             Seyfert 2 &       51 &   1.54 &         0.98 &                     4300 &                   31910 &               41.1 &              42.0 \\
	NGC4593$^c$ &             Seyfert 1 &       39 &      0 &         0.75 &                      170 &                     720 &               39.5 &              40.1 \\
	NGC4594 &        SB + Seyfert 2 &       15 &      0 &         0.28 &                      470 &                     310 &               39.1 &              38.9 \\
	NGC4636 &                 LINER &       13 &      0 &         0.26 &                      120 &                      50 &               38.4 &              38.0 \\
	NGC4691 &                    SB &       16 &   1.17 &         0.31 &                     3420 &                    1490 &               40.0 &              39.7 \\
	NGC4696 &                 LINER &       42 &      0 &         0.82 &                       30 &                      20 &               38.8 &              38.6 \\
	NGC4845 &        SB + Seyfert 2 &       18 &   3.65 &         0.34 &                     1500 &                    1700 &               39.7 &              39.8 \\
	NGC4939 &             Seyfert 2 &       44 &   0.74 &         0.86 &                      650 &                    6520 &               40.2 &              41.2 \\
	NGC4968 &             Seyfert 2 &       42 &   2.59 &         0.82 &                     2060 &                   19490 &               40.6 &              41.6 \\
	NGC4990 &        SB + Seyfert 2 &       45 &   0.66 &         0.88 &                     1870 &                    2600 &               40.7 &              40.8 \\
	NGC5128 &             Seyfert 2 &        8 &   3.79 &         0.15 &                     2250 &                    6530 &               39.2 &              39.7 \\
	NGC5135 &       PSB + Seyfert 2 &       59 &   1.46 &         1.14 &                     2600 &                    7350 &               41.0 &              41.5 \\
	NGC5252 &             Seyfert 2 &       98 &   0.54 &         1.91 &                      330 &                    1930 &               40.6 &              41.3 \\
	NGC5427 &             Seyfert 2 &       37 &   0.60 &         0.73 &                      120 &                     990 &               39.3 &              40.2 \\
	NGC5643 &             Seyfert 2 &       17 &   1.18 &         0.33 &                     1850 &                   15910 &               39.8 &              40.7 \\
	NGC5990 &  SB + PSB + Seyfert 2 &       55 &   2.70 &         1.06 &                     2000 &                    4330 &               40.9 &              41.2 \\
	NGC6328 &        SB + Seyfert 2 &       62 &   1.05 &         1.20 &                      510 &                     810 &               40.4 &              40.6 \\
	NGC6936 &        SB + Seyfert 2 &       84 &      0 &         1.62 &                       60 &                      40 &               39.7 &              39.5 \\
	NGC7172 &        SB + Seyfert 2 &       37 &   1.62 &         0.72 &                      140 &                     600 &               39.4 &              40.0 \\
	NGC7682 &             Seyfert 2 &       73 &   0.90 &         1.42 &                      480 &                    4530 &               40.5 &              41.5 \\
	PKS1306-241 &        SB + Seyfert 2 &       60 &   2.70 &         1.16 &                      300 &                    1150 &               40.1 &              40.7 \\
	PKS1521-300 &        SB + Seyfert 2 &       83 &   0.48 &         1.62 &                       40 &                      30 &               39.6 &              39.4 \\
	UGC3255 &                 LINER &       81 &   ...  &         1.57 &                     ...  &                    ...  &               ...  &              ...  \\
	\enddata
	\newline
	\tablenotetext{a}{Using the NED z and the Hubble law, with H$_0 = 70\,$km$\,$s$^{-1}$Mpc$^{-1}$.}
	\tablenotetext{b}{Diameter of aperture from which the nuclear spectrum was extracted (${\sim}4$ arcsec).}
	\tablenotetext{c}{Nuclear fluxes for these galaxies were measured after subtraction of \Ha\ and \Hb\ broad components.}
\end{deluxetable*}

\clearpage

\section{Plots of nuclear spectra}

\subsection{Seyfert~1 galaxies}
\begin{figure*}[b!]
	\begin{centering}
		\includegraphics[scale=0.90]{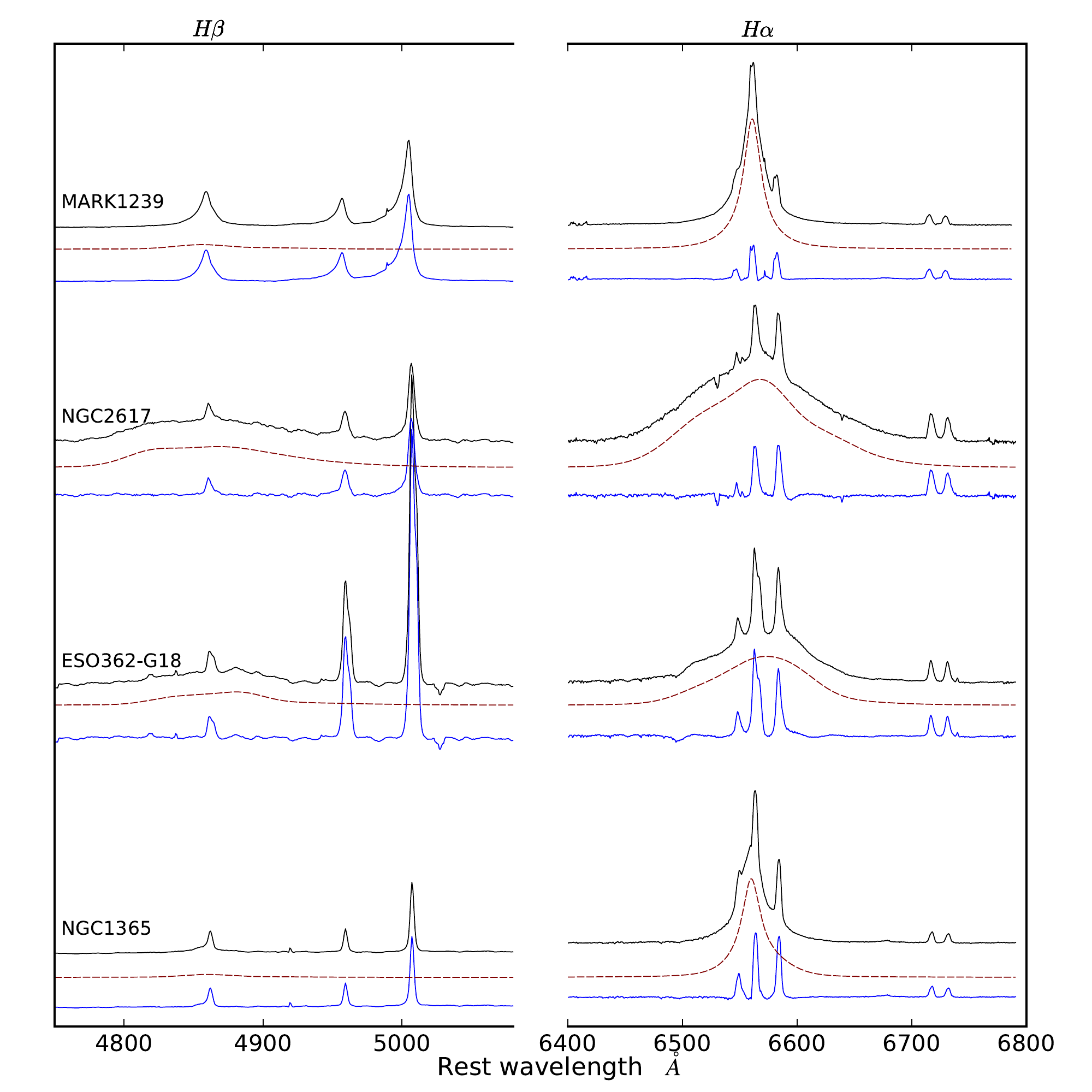}
	\end{centering}
	\caption{Seyfert~1 nuclear spectra showing the decomposition into `broad' and `narrow' spectra.  Figures~\ref{fig:nuc_spec_Sy1_A} to \ref{fig:nuc_spec_Sy1_C} show the nuclear spectra for the 14 galaxies for which broad-component subtraction was used before fitting the narrow lines (Section~\ref{sec:Sy1_fitting}).  Galaxies are ordered by RA across figures and bottom-to-top in each figure.  The decomposition into a `broad' and `narrow' spectrum was performed for each relevant spaxel independently; for each galaxy the nuclear spectra shown here were found by summing over the nuclear spaxels that had broad-component subtraction.  The left panel shows the \Hb\ and \OIII\,$\lambda\lambda\,4959,5007$ emission lines, and the right panel shows the \Ha\ line with the \NII\,$\lambda\lambda\,6548,6583$ lines on either side.  The topmost black line shows the total spectrum, the dotted maroon line shows the fitted Balmer line broad spectrum, and the solid blue line shows the narrow-line spectrum remaining after the fitted broad spectrum was subtracted.}\label{fig:nuc_spec_Sy1_A}
\end{figure*}

\begin{figure*}
	\begin{centering}
		\includegraphics[scale=0.90]{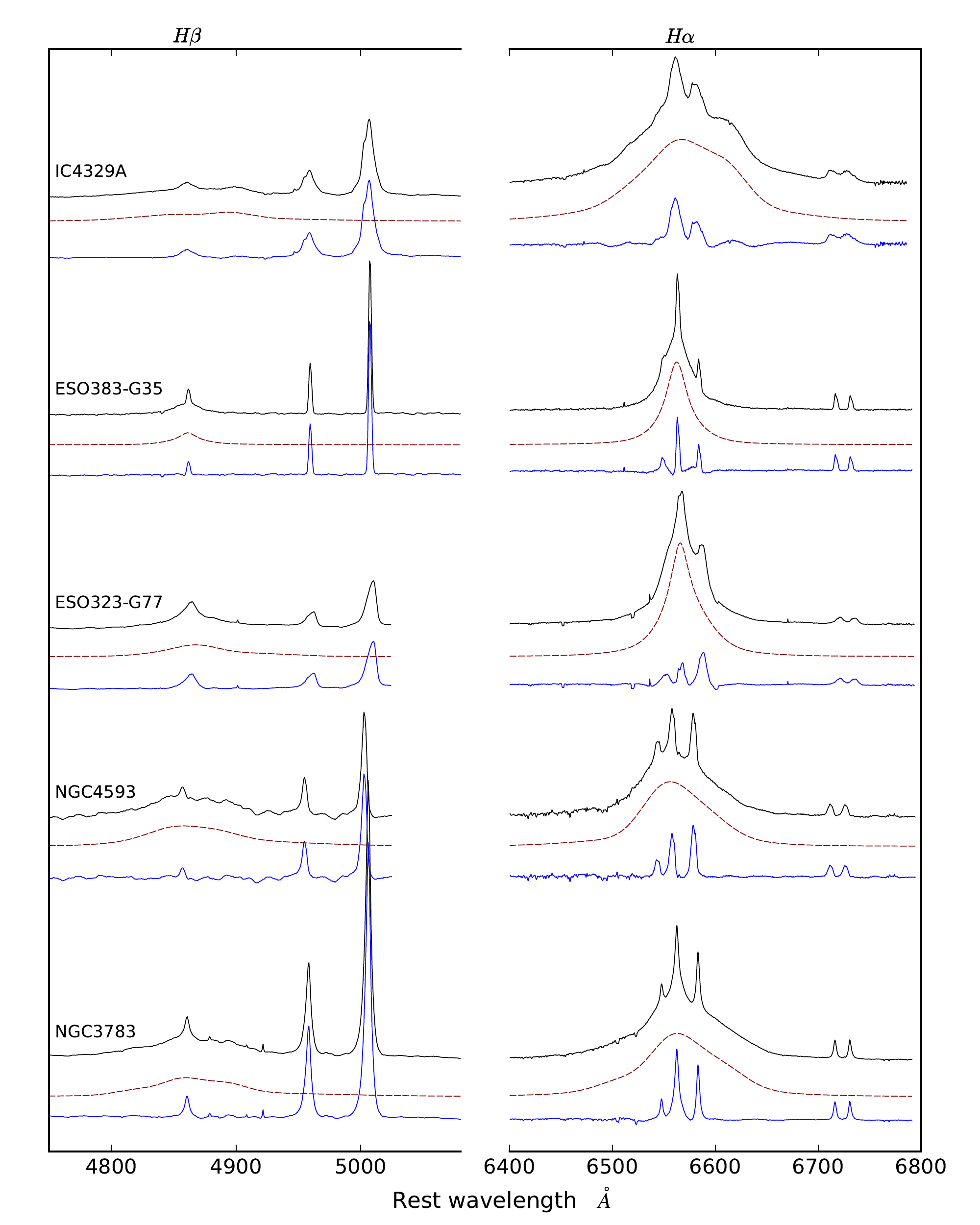}
	\end{centering}
	\caption{As in Figure~\ref{fig:nuc_spec_Sy1_A}}\label{fig:nuc_spec_Sy1_B}
\end{figure*}

\begin{figure*}
	\begin{centering}
		\includegraphics[scale=0.90]{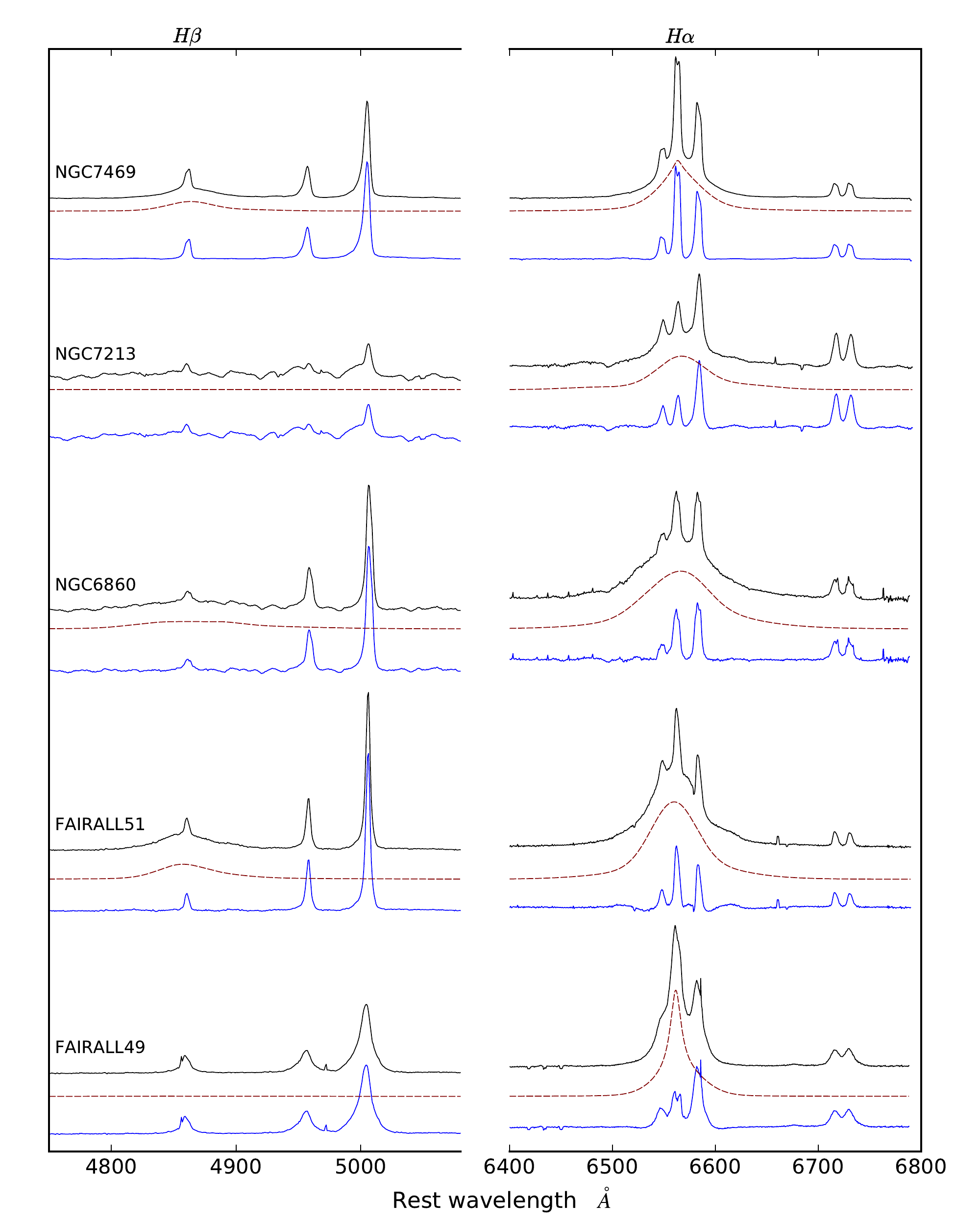}
	\end{centering}
	\caption{As in Figure~\ref{fig:nuc_spec_Sy1_A}}\label{fig:nuc_spec_Sy1_C}
\end{figure*}

Figures~\ref{fig:nuc_spec_Sy1_A} to \ref{fig:nuc_spec_Sy1_C} show the decomposition of Seyfert~1 nuclear spectra into broad and narrow spectra, and are discussed in Sections~\ref{sec:Sy1_fitting} and \ref{sec:nuc_spec_plots}.

\clearpage
\subsection{Other galaxies}
\begin{figure*}[b!]
	\begin{centering}
		\includegraphics[scale=0.85]{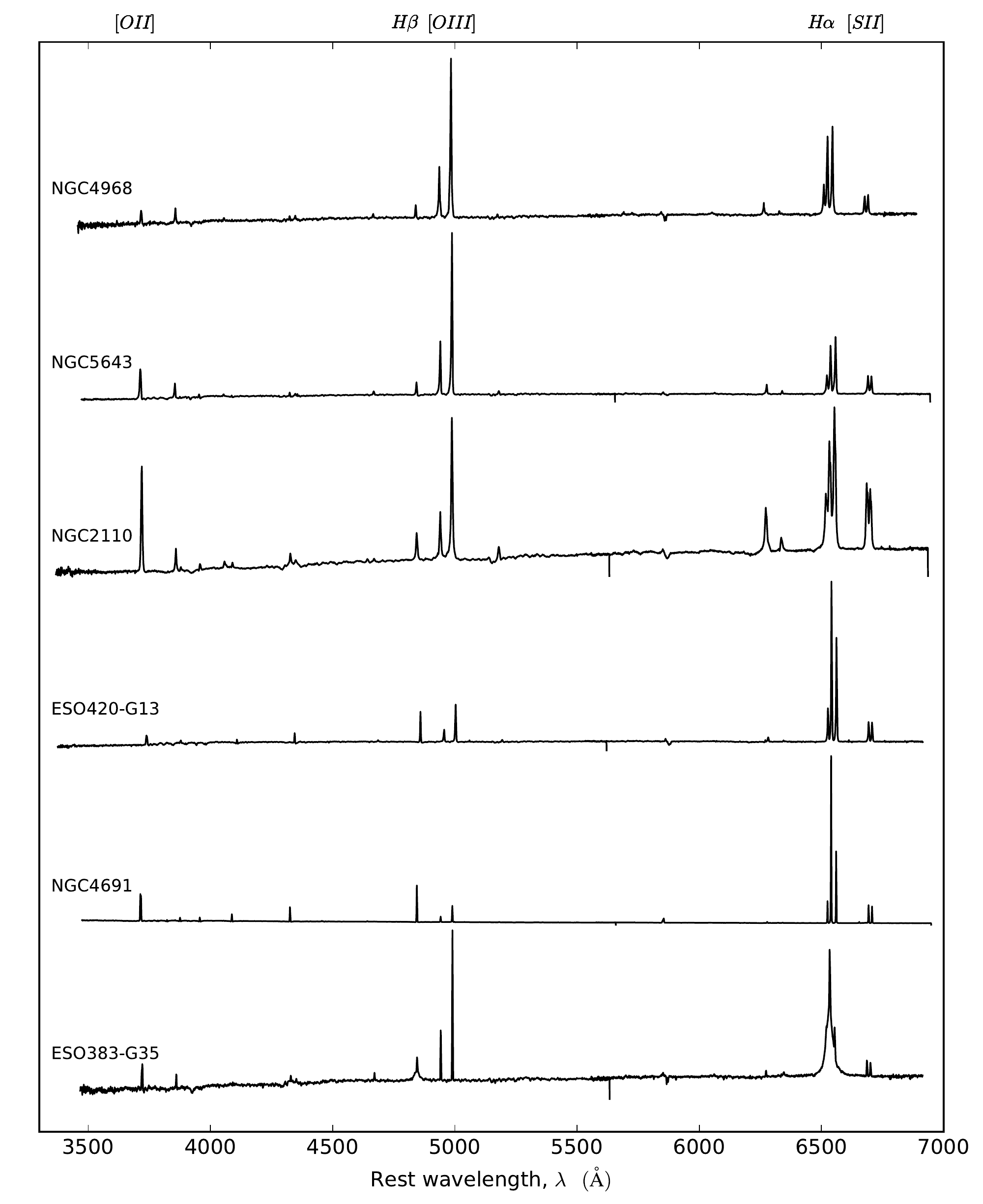}
	\end{centering}
	\caption{Nuclear spectra for a selection of six galaxies from the new S7 observations.  From top to bottom: NGC\,4968, NGC\,5643 and NGC\,2110 are Seyfert~2 galaxies, ESO\,420-G13 is a `composite' (AGN/star forming) galaxy, NGC\,4691 is star-forming (not an AGN), and ESO\,383-G35 is a Seyfert~1 with weak coronal emission.}\label{fig:nuc_spec}
\end{figure*}

Figure~\ref{fig:nuc_spec} shows nuclear spectra of several galaxies selected from the new observations.  The spectra are discussed in Section~\ref{sec:nuc_spec_plots}.

\clearpage
\section{Maps of gas excitation and kinematics}
Images showing gas excitation and kinematics for each galaxy in the S7 sample are presented in the following figures.  The figures are discussed in Section~\ref{sec:spatial_properties}.

\clearpage

\begin{figure*}
	\begin{centering}
		\includegraphics[scale=0.9]{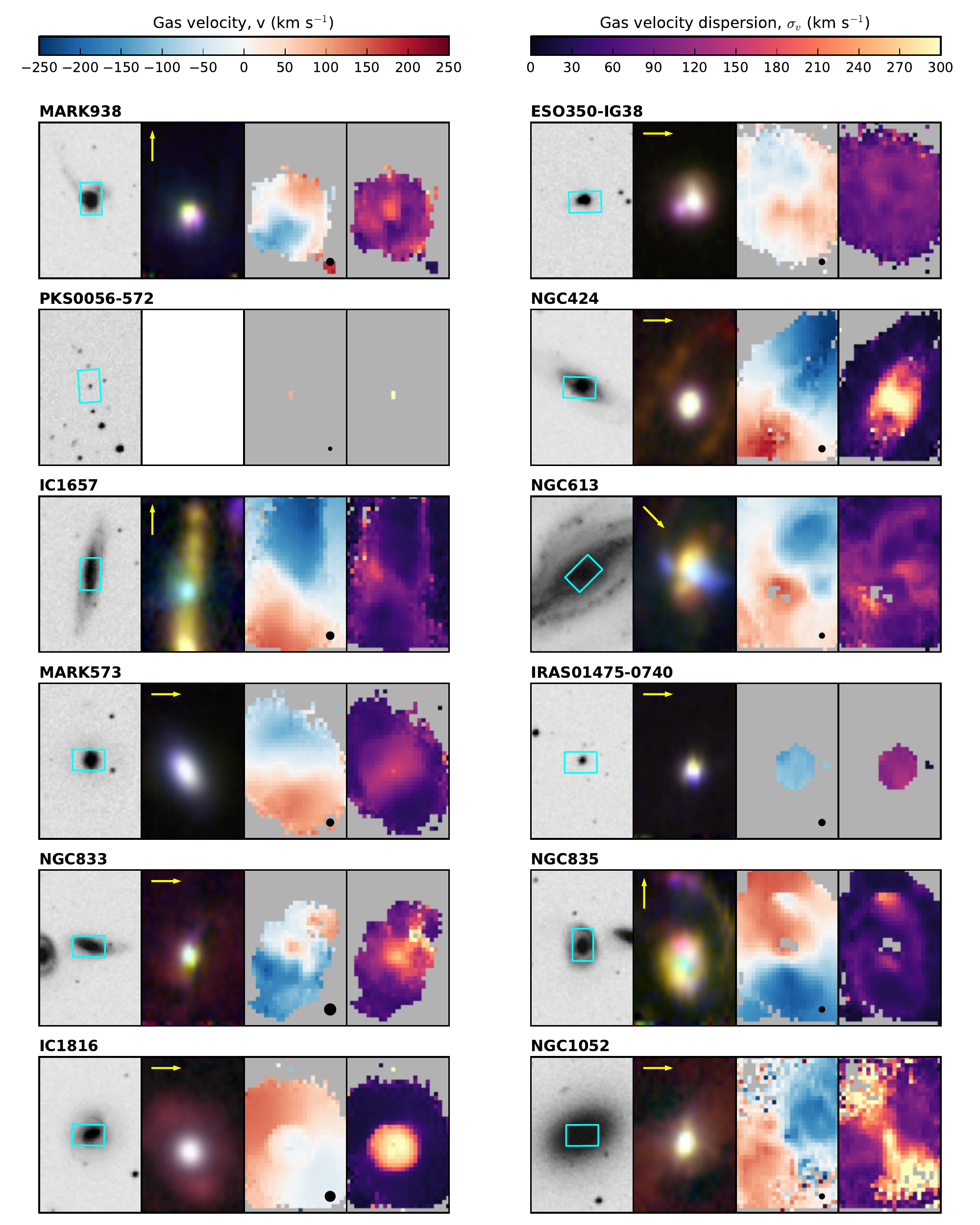}
	\end{centering}
	\caption{Maps for S7 galaxies with RAs between $00^\mathrm{h}00^\mathrm{m}$ and $02^\mathrm{h}42^\mathrm{m}$.  Galaxies are ordered by RA across rows and then down columns in Figures~\ref{fig:spatial_1} to~\ref{fig:spatial_11}.  For each galaxy, from left to right: A DSS2 r-band image showing the field of view of the S7 WiFeS observations, a three-color image showing the degree of ionization of the gas, and maps of the emission-line gas velocity and velocity dispersion for the single-component \lzifu\ fits.  Figures~\ref{fig:spatial_1} to~\ref{fig:spatial_11} are further described in Section~\ref{sec:spatial_properties}. }\label{fig:spatial_1}
\end{figure*}

\begin{figure*}
	\begin{centering}
		\includegraphics[scale=0.9]{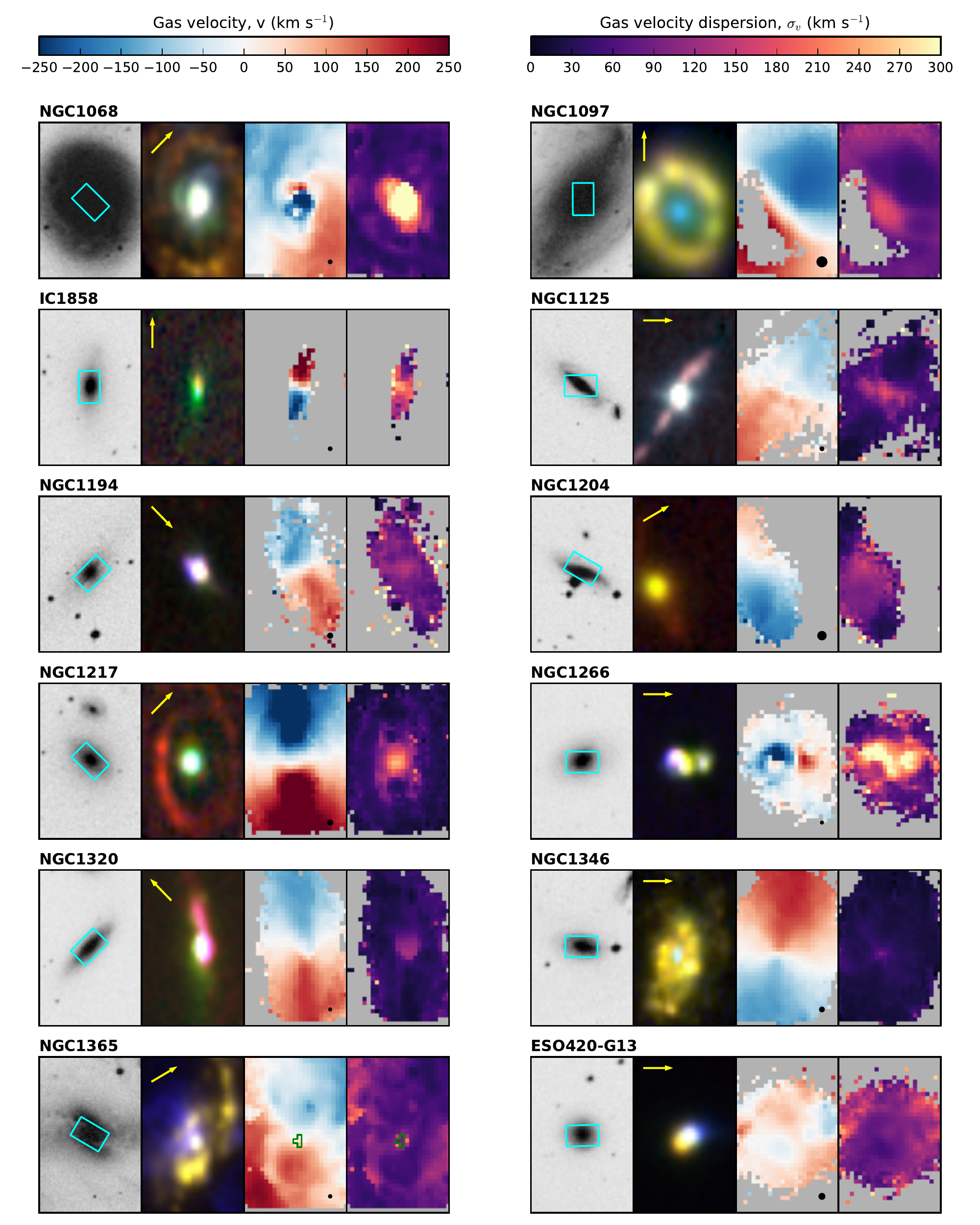}
	\end{centering}
	\caption{Maps as in Figure~\ref{fig:spatial_1}, for S7 galaxies with RAs between $02^\mathrm{h}42^\mathrm{m}$ and $04^\mathrm{h}16^\mathrm{m}$.}\label{fig:spatial_2}
\end{figure*}

\begin{figure*}
	\begin{centering}
		\includegraphics[scale=0.9]{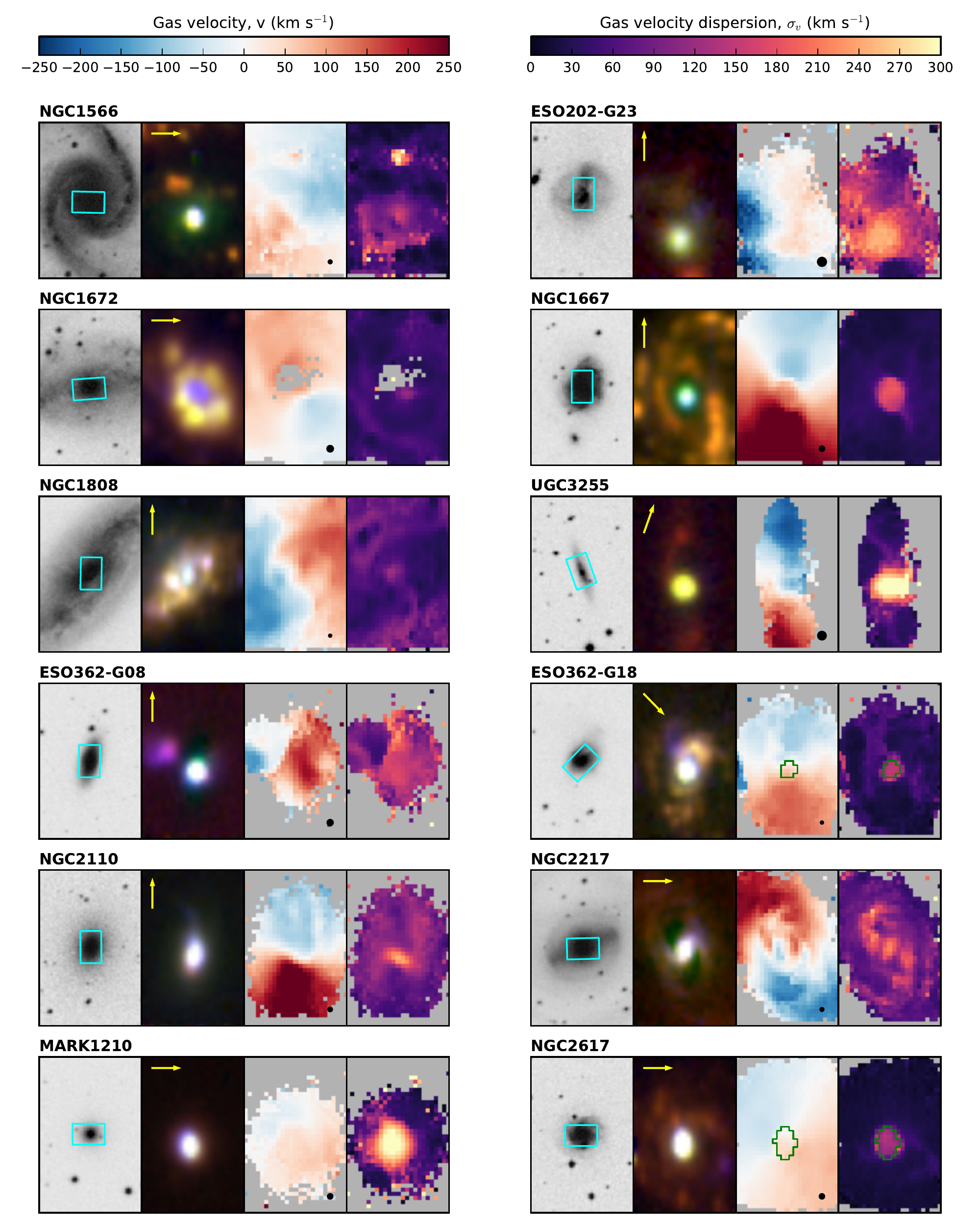}
	\end{centering}
	\caption{Maps as in Figure~\ref{fig:spatial_1}, for S7 galaxies with RAs between $04^\mathrm{h}16^\mathrm{m}$ and $09^\mathrm{h}00^\mathrm{m}$.}\label{fig:spatial_3}
\end{figure*}

\begin{figure*}
	\begin{centering}
		\includegraphics[scale=0.9]{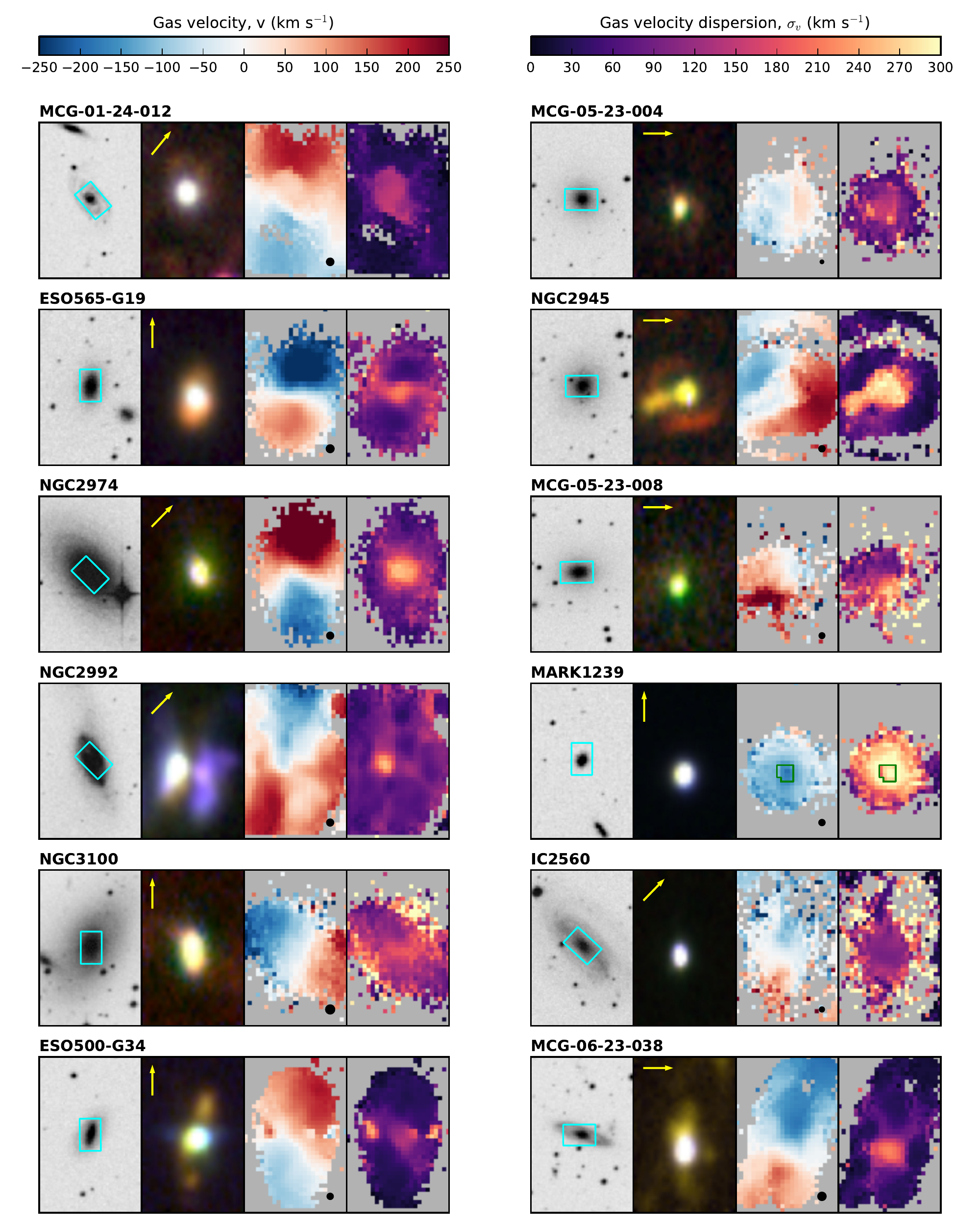}
	\end{centering}
	\caption{Maps as in Figure~\ref{fig:spatial_1}, for S7 galaxies with RAs between $09^\mathrm{h}00^\mathrm{m}$ and $10^\mathrm{h}30^\mathrm{m}$.}\label{fig:spatial_4}
\end{figure*}

\begin{figure*}
	\begin{centering}
		\includegraphics[scale=0.9]{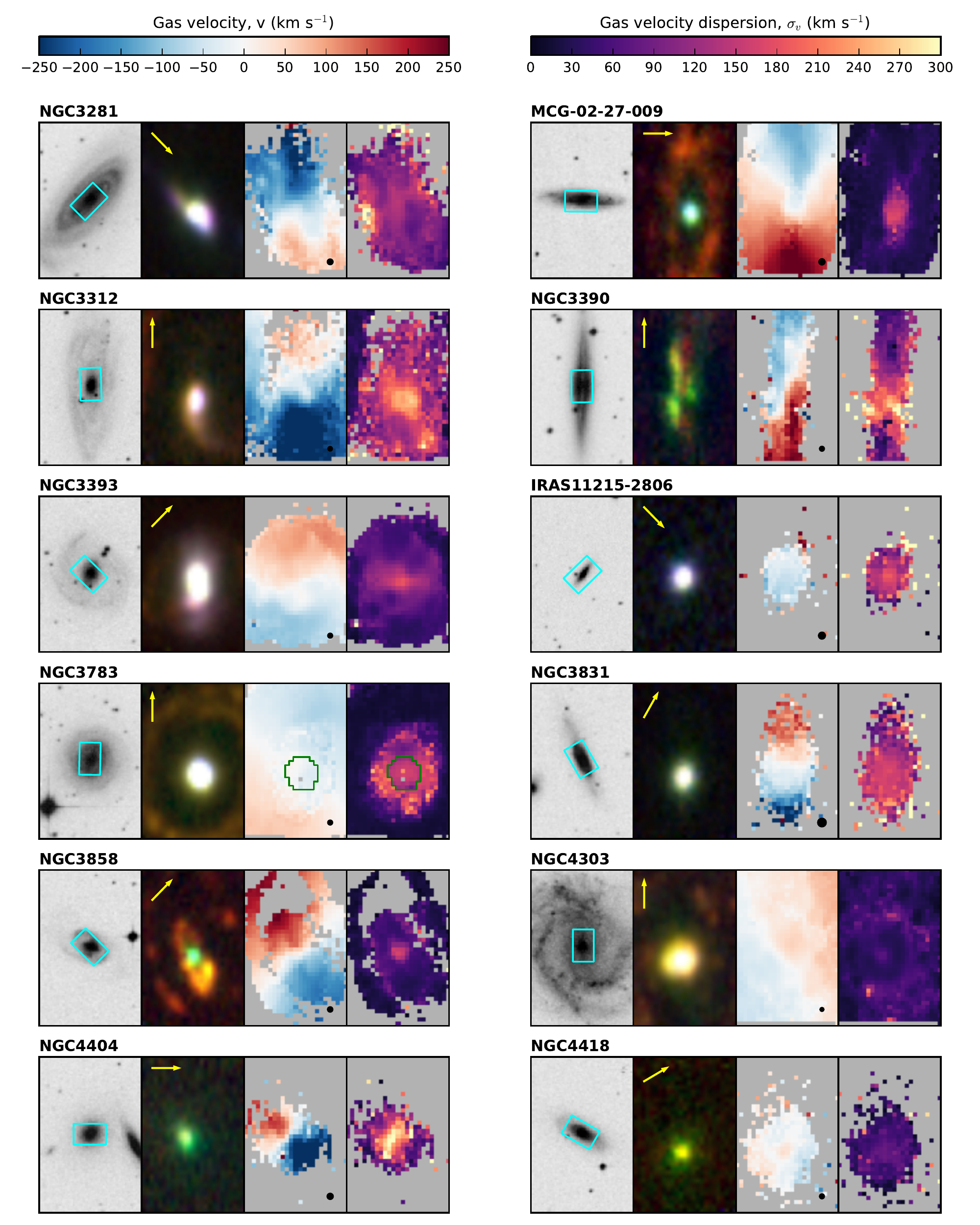}
	\end{centering}
	\caption{Maps as in Figure~\ref{fig:spatial_1}, for S7 galaxies with RAs between $10^\mathrm{h}30^\mathrm{m}$ and $12^\mathrm{h}28^\mathrm{m}$.}\label{fig:spatial_5}
\end{figure*}

\begin{figure*}
	\begin{centering}
		\includegraphics[scale=0.9]{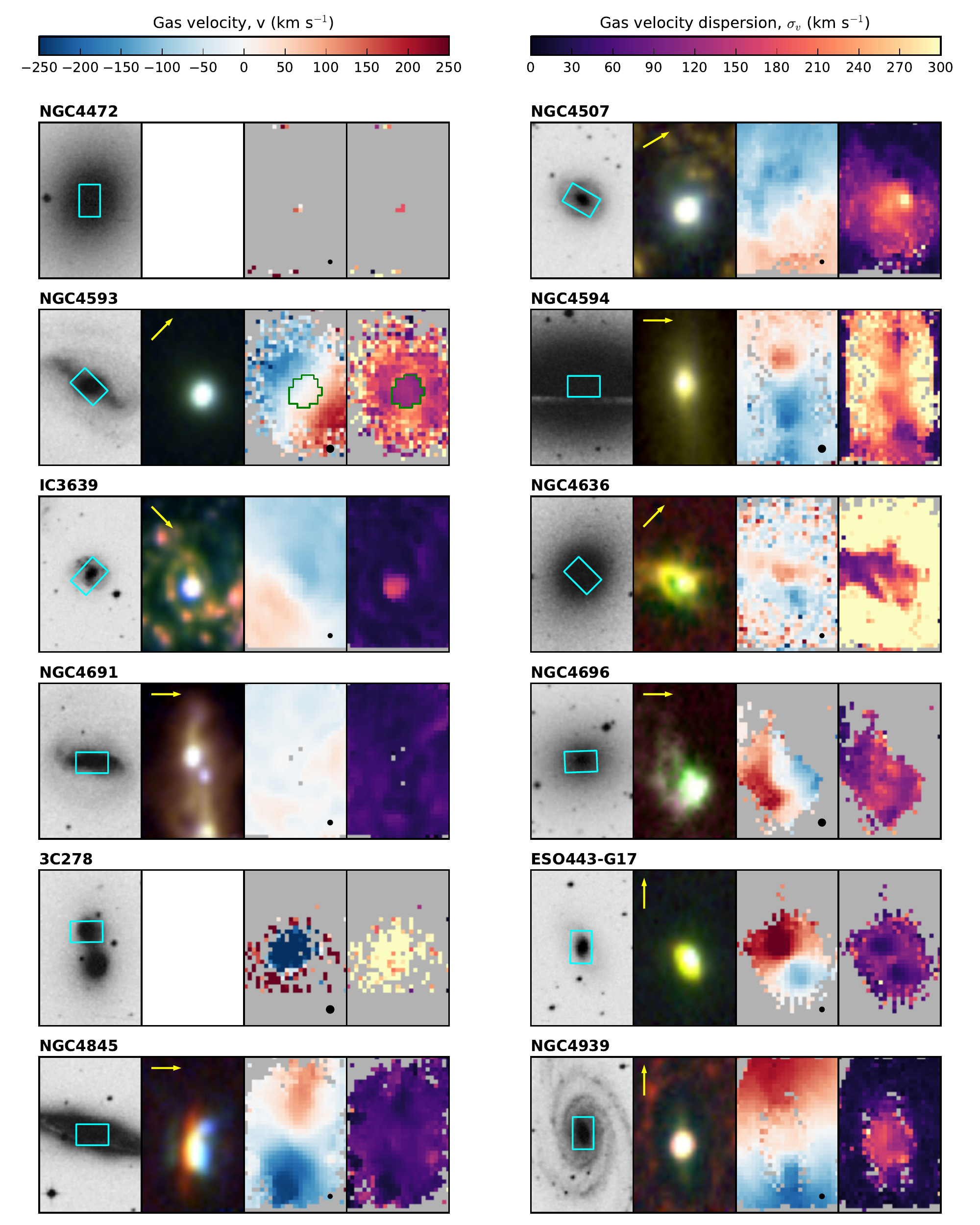}
	\end{centering}
	\caption{Maps as in Figure~\ref{fig:spatial_1}, for S7 galaxies with RAs between $12^\mathrm{h}28^\mathrm{m}$ and $13^\mathrm{h}05^\mathrm{m}$.}\label{fig:spatial_6}
\end{figure*}

\begin{figure*}
	\begin{centering}
		\includegraphics[scale=0.9]{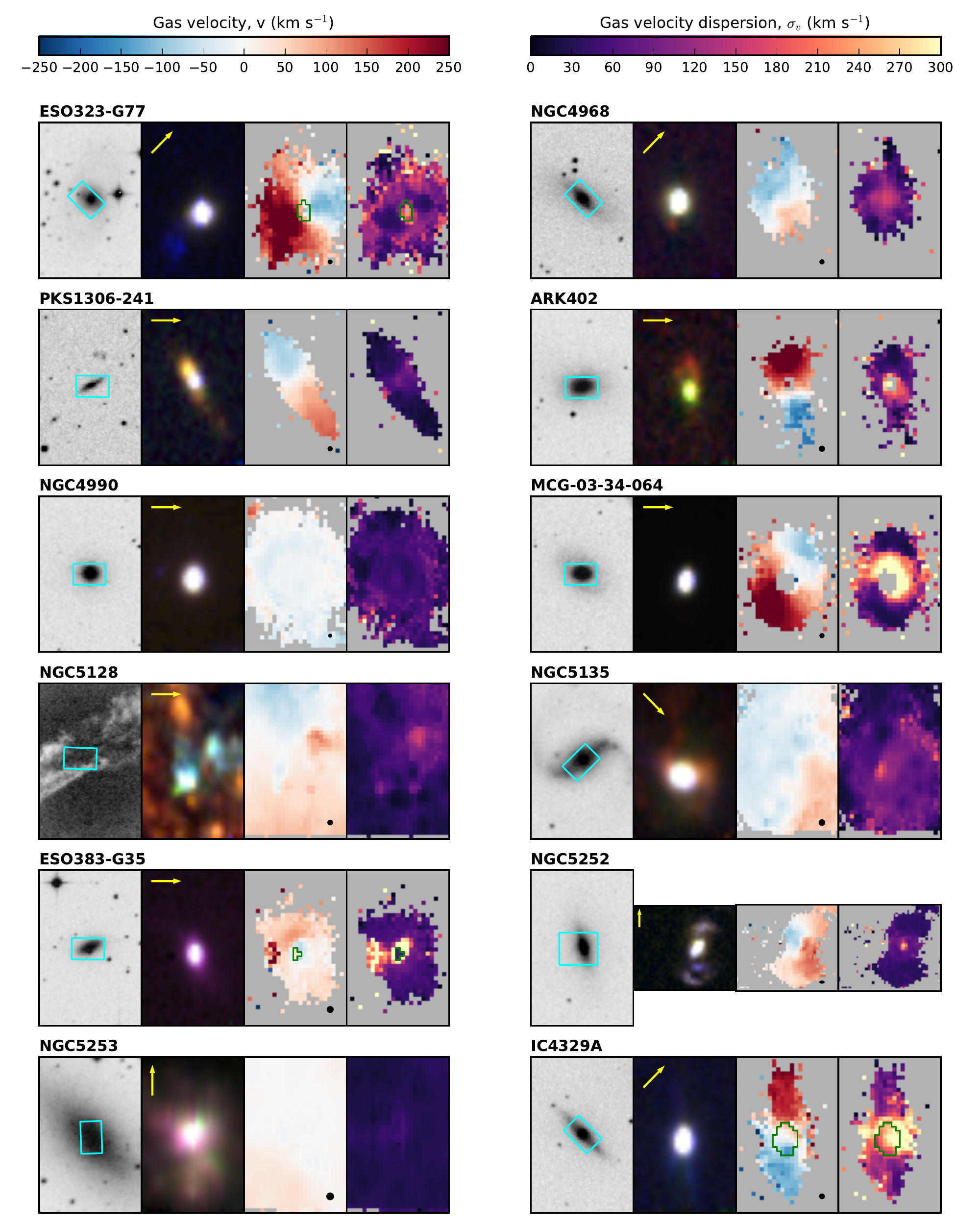}
	\end{centering}
	\caption{Maps as in Figure~\ref{fig:spatial_1}, for S7 galaxies with RAs between $13^\mathrm{h}05^\mathrm{m}$ and $14^\mathrm{h}00^\mathrm{m}$.}\label{fig:spatial_7}
\end{figure*}

\begin{figure*}
	\begin{centering}
		\includegraphics[scale=0.9]{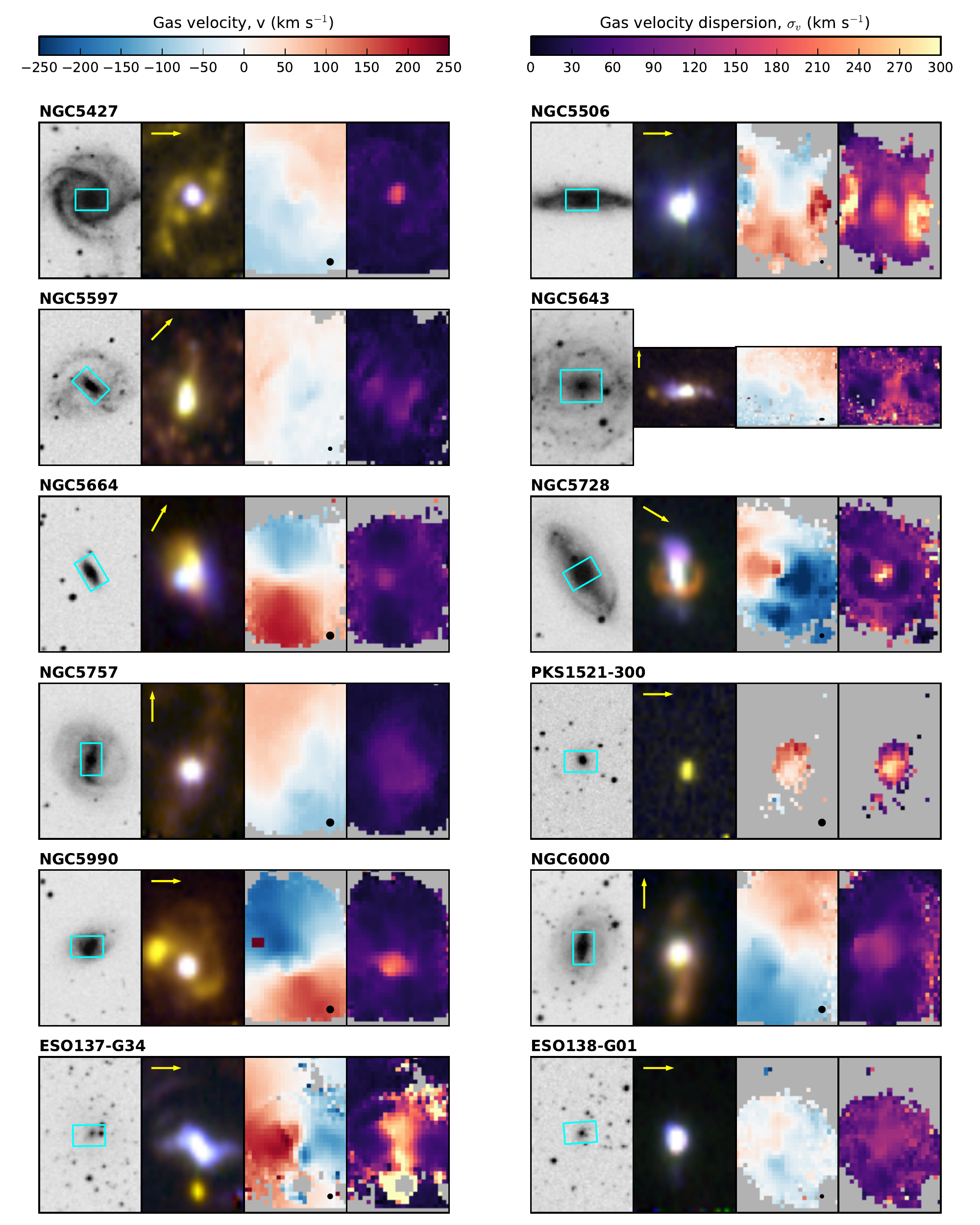}
	\end{centering}
	\caption{Maps as in Figure~\ref{fig:spatial_1}, for S7 galaxies with RAs between $14^\mathrm{h}00^\mathrm{m}$ and $16^\mathrm{h}52^\mathrm{m}$.}\label{fig:spatial_8}
\end{figure*}

\begin{figure*}
	\begin{centering}
		\includegraphics[scale=0.9]{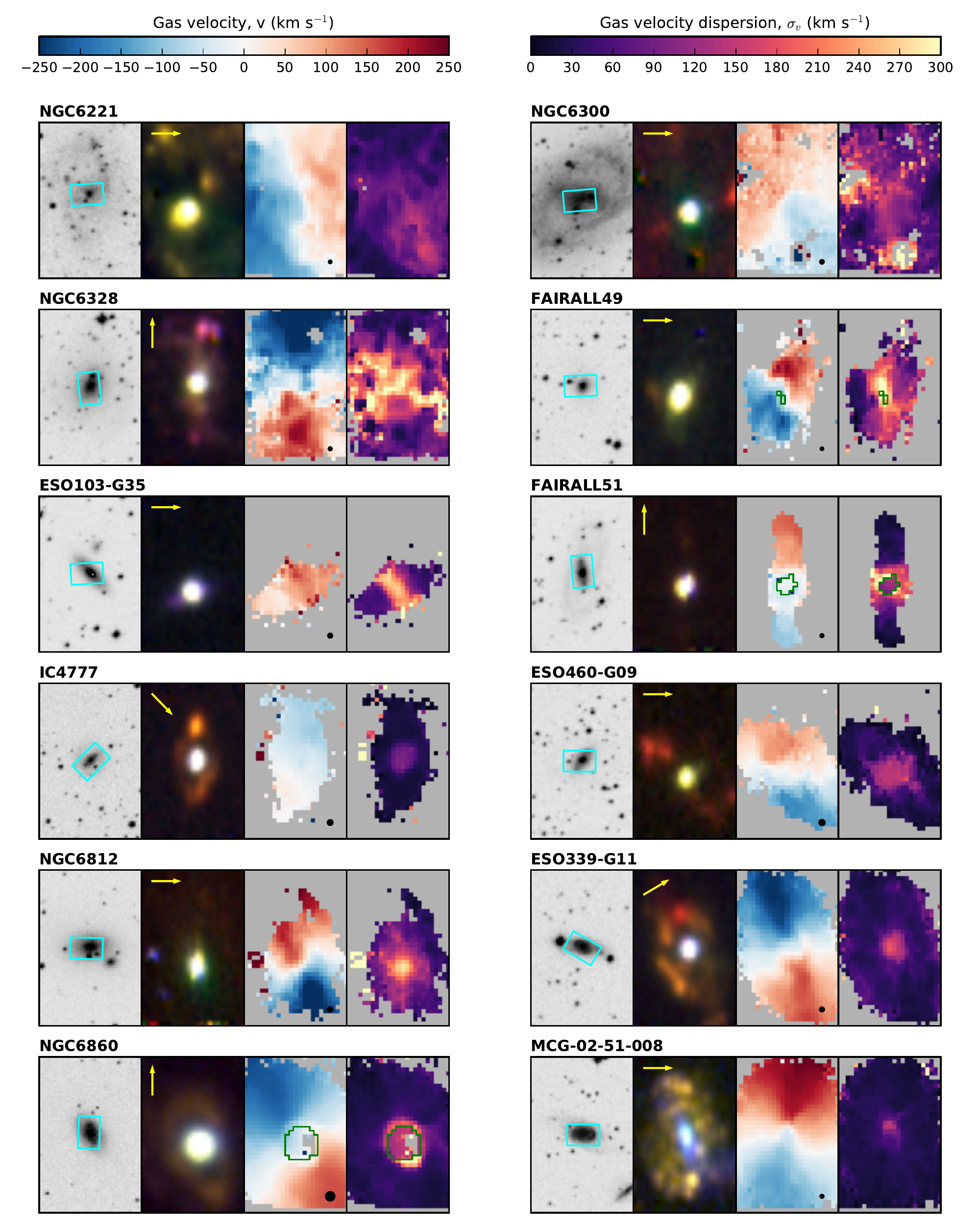}
	\end{centering}
	\caption{Maps as in Figure~\ref{fig:spatial_1}, for S7 galaxies with RAs between $16^\mathrm{h}52^\mathrm{m}$ and $20^\mathrm{h}18^\mathrm{m}$.}\label{fig:spatial_9}
\end{figure*}

\begin{figure*}
	\begin{centering}
		\includegraphics[scale=0.9]{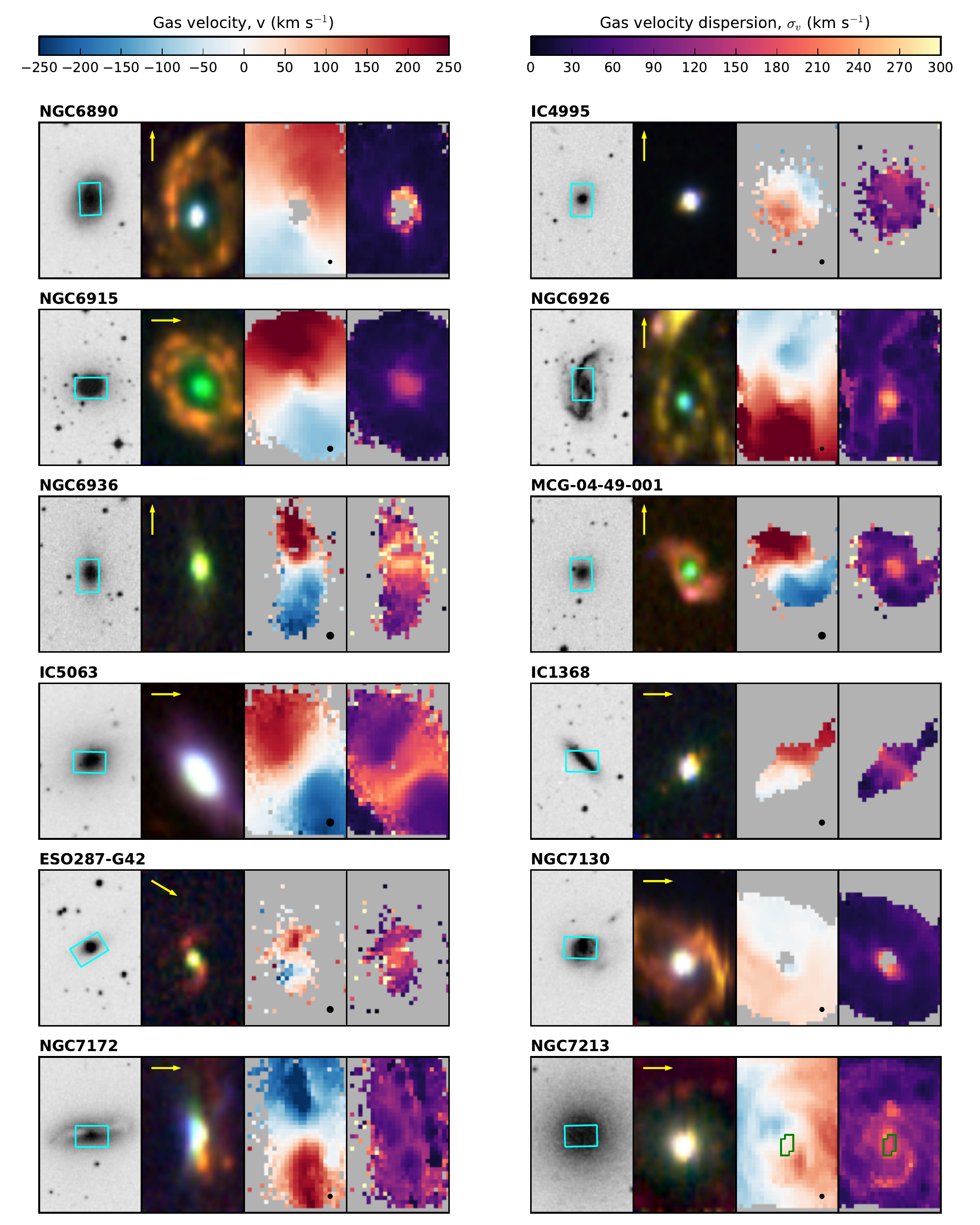}
	\end{centering}
	\caption{Maps as in Figure~\ref{fig:spatial_1}, for S7 galaxies with RAs between $20^\mathrm{h}18^\mathrm{m}$ and $22^\mathrm{h}10^\mathrm{m}$.}\label{fig:spatial_10}
\end{figure*}

\begin{figure*}
	\begin{centering}
		\includegraphics[scale=0.9]{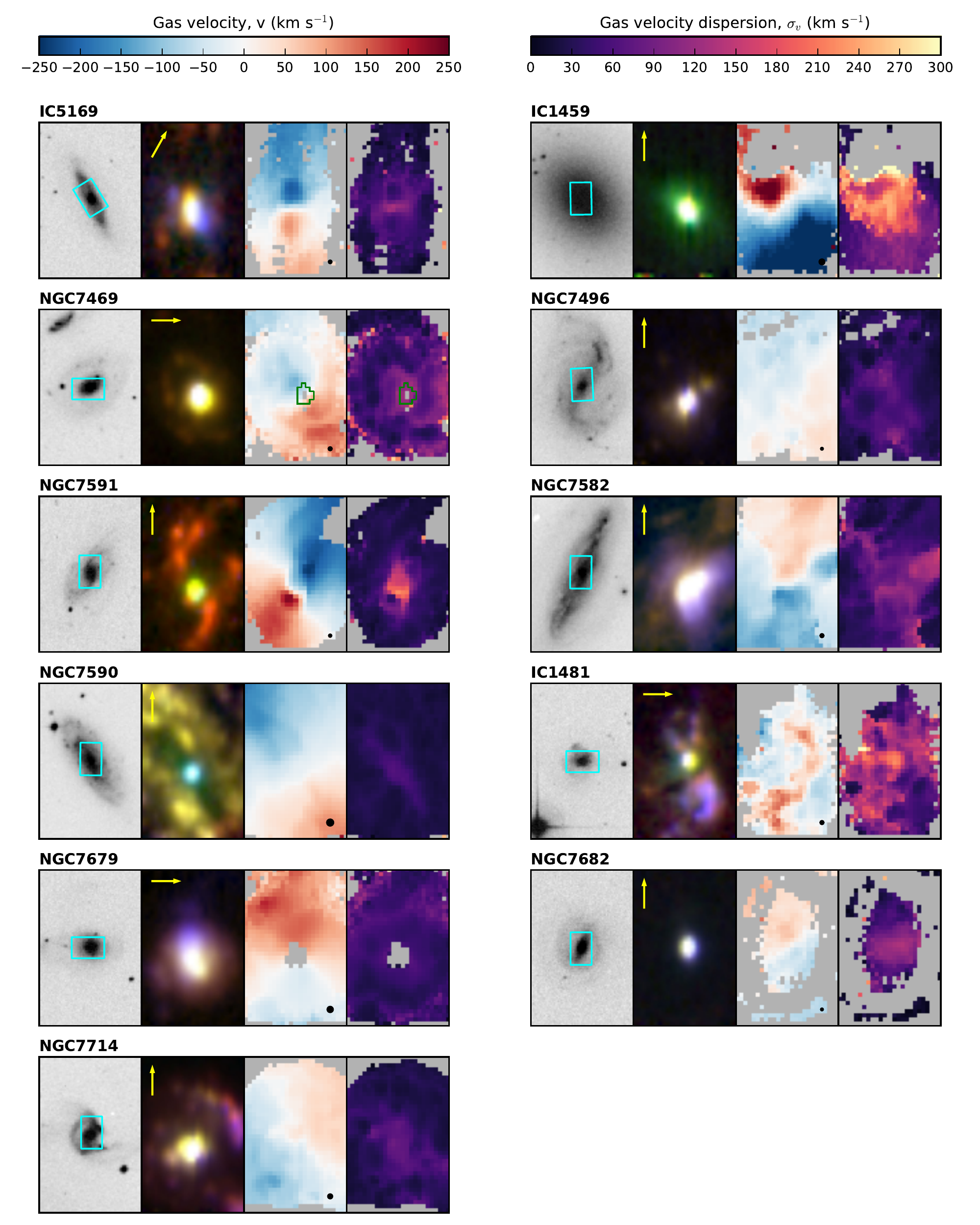}
	\end{centering}
	\caption{Maps as in Figure~\ref{fig:spatial_1}, for S7 galaxies with RAs between $22^\mathrm{h}10^\mathrm{m}$ and $23^\mathrm{h}59^\mathrm{m}$.}\label{fig:spatial_11}
\end{figure*}

\end{document}